\documentclass[nolinenumbers]{aastex631}

\shorttitle{Double radio sources with spiral host galaxies}
\shortauthors{Tate et al.}
\graphicspath{{./}{figures/}}

\begin{document}

\title{Here Be SDRAGNs - Spiral Galaxies Hosting Large Double Radio Sources}

\correspondingauthor{William Keel}
\email{wkeel@ua.edu}

\author{Jean Tate}
\altaffiliation{Deceased 2020 November 6. Jean Tate was among the most active
volunteers in Galaxy Zoo and Radio Galaxy Zoo, the one whose particular interest in SDRAGNs set the stage for these results. 
Jean's work set a high standard for the professional astronomers in the project, maintaining online
material in such good order that we were able to continue to retrieve and
understand even work left in progress. Much of the wording in the first two sections traces back to an SDRAGN draft Jean wrote before the HST observations were approved.}

\affiliation{Radio Galaxy Zoo}

\author[0000-0002-6131-9539]{William C. Keel}
\affiliation{Dept. of Physics and Astronomy, University of Alabama, Box 870324,Tuscaloosa, AL 35404, USA}

\author{Michael O'Keeffe}
\affiliation{Dept. of Physics and Astronomy, University of Alabama, Box 870324,Tuscaloosa, AL 35404, USA}
\affiliation{Current address: Social Learning Lab, Stanford University, Stanford, CA 94305}

\author{O. Ivy Wong}
\affiliation{CSIRO Space \& Astronomy, PO Box 1130, Bentley, WA 6102, Australia}
\affiliation{ICRAR-M468, University of Western Australia, Crawley, WA 6009, Australia}

\author{Heinz Andernach}
\affiliation{Th\"uringer Landessternwarte, Sternwarte 5, D-07778 Tautenburg, Germany}
\affiliation{Permanent address: Universidad de Guanajuato, Departamento de Astronomía, Callej\'on de Jalisco s/n. Guanajuato 36023, Mexico}

\author{Julie K. Banfield}
\affiliation{Research School of Astronomy and Astrophysics, Australian National University, Canberra, ACT 2611, Australia}

\author{Alexei Moiseev}
\affiliation{Special Astrophysical Observatory, Nizhny Arkhyz, Russia}

\author{Aleksandrina Smirnova}
\affiliation{Special Astrophysical Observatory, Nizhny Arkhyz, Russia}

\author{Arina Arshinova}
\affiliation{Special Astrophysical Observatory, Nizhny Arkhyz, Russia}

\author{Eugene Malygin}
\affiliation{Special Astrophysical Observatory, Nizhny Arkhyz, Russia}

\author{Elena Shablovinskaya}
\affiliation{Humboldt Research Fellow, Max Planck Institute for Radio Astronomy, Auf dem H\"ugel 69, D–53121, Bonn, Germany}

\author{Roman Uklein}
\affiliation{Special Astrophysical Observatory, Nizhny Arkhyz, Russia}

\author{Stanislav Shabala}
\affiliation{School of Mathematics and Physics, University of Tasmania, Private Bag 37, Hobart, TAS 7001, Australia}

\author{Ray Norris}
\affiliation{CSIRO Space \& Astronomy, PO Box 76, Epping, NSW 1710, Australia}
\affiliation{School of Science, Western Sydney University, Locked Bag 1797, Penrith, NSW 2751, Australia}

\author{Brooke D. Simmons}
\affiliation{Physics, Lancaster University, Lancaster LA1 4YB, UK}

\author{Rebecca Smethurst}
\affiliation{Department of Physics, University of Oxford, Denys Wilkinson Building, Keble Road, Oxford OX1 3RH, UK}

\author{Ivan Terentev}
\affiliation{Radio Galaxy Zoo}

\author{Chris Molloy}
\affiliation{Radio Galaxy Zoo}

\author{Victor Linares}
\affiliation{Radio Galaxy Zoo}




\begin{abstract}

We present a sample of large double radio sources hosted by spiral galaxies 
(Spiral Double Radio Active Galactic Nuclei, SDRAGNs). Candidates were selected during Radio Galaxy Zoo, 
and refined using Sloan Digital Sky Survey images. The most promising 
were targeted in  
the Zoo Gems Hubble Space Telescope program, yielding images for 36 
candidates. We assess the likelihood of each spiral galaxy being the genuine host of the radio emission finding 15 new 
high-probability SDRAGNs. 
SDRAGN hosts are seen preferentially close to edge-on. SDRAGNs 
predominantly show FR II radio structures and optical pseudobulges. Accounting for sample selection effects, 
the radio-jet axes lie preferentially near the poles of the galaxy 
disks; we find a constant probability distribution for intrinsic pole-jet angles $\phi <30^\circ$ 
ramping to zero at $\phi = 60^\circ$. We have obtained optical spectra for all these new SDRAGNs.  
Among both previous and new SDRAGN samples, 8/25 show Seyfert 2 signatures, 6/25 show central 
star formation, and 5/25 show LINER emission strong enough to indicate AGN or shock ionization, 
broadly similar to radio galaxies in elliptical hosts with the addition of star formation 
(diluting or masking weak AGN signatures). SDRAGNs include FR II sources seen at unusually low radio 
power, and preferentially occur in significant galaxy overdensities on 1-Mpc scales. 
Our ``false alarms" -- systems where HST data show the spiral to not be 
the actual host galaxy -- include radio sources seen through large parts of foreground spiral disks, 
potentially useful for Faraday-rotation studies of disk magnetic fields.

\end{abstract}



\section{Introduction} \label{sec:intro}

Giant, massive elliptical galaxies are by far the most common hosts of large-scale, double-lobed radio sources in the local Universe,
as shown in numerous studies from \cite{Matthews1964}  to \cite{Best2005}. Many of these nominally-elliptical hosts
show morphological features -- twisted dust lanes, shells, tails, boxy or twisted and non-concentric isophotes -- indicating recent 
interactions or merging events, which give plausible ways to channel material to the nucleus. This has been found statistically  (\citealt{Heckman86}, 
\citealt{RocheEales}) and
in detailed study of such individual systems as 3C 293 (\citealt{vBreugel84}, \citealt{Emonts}), 
3C 305 (\citealt{Sandage66}, \citealt{Heckman82}, \citealt{deKoff2000}, \citealt{Jackson03}), 
3C 40 (NGC 547) \citep{Fasano}, 3C 223.0, 3C 285 \citep{RocheEales}, and the numerous studies of Centaurus A (NGC 5128).
We follow \cite{Leahy} in calling these Double Radio Sources Associated with Galactic Nuclei (DRAGNs).

The strong contrast in host galaxies between active galactic nuclei (AGN) accompanied by large double radio sources and those without such sources,
though they may have core emission and jets on smaller scales (that is, within the galaxies), has focused attention on the handful of
spiral galaxies hosting large double sources. Beyond the way they stand out in surveys of host galaxies, they might share
some set of (rare?) properties which illuminates the conditions needed to produce such large radio jets and lobes.
By analogy with DRAGNs, we follow \cite{Mulcahy} and \cite{Mao0313} in calling spiral hosts of such double radio sources SDRAGNs.
This possibility motivated us to seek possible spiral hosts among radio sources identified in the Radio Galaxy Zoo project (RGZ, \citealt{RGZ}),
and to include some of the best candidates in a {\it Hubble Space Telescope} (HST) program of short-exposure imaging, to clarify these galaxies' structures. 
This paper reports the results, beginning with constructing a
candidate list from RGZ, winnowing the candidate list with ground-based imaging surveys, selection for HST imaging,
and the results of those HST observations and supporting optical spectroscopy.

We concentrate on galaxies whose radio emission extends well beyond the usual bounds of the galaxy's starlight. 
Small-scale double radio sources are seen around Seyfert nuclei in spiral galaxies  (\citealt{Nagar}, 
\citealt{KSDM}), but on scales much smaller than
typically associated with DRAGNs. Among the 12 Seyferts at $z<0.035$ showing such double sources from
data in \cite{Xanthopoulos}, \cite{Nagar}, \cite{Mutie}, and \cite{Pedlar}, the largest projected
size is 8.6 kpc on NGC 526A, with a median of 0.97 kpc. These sources often display bending of the jets
and arcuate structure to the lobes, both suggesting strong interaction with the dense interstellar medium of spiral hosts. 
These sources contrast with the sizes of typical DRAGNs, and with the previously-identified SDRAGNs. During the RGZ
search for SDRAGNs, it proved to be a useful heuristic, in the absence of redshifts for many host galaxies, to 
look for double radio sources projected outside the extent of the host galaxies as shown on the optical and near-IR images. 
The surface-brightness mapping for display is consistent across all images from each survey, and typical detection
radii for even the most luminous spirals seldom exceed 15 kpc. The gap between the largest double sources seen in Seyferts 
(8.6 kpc total extent end-to-end) and the smallest RGZ SDRAGNs (46 kpc) supports the value of this simple criterion.
It also eliminates  two-sided radio
 emission off the disk plane from outflows in spirals driven by either AGN or star formation (\citealt{DuricSeaquist}, 
 \citealt{Harnett}, 
 and \citealt{Gallimore}); the
 double sources we examine are on much larger scales and much more tightly collimated. This dichotomy has
 driven speculation that the dense cool interstellar medium (ISM) prevalent in the disks and nuclei of spirals can disrupt even
 relativistic jets before they pass from the ISM to the intergalactic or circumgalactic medium outside, where
 the jet interacting with these more rarefied surroundings can give rise to the large-scale lobes which are the 
 hallmarks of traditional radio galaxies.
 
Several broad possibilities might be envisioned for how (only) a few spirals produce double radio sources spanning hundreds 
of kpc, some discussed in previous work. They
could result from unusually large masses of the central supermassive black holes (SMBHs) for spirals, which
might accompany unusually luminous classical bulges. Another straightforward possibility is that SDRAGNs result
from fortuitous alignment of the jets with the poles of the host-galaxy disks, so they encounter minimal
column densities of dense ISM (a situation which might be more common in spirals lacking classical bulges and
less likely to have undergone major merges which could change the spin of the SMBH; 
\citealt{Beckmann}), which would yield galaxies with pseudobulges rather than classical bulges.
They could also be favored by particular density ranges in the circumgalactic medium, 
so observable radio lobes are produced when
jets interact with this material. This last possibility might be probed through the group or tidal environments of SDRAGNs. All these ideas can be confronted with 
detailed study of larger samples, such as we describe here.

As this manuscript was in preparation, an analysis involving much of the same HST data appeared by \cite{Wu2022}. We
will highlight comparison of our results wherever relevant, and adopt their GALFIT decompositions of the host-galaxy structures.
Our final list of high-probability SDRAGNs mostly overlaps with that of \cite{Wu2022}, but there are significant differences in our results
due to newer HST images, improved radio data including low-frequency surveys, availability of redshifts from optical spectra, and different morphological
assessments for 2 systems. We identify 15 high-probability SDRAGNs from the RGZ+HST sample.

\section{Sample selection and observations}

\subsection{Known SDRAGNs}\label{prior}

The term SDRAGN is not completely well-defined; generally it refers to a host galaxy with spiral optical morphology and a pair of radio lobes approximately symmetrically located on opposite sides of the host, and extending well beyond its optical boundary. In other words, we seek classical DRAGNs, but with a spiral galaxy as host. SDRAGN is an empirical
rather than physical definition; for example, hosts with double lobes oriented near the line of sight may have them overlap with the host emission and each other, so from our perspective they may be indistinguishable from radio emission entirely within the optical host, e.g. PKS 1814$-$637 \citep{Morganti} and PKS 1413+135 \citep{Lamer}. Among well-known SDRAGNs, the lobes may be completely detached (\citealt{Hota}) or connected to the core by a bridge of radio emission (e.g. J1649+2635, \citealt{Mao}). We seek
evidence of actual spiral structure; the number of lenticular (S0) hosts of large double radio sources is a separate question, perhaps less
complicated by the dense, cold interstellar material in spirals but often difficult to distinguish from ellipticals with extended stellar halos,
as in Fornax A (NGC 1316). Ambiguity remains in, for example, merger remnants, where tidal
features may mimic spiral arms, as seen in such systems as CGCG 292-057 \citep{cgcg292057} and NGC 5972 (\citealt{ngc5972}, reinforced in that case by AGN photoionization of gas in the tails). Similarly, dust lanes acquired during a merger may give a galaxy some features of
a disk (as in 3C 293 and 3C 305 shown by \citealt{deKoff2000}, where prominent dust lanes cross the central regions at steep angles to the major axis of the starlight distributions). For these reasons, we seek reasonably normal spiral patterns to be as certain as possible of the morphology of these systems. Two of the systems reported in the catalog-matching results of \cite{J0836paper} are normal spirals, while the other two show
tidal tails and loops indicating recent major mergers (even if a spiral was involved), so we do not consider the latter galaxies to be SDRAGNs here.

For comparison with the new RGZ sample we observed, data on the few previously identified SDRAGNs
are collected in Table \ref{table-previous}. In this case ``previous" includes objects reported in the literature (that is, SDRAGNs found outside this RGZ-HST program) while our analysis was in progress
up to 1 August 2025. Along with names used in the literature, we include a short
coordinate-based identifier of the same form (JHHMM+DDMM) used for the Zoo Gems objects. Where possible, we remeasured available data for consistency with the
new sample. For 0313$-$192 (J0315$-$1906, catalogued after the initial detection with J2000 coordinates as 2MASX J03155208$-$1906442 and WISE J031552.09$-$190644.2, 
{\emph using
catalogs from the 2-Micron All-Sky Survey Extended Source listing described by \citealt{2MASX} and Wide-field Infrared Survey Explorer as reported by \citealt{WISE}}), we retrieved 
new versions of the HST images from \cite{HST0313} to take
advantage of improved astrometric accuracy (section \ref{sec-astrometry}), and use the same VLA data reported by
\cite{Ledlow}. We also retrieved the archival HST images obtained by \cite{Bagchi2024} for J2345$-$0449 (2MASX J23453268$-$0449256)
and by \cite{Capetti} for J0725+2957 (B2 0722+30).
For the other objects, we use optical FITS images from the Legacy Surveys \citep{LegacySurveys}, using the $g$ band for closest comparison with the HST F475W data, and radio data from the
surveys listed in Section \ref{radiosurveys}. We obtained new images for J2318+4314 (MCG +07-47-10) which is not covered by current deep optical surveys, as described in Section \ref{JKTdata}). For some of the previous SDRAGNs, the lobe structure is not well detected in the
VLA Sky Survey (VLASS) data \citep{VLASS}, so we adopt radio properties from the discovery papers, some measured at lower frequencies than the VLA surveys. Implicitly, this suggests that the new sample
from RGZ has lobe emission which is more prominent than the previous SDRAGNs as a group, at least as filtered through the VLASS surface-brightness
and spatial-frequency limits. It also illustrates that increased surface-brightness sensitivity and longer-wavelength 
observations will show more of these objects, since the lobes tend to have steep spectra and be better detected 
at frequencies lower than the 3 GHz VLA survey band. Notable examples are 2MASX J23453268$-$0449256, first identified 
as a giant radio galaxy with a spiral host by \cite{Andernach2012}, the even larger source surrounding NGC 6185 (J1633+3520; \citealt{Oei}), for which the lobes are undetected in the VLASS data, and 
J0408-6247 (NGC 1534), where the lobes were first detected at 185 MHz 
\citep{HurleyWalker}. 

Optical spectra for 
the three galaxies with designations from the Sloan Digital Sky Survey (SDSS) were retrieved from the SDSS DR15 SkyServer, and for J0354-1340, 
J1350$-$1634, 
and J2345$-$0449, from the 6dFGS archive (\citealt{Jones2004}, \citealt{Jones2009}). We recovered the optical spectrum of J0315$-$1906 from Fig. 7 
of \cite{Ledlow1998}, using the {\tt automeris} web tool\footnote{https://automeris.io}. For J1633+3520 (NGC 6185), we obtained the summed spectra from \cite{Jansen2000}. 
For J0725+2957 (B2 0722+30), we use the William Herschel Telescope (WHT) long-slit data from \cite{Emonts}. We obtained a new optical spectrum of J2318+4314 (MCG +07-47-10; Section 2.6). 
We do not have an optical spectrum of J0408$-$6247 (NGC 1534).


Double radio sources are often characterized both by projected peak-to-peak lobe separations (Sep) and largest (detected) angular size (LAS) or its equivalent in linear units the largest linear size (LLS), the latter value being more sensitive to very extended emission but often dependent on the observational sensitivity for lobes without distinct terminating hot spots, an issue which is particularly 
important for sources of Fanaroff-Riley (FR, \citealt{FanaroffRiley}) class I.
Because of the wide span of dynamic range in the radio data we use, we favor comparisons using primarily the lobe separation rather than LLS, while recognizing that some of our sources show significant emission beyond the lobe peaks. Both
lobe separations and LAS are transformed into projected linear units, where redshifts are known, using a ``consensus" cosmology 
with ${\rm H}_0=70$ km s$^{-1}$ Mpc$^{-1}$, $\Omega_M=0.286$, and flat geometry; distance calculations used the
Javascript tool of \cite{CosmoCalc}. Instead of the angular LAS, we tabulate largest linear size (LLS) for the projected extent 
transformed to linear units (kpc). For some of the previously-identified objects where the discovery papers listed
only LAS or LLS values, we estimate the lobe separations using published radio maps in the references listed. 

Table \ref{table-previous} includes our derived viewing geometries - major-axis position angle and inclination from
edge-on - for  the previously reported SDRAGNs. Throughout this paper, we use this convention for disk inclination measuring from edge-on,
so the error behavior is more intuitive (smaller errors near zero) and the analysis for disk/jet angles likewise becomes more transparent. 
When dust in the disk is detected, we consider the outer edge of the
dust lane to outline a circle in the disk plane for
one determination, and the offset from the centerline of the dust lane to center of the bulge isophotes as another measure. 
Otherwise, we use isophotes of the disk and outer regions, correcting for finite disk thickness as in, for example,
\cite{Tully} with the listed uncertainties including a range of values for the ratio $r_0$ between disk thickness and radius from 0.1--0.2. 
Three systems are so nearly face-on
that we cannot rely on the outer isophotes or resonance rings to be circular at the requisite level, so the listed values
reflect these uncertainties. In J1649+2635 and J1633+3520 (NGC 6185), the derived inclination ranges incorporate the criterion that spiral arms be monotonic in
radius versus angle, without requiring that they be logarithmic. We also list Hubble types which we derived by inspection of the
HST or Legacy Survey images, for uniformity with the new sample. J0725+2957 has FR I morphology (so the lobe separation is poorly defined), and is
substantially smaller in LLS than the others. Its radio structure does appear better collimated than starburst-wind sources, so we include it among SDRAGNs here.

We attempt to distinguish genuine spirals from galaxies having prominent stellar disks without spiral structure (S0 or lenticular systems). 
Some hosts of large-scale double sources are found to be S0s {such as NGC 612, \citealt{VCV2001}, PKS 1814$-$627, \citealt{Morganti}, and ESO 514-G03,  \citealt{GopalKrishna}) but since their ISM content is so different from
spirals, they are outside the scope of this project. Based on the results of \cite{Vietri}, it is likely that additional NLSy1 galaxies
will prove to host SDRAGNs; they are routinely catalogued now at redshifts $z>0.4$, too distant for typical ground-based images to show 
spiral structure, and a significant fraction of them have radio jets, some aligned with large-scale double sources \citep{Rakshit}.

Some of the radio lobes in these previously-identified SDRAGNs are not detected in VLASS data, especially those
discovered using lower-frequency data such as from the Giant Metrewave Radio Telescope (GMRT). This hints that lower-frequency data will
be fruitful in finding more (as is true for traditional DRAGN sources as well). For example, Speca shows potential ``relic lobes" spanning nearly a Mpc in projection when observed at 325 MHz \citep{Hota}, 
while J1350$-$1634 extends beyond 2 Mpc \citep{Sethi}. 


\begin{deluxetable*}{llccccccccl} 
\tablecaption{Previously identified SDRAGNs\label{table-previous}}
 \tablehead{
\colhead{Name} & \colhead{Identification}  &  \colhead{$z$} & \colhead{Sep} & \colhead{LLS} & \colhead{Optical spectrum} & \colhead{PA$^\circ$} & 
\colhead{Inclination$^\circ$} & \colhead{Type} & \colhead{References}}
\startdata
J0315$-$1906 & 2MASX J03155208$-$1906442 & 0.0677 & 126 &  350 &  Sy 2 & $87 \pm 1$ & $ 0.0 \pm 0.5$ & Sb &  1,2  \\ 
J0354$-$1340 & 6dFGS gJ035432.8$-$134008 & 0.0772 &  172 & 177 & NLSy1 & $ 90 \pm 20$ & $80 \pm 10$ & (R)SBa & 3 \\
J0408$-$6247 & NGC 1534 &        0.0178   &  755 & 755 & \nodata & $ 79\pm 3$  & $21\pm 2$  & Sa & 4  \\ 
J0725+2957 & B2 0722+30 & 0.0197 & 2.5 & 12.5 & LINER & $142 \pm 2$ & $9.8 \pm 1.5$ & Sc & 5, 6\\
J0836+0532 & SDSS J083655.86+053242.0 & 0.0993 & 280 & 280 &  Sy 2  & $ 70 \pm 30$ & $80\pm 10 $ & Sc(s) &7 \\ 
J1350$-$1634 & LEDA 896325 & 0.0877 & 517 & 832 & Sy 1.8? & 161 & $27.5 \pm 2.0$ & Sb & 8 \\
J1409$-$0302 & SDSS J140948.85$-$030232.5 (Speca) & 0.1376 & 312 &1300 & LINER  & $21\pm 2$ & $22\pm 2$ & Sab & 9\\ 
J1633+3520 & NGC 6185             & 0.0343      &  1844 & 2450  & LINER & $4\pm 4$ & $36\pm 3$& Sa & 10 \\
J1649+2635 & SDSS J164924.01+263502.5 & 0.0545 & 60 & 85 &  LINER  & $120 \pm 12$ & $57 \pm 5$ & Sc & 11, 12\\ 
J2318+4314 & MCG +07-47-10                 &  0.0169    &  168  & 207 & SF  &   $235 \pm 5$  &   ($80\pm10$)   &  SBc & 13 \\ 
J2345$-$0449 &  2MASX J23453268$-$0449256 & 0.0755 &  1320 & 1600 & LINER  & $98 \pm 3$ & $24\pm 2$ & Sb & 14, 15 \\
\enddata
\tablerefs{1. \cite{Ledlow}. 2. \cite{HST0313}. 3.  \cite{Vietri}.  4.  \cite{HurleyWalker}.  5.  \cite{Capetti}. 6.  \cite{Emonts}.  
7.  \cite{J0836paper}. 8. \cite{Sethi}  9. \cite{Hota}. 10.   \cite{Oei}.  11. \citealt{Mao}. 12. \citealt{J0836paper}. 13.   \cite{Mulcahy} .
14.  \cite{Bagchi}. 15. \cite{Bagchi2024}.}
\tablecomments{``Sep" denotes the peak-to-peak separation of radio lobes, projected on the sky, in kpc. LLS (largest linear size)
is the largest projected size from available radio data, also in kpc, omitting relic or outer double structure detected at low frequencies. 
The column ``Optical spectrum" shows the emission-line classification of each nucleus where SF indicates emission lines 
from star formation and NLSy1 is a narrow-line Seyfert 1. BTA spectroscopic data 
for J2318+4314 (MCG +07-47-010) are new, described in section \ref{newspectroscopy}; SDSS data were retrieved from the Sky Server, 
and all other spectra are described in the text. Position angle PA of the galaxy major axis is in degrees from north through east. Disk inclination is in degrees from edge-on, so 90$^\circ$ is face-on. Type is our estimate of each galaxy's Hubble morphological type.}
\end{deluxetable*}


\subsection{Radio Galaxy Zoo selection of potential spiral host galaxies}

The original RGZ (\citealt{RGZ}, \citealt{RGZDR1}) selection of host galaxies used the Faint Images of the Radio Sky at Twenty-cm
(FIRST; \citealt{FIRST}) survey along with Australia Telescope Large Area Survey (ATLAS; \citealt{ATLAS1}, \citealt{ATLAS2}) data, with host-galaxy
identifications starting from {\it WISE} or {\it Spitzer} data complemented by SDSS images. In detail,
the web interface presented volunteers with a $3 \arcmin \times 3\arcmin$ field combining data from 3.4$\mu$m WISE, FIRST images, and radio contours. A slider allowed each field to be viewed as a combined WISE/radio image, from radio (FIRST image plus contours) only to WISE only. Once a volunteer finished the classification task, an option allowed discussion of the field, in the discussion forum\footnote{https://radiotalk.galaxyzoo.org, content frozen as of May 2019.}. Each RadioTalk page for a field includes links to larger ($9\arcmin  \times 9\arcmin$) FIRST and WISE images, images from NVSS \citep{NVSS}, and optical observations from SDSS Data Releases 10 \citep{SDSS-DR10},  12 \citep{SDSS-DR12}, and 13 
\citep{SDSS-DR13}. There is also a 140-character Comment box. RadioTalk includes a set of discussion boards where more extended commentary on individual sources and classes of sources took place. 
Selection of SDRAGN candidates used only sky area covered by FIRST data.

A group of RGZ volunteers specifically noted potentially spiral hosts and flagged them for discussion on the project
forum. Experience eventually showed that  the SDSS {\it fracdev} parameter, the fraction of galaxy light in a de Vaucouleurs-profile bulge, was a useful quantitative way to evaluate candidates as well. At the project's end, there were 215
SDRAGN candidates, initially divided into three broad likelihood categories, and a few
additional ones from literature sources. These are all listed in Table \ref{tbl-candidates} in Appendix \ref{AppendixC}, to document the starting sample,
preserve the initial priority categories, and tabulate quantities important in understanding the selection of both
the overall sample of RGZ SDRAGN candidates and those targeted for HST imaging.

\subsection{Radio data}\label{radiosurveys}		
Our analysis in this work uses radio data from the Karl G. Jansky Very Large Array (VLA), specifically the
FIRST survey and ongoing VLA Sky Survey (VLASS; \citealt{VLASS});
the Low-Frequency Array (LOFAR; \citealt{vanHaarlem}), via the LOFAR Two-meter Sky Survey
(LoTSS; \citealt{LOTSS1}, \citealt{LOTSS2}, \citealt{LOTSS3}); and Australian Square-Kilometre Array Precursor (ASKAP), 
using the Rapid ASKAP Continuum Survey (RACS, \citealt{RACS}) near 900 MHz.
These are complementary in angular resolution, sensitivity,
frequency, and spatial frequencies sampled. For the VLASS where observations continue, we used the ``QuickLook" median
of observations in epochs 1-3\footnote{Available from   http://archive-new.nrao.edu/vlass/HiPS/MedianStack/Quicklook}, which mitigates some artifacts compared to single-epoch products.

\subsection{Zoo Gems target selection and observations}
As described in \cite{ZooGems}, the list of SDRAGN candidates from RGZ (Appendix B) was winnowed for inclusion in the
Zoo Gems program of short-exposure HST images, by having members of the public vote on which objects from a list
(already screened for scientific suitability) should be carried forward into a limited target list for HST observations. At this
stage, participants had the option of blinking between the optical SDSS image of a field and a version with radio contours
overlaid. All potential SDRAGNs had been examined by the first author, to confirm evidence of disk structure in SDSS images
before inclusion in the RGZ sample.

Table \ref{tbl-candidates} in Appendix \ref{AppendixC} lists the whole sample of SDRAGN candidates from RGZ, 
which ones were selected for the
HST target list or actually observed, and the magnitude and axial-ratio quantities we considered for observational selection.	
The HST observations themselves formed part of a pool to be scheduled as fillers chosen for sky location between higher-priority observations, so they were taken from the overall 
Zoo Gems list essentially at random. As fillers,
these observations were short - pairs of 337-second exposures with a small dither offset in between. Among the 299
Zoo Gems input targets, 66 were SDRAGN candidates, of which 36 were observed and 30 were not.

As required for the gap-filler programs, these HST images used the Advanced Camera for Surveys (ACS;
\citealt{Ford1998}, \citealt{Ryon}).
To minimize effects of deteriorating charge-transfer efficiency (CTE) in the ACS images, the target galaxies were located near one corner of the
instrument field (29-40\arcsec \  from the nearer edge of the two dithered exposures, and 46\arcsec \ for the nearby system UGC 1797). As a result, one radio lobe often falls outside the HST field of view, as reflected in some of our image overlays. For
the image versions we show here, the numerous cosmic-ray residuals remaining after pipeline combination of the two offset
images for each object were patched interactively in regions around the target galaxy, by interpolation from surrounding pixel values
using the IRAF {\it imedit} task.

We chose the F475W filter, similar to the SDSS $g$ passband, to maximize the contrast of any (usually blue) spiral arms in the
galaxies. For larger redshifts, F555W might have given similar contrast and better detector efficiency, but fewer than half
the sample had known redshifts at the outset.

Table \ref{table-HSTdata} lists the 36 candidate SDRAGN galaxies observed by HST\footnote{Available at \dataset[doi:  10.17909/rvae-bn13]{\doi{ 10.17909/rvae-bn13}}}, 
now that the Zoo Gems program has completed, along with 
spectroscopic redshifts where they are known. Optical coordinates listed are from the HST images, for the nuclei of
possible SDRAGN hosts (Section \ref{sec-astrometry}). The precision of these coordinates is 0.02\arcsec for bright symmetric bulges, while for
obscured nuclei located by symmetry of surrounding dust-free isophotes, the precision may be as poor as 0.2\arcsec. Table \ref{table-crossids} lists alternate 
identifications, including the
RGZ internal designations used in online discussions, and the short names we use for convenience.
Among these galaxies, four (J0016+0226, J1516+0517, J1457+2832, and J613+3018) were observed recently and were not included by \cite{Wu2022}. Our morphological
classification of J0958+5619 differs from that by \cite{Wu2022}, who assign it as a spiral. Based on
 the dust lane cutting across the
galaxy core nearly perpendicular to the possible disk, we consider it more likely that this is the late aftermath of a merger with clumpy debris in the outer regions, some of which is highly ionized and affects the structure in the HST image through the passband's inclusion of [O II] emission (Appendix \ref{AppendixEELR}).

		
\begin{deluxetable*}{llcccc} 
\tablecaption{HST observations\label{table-HSTdata}}
\tablehead{
\colhead{Data set}       & \colhead{HST Object Name}  & \colhead{$\alpha_{2000}$} & \colhead{$\delta_{2000}$} &\colhead{Date observed} & \colhead{$z$}}  
\startdata
JDS45K010 & SDSS J001627.47+022602.1 & 00 16 27.476	 & +02 26 02.25	& 2022-09-23 & 0.0583 \\ 
JDS44T010 & SDSS J020904.75+075004.5 & 02 09 04.755 & +07 50 04.53 & 2021-10-22 & 0.2552 \\ 
JDS43Y010  & UGC 1797                            & 02 19 58.732 & +01 55 49.16 & 2018-07-03 & 0.0410  \\ 
JDS44I010 & SDSS J080259.73+115709.7	 & 08 02 59.738 & +11 57 09.70 & 2021-01-10 & -----	 \\ 
JDS44J010 & SDSS J080658.46+062453.4 & 08 06 58.476 & +06 24 53.15 & 2020-05-28  & 0.0834 \\ 
JDS45G010 & SDSS J081303.10+552050.7 & 08 13 03.125 & +55 20 50.66 & 2019-12-25 & 0.2645 \\ 
JDS44G010 & SDSS J082312.91+033301.3 & 08 23 12.915 & +03 33 01.39 & 2021-03-15 & 0.0897 \\ 
JDS44K010 & SDSS J083224.82+184855.4 & 08 32 24.828 & +18 48 55.45 & 2021-09-28 & 0.1138 \\ 
JDS45B010 & SDSS J083351.28+045745.4 & 08 33 51.292 & +04 57 45.35 & 2021-03-18 & ------ \\ 
JDS45I010 & SDSS J084759.90+124159.3 & 08 47 59.901  & +12 41 58.96 & 2021-10-02 & 0.1745 \\ 
JDS41L010 & B3 0852+422                         & 08 55 49.153 & +42 04 20.10 & 2021-03-14 & 0.1804 \\ 
JDS45W010 & SDSS J090147.17+164851.3 & 09 01 47.173 & +16 48 51.32 & 2021-11-13 & ----- \\ 
JDS45L010 & SDSS J090305.84+432820.4 & 09 03 05.861 & +43 28 20.41 & 2021-03-15 & 0.3736 \\ 
JDS43Z010 & B3 0911+418                         & 09 14 45.557 & +41 37 14.04 & 2021-12-29 & 0.1404 \\ 
JDS45H010 & SDSS J091949.07+135910.7 & 09 19 49.085 & +13 59 10.78 & 2018-05-15 & ------ \\ 
JDS47L010 & SDSS J092605.17+465233.9 & 09 26 05.183 & +46 52 33.68 & 2021-12-29 & 0.2181 \\ 
JDS44R010 & B2 0938+31A                       & 09 41 03.617 & +31 26 18.56 & 2020-02-13 & 0.3940 \\ 
JDS47H010  & SDSS J095605.87+162829.9	  & 09 56 05.861 & +16 28 30.13 & 2020-02-14 & 0.2745 \\ 
JDS44P010 & SDSS J095833.44+561937.8 & 09 58 33.433 & +56 19 37.80 & 2020-10-17 & 0.2422 \\ 
JDS45T010 & SDSS J112811.63+241746.9 & 11 28 11.617 & +24 17 46.82 & 2019-02-23 & 0.1153 \\ 
JDS47K010 & SDSS J113648.57+125239.7 & 11 36 48.578  & +12 52 39.48 & 2021-03-14 & 0.0345 \\ 
JDS45E010 & SDSS J130300.80+511954.7 & 13 03 00.803 & +51 19 54.70 & 2021-10-01 & 0.1164 \\ 
JDS45A010 & IC 4234                                 & 13 22 59.871 & +27 06 59.00 & 2020-04-09 & 0.0344 \\ 
JDS44X010 & SDSS J132809.31+571023.3 & 13 28 09.247 & +57 10 23.97 & 2019-05-14 & 0.0211 \\ 
JDS47J010 & SDSS J134900.13+454256.5 & 13 49 00.137 & +45 42 56.41 & 2019-11-13 & ------\\ 
JDS45Z010 & B3 1352+471                     &  13 54 36.019 & +46 57 01.26 & 2019-04-28 & ------ \\ 
JDS44A010 & 4C +28.38	                           & 14 57 53.810 & +28 32 18.77 & 2023-08-25 & 0.1440 \\ 
JDS44D010 & SDSS J150903.21+515247.9 & 15 09 03.224 & +51 52 47.94 & 2020-01-14 & 0.5789 \\ 
JDS43X010 & SDSS 151659.24+051751.5 & 15 16 59.245 & +05 17 51.58 & 2023-08-25 	& 0.0512 \\
JDS44H010 & B2 1611+30			    & 16 13 58.621 & +30 18 09.51 & 2023-09-25 	& 0.1522 \\
JDS45J010 & SDSS J163300.85+084736.4 & 16 33 00.862 & +08 47 36.62 & 2019-07-11 & 0.2247 \\ 
JDS45F010 & SDSS J163624.97+243230.8 & 16 36 24.984 & +24 32 30.78 & 2021-06-05 & 0.1016 \\ 
JDS44Z010 & B2 1644+38                       &  16 46 28.433 & +38 31 15.73 & 2019-07-19 & 0.1075 \\ 
JDS44C010 & SDSS J16562058+6407529	 & 16 56 20.685 & +64 07 52.90 & 2018-08-24 & 0.2121 \\ 
JDS43V010 & SDSS J172107.89+262432.1 & 17 21 07.917 & +26 24 31.93 & 2019-08-22 & 0.1696 \\ 
JDS45V010 & SDSS J214110.61+082132.6 & 21 41 10.612 & +08 21 32.54 & 2019-12-11 & 0.3946 \\ 
\enddata
\end{deluxetable*}

\subsection{Geometric selection effects}\label{selectioneffects}

Selection effects may enter the input sample of candidate SDRAGNs involving both optical and
radio properties. In optical images, spiral arms are more easily recognized when a galaxy
is seen nearly face-on, while the SDSS exponential-disk fitting becomes more sensitive
when the disk is nearly edge-on to separate it from a rounder bulge component. The radio
structure may not be recognized as a double if the two lobes are projected so close together
that they blend in available data, and similarly, if seen too close to the radio axis we
might not recognize a large-scale double source as extending beyond the host galaxy.

The combined set of observed and targeted Zoo Gems SDRAGN candidates implicitly carries the
geometric and magnitude selection factors operating both in selection as an RGZ candidate SDRAGN, and for
a galaxy's inclusion in the HST target list.
We examine the optical selection using the set of all galaxies considered for inclusion 
in the Zoo Gems SDRAGN category (213 in the SDSS sky coverage) as well as the set of these 
actually voted in to the
target list (66, of which 64 are within the SDSS). We extracted SDSS ``model magnitudes" in $r$ 
along with the axial ratio $b/a$ (SDSS {\it PhotObj} parameter {\it expAB\_r}) 
of an exponential-disk fit and the nominal bulge fraction {\it fracdev\_r}, using the measured
axial ratio rather than the more physical inclination angle since the axial ratio is available for
all candidates. We collect similar data for the heterogeneously selected group of previous SDRAGNs
as well, using SDSS or Legacy Survey $r$ magnitudes where available and measuring new
axial ratios when necessary. The SDSS axial ratios are good
representations of the structures seen in HST images except for the case of J1721+2624, where the SDSS fit does not
capture outer spiral structure tracing an ellipse with axial ratio 0.51.

As a comparison, we take the empirical distribution of axial ratios for spirals in the SDSS
presented by \cite{PadillaStrauss}. For this entire set of 213 candidates, the
axial-ratio distribution is consistent with that from \cite{PadillaStrauss}, but there are
significant differences with galaxy magnitude (Figure \ref{fig-magtilt}). Galaxies brighter than $r=16$ have mean $b/a=0.69$, 
while for $r > 20$ the mean drops to 0.41 with monotonic behavior among 2-magnitude bins. 
In particular there is a relative deficit of bright and edge-on candidate galaxies (or
an excess of bright and face-on objects). The nine
previously-identified SDRAGNs are all brighter than $r=16.1$, and have mean $b/a = 0.57$, median $0.45$.

One might see these statistics, and the lack of highly-inclined bright and nearby galaxies in Figure \ref{fig-magtilt},
as evidence that each of the obvious selection effects above is operating
for galaxies at opposite ends of the magnitude range we sample. It therefore makes sense to 
examine the relation between orientations of radio sources and galaxy disks taking account of
this external selection; the bright objects are preferentially selected in disk orientations which are
least informative about the disk/jet angle, while the faintest galaxies carry the most information by 
reducing the range of one angular variable. We incorporate these factors in our analysis of the
orientations of the radio-source axes with respect to the galaxy disks 
in section \ref{radioproperties}.

\begin{figure*}
\includegraphics[width=120.mm,angle=90]{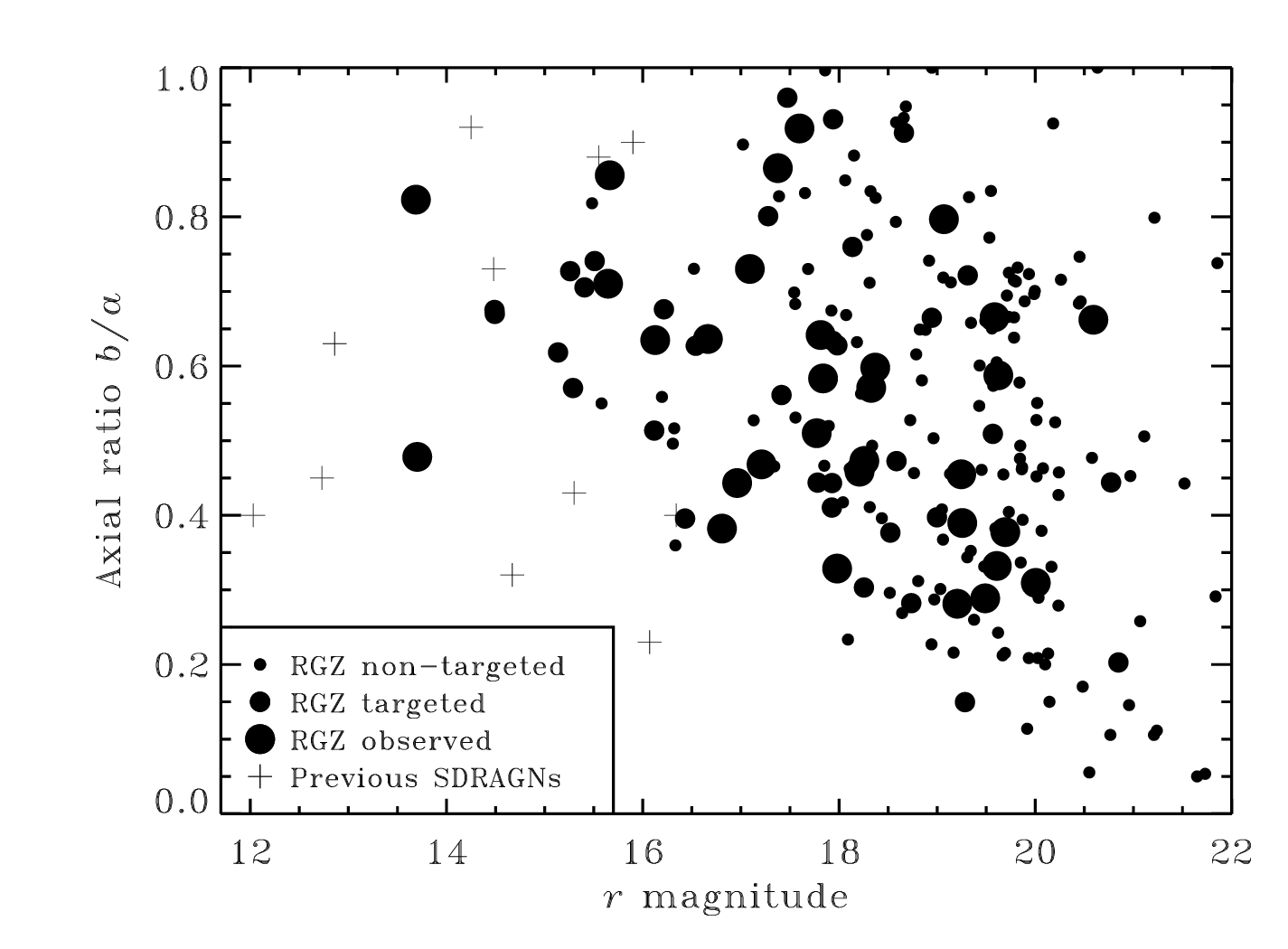}
\caption{Magnitude in $r$ versus SDSS axial ratio {\it b/a} for subgroups of the SDRAGN sample,
illustrating selection effects with apparent magnitude. Small filled circles are galaxies in the sample
from which the observing list was drawn but not selected for the HST target list, intermediate filled circles are galaxies in the HST target list
but not observed, and large filled circles represent the objects observed and analyzed here. The eleven 
previously-identified SDRAGN systems are represented by plus signs.
The brightest RGZ galaxies (12/14 of which are in the HST target sample) are seen systematically more face-on
than a random set of SDSS disk systems, while the faintest ones are systematically more edge-on.}
\label{fig-magtilt}
\end{figure*}

\subsection{Optical spectroscopy}\label{newspectroscopy}

The most promising SDRAGN candidates without previously published redshifts, including one from the previous list in Table \ref{table-previous},
have been observed spectroscopically using the SCORPIO-1 (\citealt{SCORPIO-1}, used for J0855+4204) or SCORPIO-2 \citep{SCORPIO-2} systems in their long-slit modes,
at the prime focus of the 6-meter telescope (BTA) of the Special Astrophysical Observatory, Russian Academy of Sciences. These
were primarily in a backup program for poor weather conditions, so redshifts and emission-line equivalent widths and ratios
could be measured but not always absolute flux values. The resulting redshifts are included in Table \ref{table-HSTdata}. These new
spectroscopic observations are listed in Table \ref{table-btaspectra}, including the position angle of the slit (PA, north through east) and emission-line classification of the nucleus, as in Section \ref{spectra} (where we use SF for emission lines indicating star formation, composite to denote objects in the uncertainty range of the SF/AGN boundary, and Sy 1.9 to indicate a Seyfert narrow-line spectrum with a broad-line component detected at H$\alpha$). Among these galaxies, J0958+5619, while not a spiral as shown by the HST image, shows extended emission-line regions, 
including He II off the nucleus,
spanning 22\arcsec (85 kpc) end-to-end. We document these features and their kinematics in Appendix \ref{AppendixEELR}. We now
have spectra and redshift values for all the 
new SDRAGNs we identify in sections \ref{sec-cored} and \ref{nocores}.

\begin{deluxetable*}{lccccc} 
\tablecaption{BTA spectroscopy of nuclei\label{table-btaspectra}}
\tablehead{
\colhead{Object} & \colhead{Date} & \colhead{Exposure (s)} & \colhead{Nucleus} & \colhead{$z$} & \colhead{PA$^\circ$}}
\startdata
J0209+0750 & 2022-09-19 & 2400 & Sy 2 & 0.2552 & 122 \\ 
J0219+0155 & 2023-07-19  & 1200 & LINER &  0.0410 & 0 \\ 
J0806+0624  & 2023-03-14 &1800 & SF & 0.0834 & 145 \\
J0813+5520 & 2022-12-30 & 3600 & SF & 0.2645 & 97 \\ 
J0855+4204  & 2024-08-11 & 3600 & Quiescent & 0.1804 & 177 \\ 
J0926+4652 & 2022-06-03  & 3000 & Sy 2 & 0.2181 & 161 \\
J0941+3126 & 2022-02-07  & 1800 &  Sy 2? & 0.3940 & 44 \\ 
J0956+1628 & 2022-02-07 & 1800 &  Sy 2 & 0.2745 &  121 \\ 
J0958+5619 & 2016-02-15  & 1800 & Sy 2 & 0.2422 & 90 \\ 
J1128+2417 & 2023-01-18   & 2400 &  H II & 0.1153  & 139 \\ 
J1303+5119 & 2023-01-16  & 3600 &  H II? & 0.1164 & 60 \\
J1328+5710 & 2023-01-20 & 1800 &  H II & 0.0211 & 114 \\ 
J1613+3018  & 2023-09-18  & 1200 & H II & 0.1522 & 60  \\ 
J1633+0847 & 2022-03-09 & 1800 &  Sy 1.9 & 0.2247  & 134 \\
J1636+2432 & 2023-07-18  & 4500 & SF & 0.1016 & 40 \\
J1656+6407 & 2023-03-27 & 3600 & SF & 0.2121 & 60 \\ 
J2141+0821 & 2022-11-30 & 1200 &  Sy 2? & 0.3946 & 110 \\
J2318+4314  & 2022-10-22 & 1500 & SF & 0.0169 & 139 \\ 
\enddata
\tablecomments{``Nucleus" indicates the emission-line classification of the galaxy nucleus, as in section \ref{spectra}. PA gives the position angle of the spectrograph slit, north through east, in degrees.}
\end{deluxetable*}

\subsection{Ground-based imaging}\label{JKTdata}

Among the previously-reported SDRAGNs in Table \ref{table-previous}, J2318+4314 (MCG +07-47-010) lies outside the region imaged by either the SDSS or Legacy Surveys. For a more detailed view of the host-galaxy
structure, we obtained new $B$ and $R$ images using the 1m Jacobus Kapteyn Telescope (JKT) on the island of La Palma,
remotely operated as part of the SARA Observatories \citep{Keel-SARA}. From observations on 4 June 2022, we stacked $6 \times 300$-s exposures in B and $12 \times 180$-s exposures in R. Image quality in stacked sums was FWHM=1.2\arcsec ($R$) and 1.3\arcsec ($B$). These images are shown in Fig. \ref{fig-MCG74710}, illustrating the
prominent grand-design spiral pattern, asymmetric outer structure.of the system and an inner bar reaching to radius 1.6\arcsec (0.5 kpc).

\begin{figure*}
\includegraphics[width=120.mm,angle=90]{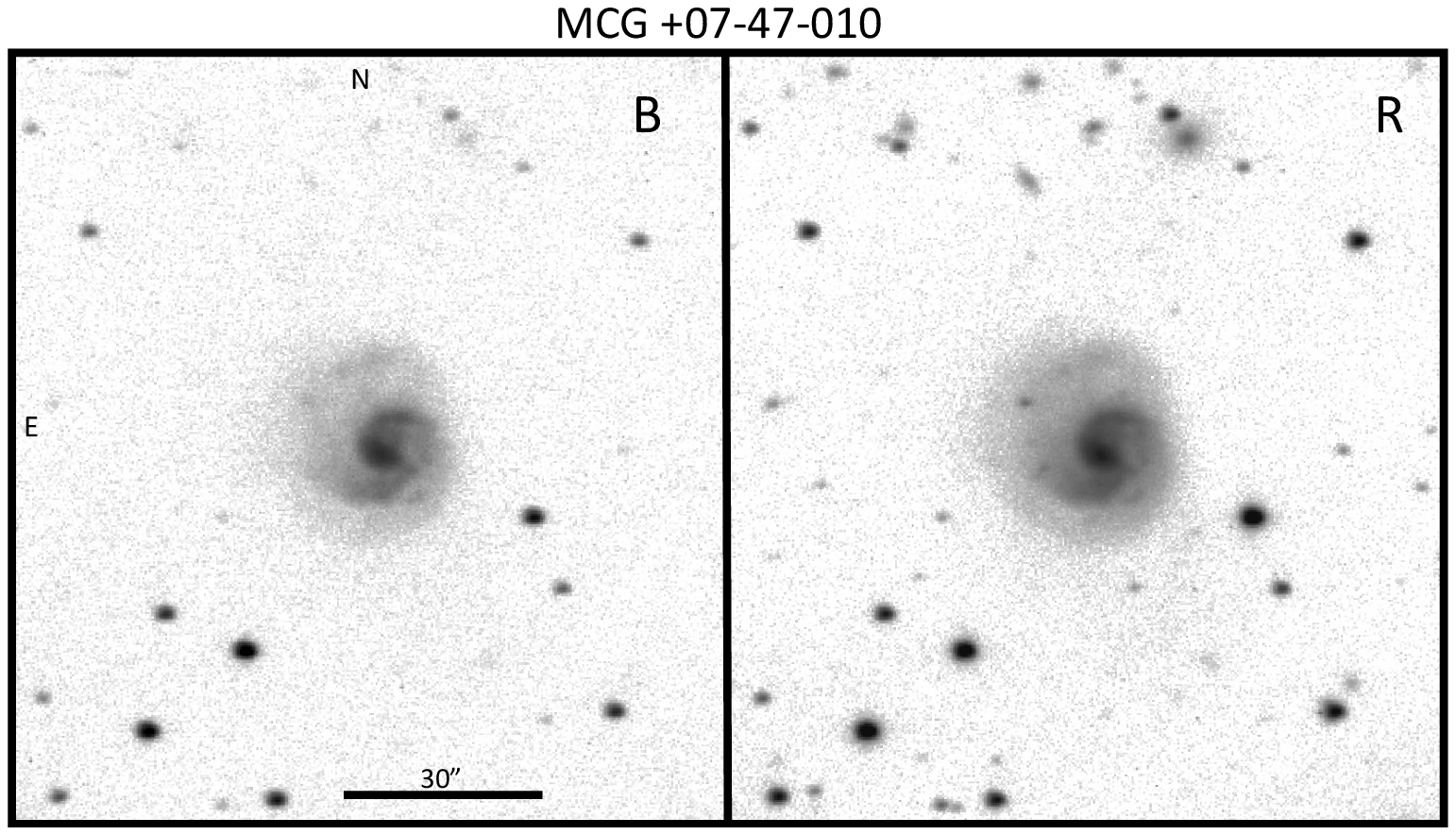}
\caption{Previously identified SDRAGN J2318+4314 (MCG +07-47-010), in $B$ and $R$ images from the Jacobus Kapteyn Telescope (JKT). 
Each is displayed with a logarithmic intensity mapping. North is at the top and east to the left; the field shown spans 
160\arcsec\ E-W by 172\arcsec\ N-S.}
\label{fig-MCG74710}
\end{figure*}

\section{Results} 

\subsection{Host identifications: astrometry}\label{sec-astrometry}
	
\subsubsection{Sources with radio cores}\label{sec-cored}
	
Where radio cores are detected, we use astrometric matching to assess how likely each observed spiral is to be the
actual AGN host. The absolute coordinate accuracy of HST image products retrieved from MAST has been considerably improved by incorporating {\it Gaia}
results both in coordinates of guide stars and directly in the images when Gaia objects are included, as set out by \citealt{GaiaHST}
and by J. Mack\footnote{https://outerspace.stsci.edu/display/HAdP/Improvements+in+HST+Astrometry}. Where dust lanes affect
the peak position of a nucleus in the HST images, we use the center of surrounding isophotes to get a better estimate of
the nucleus location.

In Table \ref{table-astrometry} we compare the positions of optical nuclei of those candidate SDRAGNs having both radio cores
and spiral morphologies, 
as derived from the HST images, with
available radio data having enough sensitivity and resolution to distinguish core sources from surrounding structure.
This comparison
omits galaxies which are either not spirals or so far from the radio positions ($>1$\arcsec) as to be ruled out as the AGN hosts.
This astrometry was done using the VLASS products combining first- and second-epoch observations, which corrected astrometric biases in the
first-epoch quick-look reductions. For cores detected in the VLASS but not listed in the VLASS source catalog, we derive positions from their peaks in the FITS cutout images. We compare the positions of
the four core sources also included in the FIRST source catalog. 
We compare the optical and radio positions in Table \ref{table-astrometry}, listing only the 
seconds and arcseconds fields of the radio coordinates for brevity. 
As a heuristic guide to the expected astrometric precision of the radio surveys, we follow \cite{Condon1997} in
using the estimate FWHM/ ($ 2 \times$ SNR), except for values from the FIRST source 
catalog\footnote{Version 14Dec17, at https://sundog.stsci.edu/first/catalogs/} which recommends positional uncertainty 
given by (observed source size) $\times 1/ ({\rm SNR} +0.05)$. The largest position offsets seen here
for isolated core sources, 0.5\arcsec, occur for
systems with an obscured nucleus whose location is inferred from unobscured isophotes; this offset is less than 
1.7 times the combined HST and radio uncertainties. We use values from the VLASS source catalog at 
CIRADA\footnote{https://cirada.ca/vlasscatalogueql0} when
the source is included; otherwise, for lower-S/N sources, we measured positions from the VLASS image cutouts.
In almost all cases, the error contribution from the S/N in the VLA surveys is larger than the expected contribution from HST data, even in those cases where the position of the obscured nucleus must be estimated from surrounding isophotes.
In B3 0852+422, we see an inner 7.9\arcsec\ 
double source with excess emission between the lobes, which we identify with the radio core,
whose position is uncertain at the 1.0\arcsec\ level due to confusion with the small-scale lobes. The VLASS data for B2 1644+38 are affected by 
sidelobe artifacts, giving a similar uncertainty. Globally, the mean position offset between optical nuclei and radio cores
for all VLASS sources in Table \ref{table-astrometry} is 0.42\arcsec, and 
for sources in the FIRST catalog, 0.31\arcsec. We might expect an additional physical contribution to these offsets when, for example, 
small-scale jets occur but are not resolved in the radio surveys, so the radio core location is more like a centroid of this more complex structure. The thirteen galaxies in Table \ref{table-astrometry} form our
highest-probability sample of new SDRAGNs, for which confusion with distant background systems is least likely.

\begin{deluxetable*}{lcccccccc} 
\tablecaption{Optical and radio core positions\label{table-astrometry}}
 \tablehead{
\colhead{Object}   & \colhead{HST: $\alpha_{2000}$} & \colhead{$\delta_{2000}$} & \colhead{VLASS $\alpha$} & \colhead{$\delta$} & \colhead{Offset$^{\prime\prime}$} & 
\colhead{FIRST  $\alpha$} & \colhead{$\delta$} & \colhead{Offset$^{\prime\prime}$}}
 \startdata
J0209+0750  &   02 09 04.755	& +07 50 04.53	& 04.739  & 04.23   & 0.38 & ... & ... & ... \\ 
J0219+0155 (UGC 1797)                              & 02 19 58.732&	+01 55 49.16	& 58.756 & 49.08 &  0.38 & ... & ... & ...\\
J0847+1241	 & 08 47 59.901 & +12 41 58.96 & 59.89 & 59.4 & 0.47 & 59.889 & 59.45 & 0.52 \\ 
J0855+4204     & 08 55 49.153 & +42 04 20.10  & 49.085 & 20.67 & 0.94 & ... & ... & ... \\ 
J0926+4652 & 09 26 05.183 & +46 52 33.68 & 5.186 & 33.66 & 0.04  &  ... & ... & ... \\ 
J0941+3126  & 09 41 03.617	& +31 26 18.56	& 03.620 & 18.22	& 0.34 & ... & ... & ...\\ 
J0956+1628   & 09 56 05.861 & +16 28 30.13 & 5.855 & 30.27 & 0.15 & ... & ... & ...\\ 
J1322+2706 (IC 4234)                  & 13 22 59.871 & +27 06 59.00 & 59.828 & 59.06 & 0.57 & 59.882 & 59.08 & 0.17 \\
J1516+0517   & 15 16 59.245 & +05 17 51.58 &59.254 & 51.77 & 0.23  & ... & ... & ...\\ 
J1633+0847 & 16 33 00.857 & 	+08 47 36.56	& ...  & ... & 	...	& 00.832  &   36.60	& 0.37 \\ 
J1646+3831            & 16 46 28.433 & +38 31 15.73  & 28.373 & 16.43 & 1.00 & ... & ... & ...\\ 
J1656+6407 &  16 56 20.685	& +64 07 52.90	& 20.641 & 52.89  & 0.29 & ... & ... & ...\\ 
J1721+2624  & 17 21 07.917 & 	+26 24 31.93	& 07.931 & 32.24	&  0.36 &    07.914 & 32.10 & 0.17 \\ 
\enddata
\end{deluxetable*}

\subsubsection{Sources without detected radio cores}\label{nocores}

Secure identification of a host galaxy is inevitably less certain when no radio core has been detected,
since double sources may be asymmetric in intensity, distance from the core, and direction from the core. 
A useful (if difficult-to-quantify) heuristic using radio contour maps has been to accept the identification if the
optical nucleus lies between the lowest contours of the radio lobes at their closest approach to one another -
that is, the minimum-intensity location along the ridgeline of radio emission between the lobes.
This is less useful when the two lobes are separated by many times their sizes. An independent estimate of the
plausible degree of source bending is sometimes possible from the host's location with respect to
galaxy environment; wide-angle tail (WAT) sources are most common within dense regions rather than
cluster outskirts or even lower-density regions \citep{Garon}.

Among the Zoo Gems SDRAGN candidates, we find five candidates for genuine SDRAGNs where no
radio core is detected in VLASS data: J0016+0226, J0806+0624, J1128+2417, J1613+3018 (B2 1611+38), and J1636+2432.
as detailed in the notes on individual sources, two of these have more compelling faint, red host
candidates nearer the axis between lobes than the candidate SDRAGNs, two spirals remain
the best detected host candidates (for J0806+0624 and J1636+2432), and we find the
remaining case, J0016+0226, to remain ambiguous on both grounds. None of their optical spectra
show significant AGN components, which is to be expected given the lack of detected radio
cores but deprives us of a possible additional clue as to the host identifications. In our further analysis,
we include J0806+0624 and J1636+2432 among SDRAGNs; together with the cored objects from the previous section (Table \ref{table-astrometry}),
they make up our sample of 15 RGZ SDRAGNs.


\section{Galaxy identification and properties}

Fig. \ref{fig-sdragnmontagebw} shows the fifteen spirals we identify from sections \ref{sec-cored} and \ref{nocores} as the RGZ SDRAGN
sample. Contours from the VLASS and LoTSS data are overlaid on SDSS color images to show the location, extent, and structure of the radio
emission. Due to the widely differing sizes of the radio sources, the angular scale varies between objects, as indicated. In the case of J1322+2706 (IC 4234), 
structure within the source makes contour displays difficult to interpret, so we show its LoTSS low-resolution data in grayscale in Fig. \ref{fig-ic4234} along
with a schematic depiction of two elliptical and filamentary lobes partially overlapping along our line of sight.

\begin{figure*}
\includegraphics[width=166.mm,angle=0]{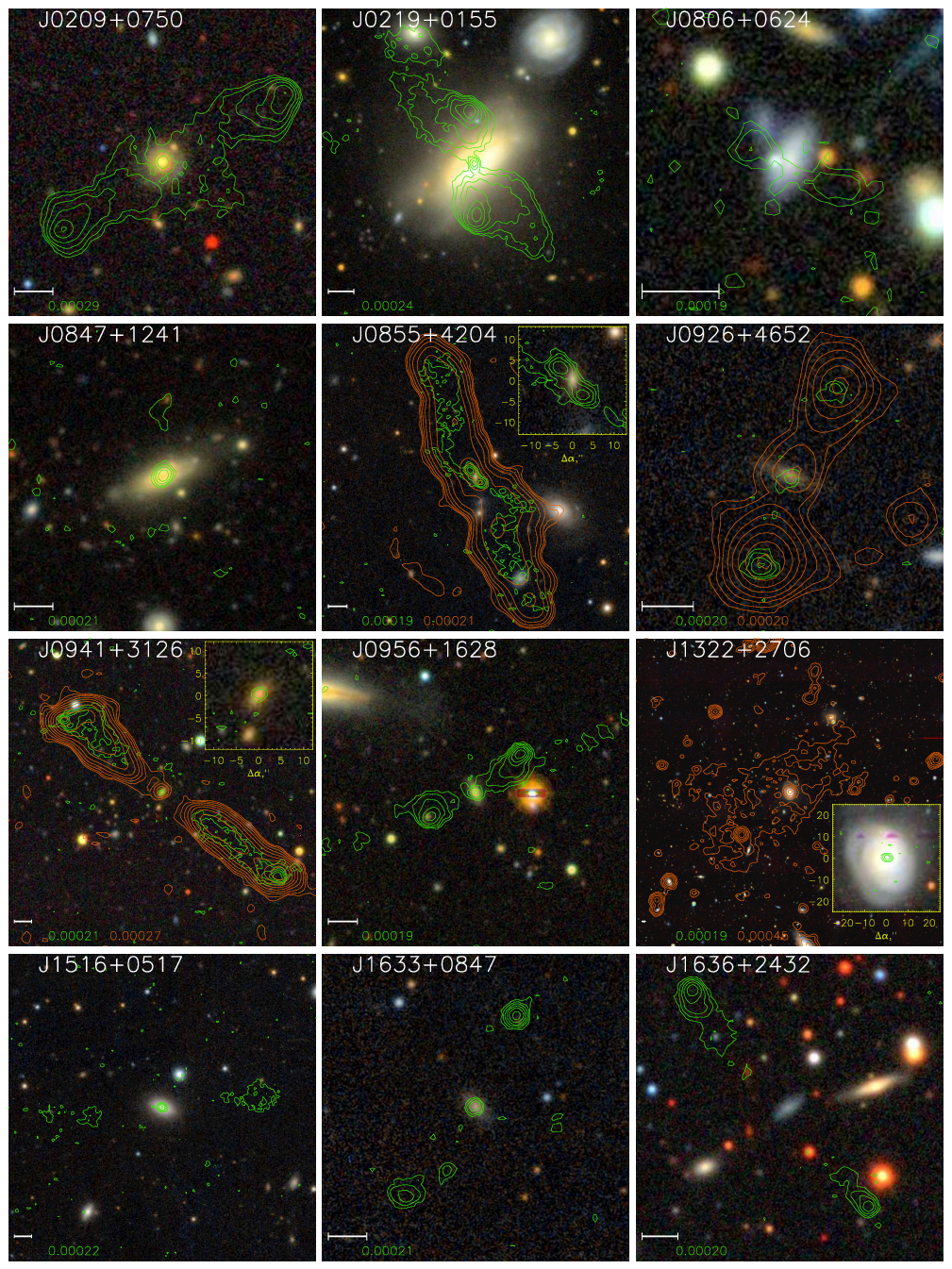}
\caption{Montage of radio survey contours overlaid on 
SDSS $gri$ images 
for the fifteen probable SDRAGNs (with and without radio cores)
from the RGZ-Zoo Gems sample. North is at the top
and east to the left; horizontal scale bars indicate 5\arcsec. Some double sources are so large that 
the galaxy cannot be well shown (and core contours are affected by smoothing at this scale); for these we include
an enlarged ``core" region as well.
Radio data are from the VLA Sky Survey median of epochs 1--3 
(green contours), and LoTSS (orange contours). 
Radio contours are spaced by factors of 2 in brightness, starting with the values in Jy per beam shown in the numbers in matching colors at the bottom of each panel. These
represent values 1.8-2.5 times the RMS noise level, set to minimize appearance of interferometric artifacts, over
ranges appropriate for showing important radio structures while reducing clutter. The SDSS images are shown in the 
\cite{Lupton} 
``sinh" mapping to show detail
over a wide dynamic range. 
Coordinate-based designations are as in Table 
\ref{table-crossids}.
}
\label{fig-sdragnmontagebw}
\end{figure*}

\addtocounter{figure}{-1}

\begin{figure*} 
\includegraphics[width=168.mm,angle=00]{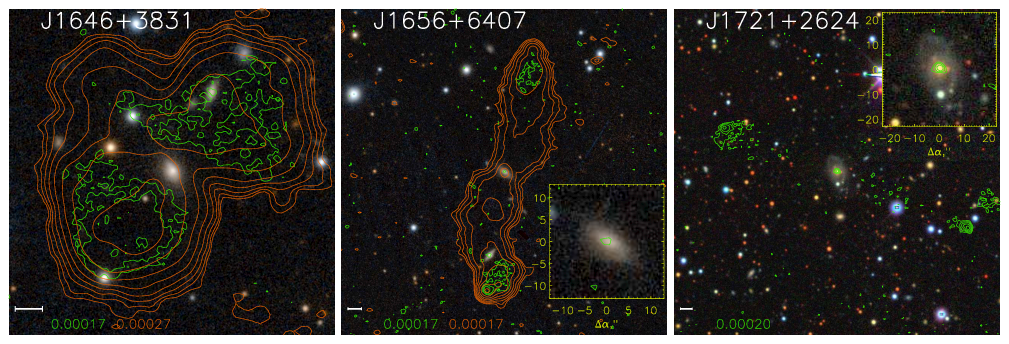} 
\caption{(continued)} 
\end{figure*} 

\begin{figure*}
\includegraphics[width=150.mm,angle=90]{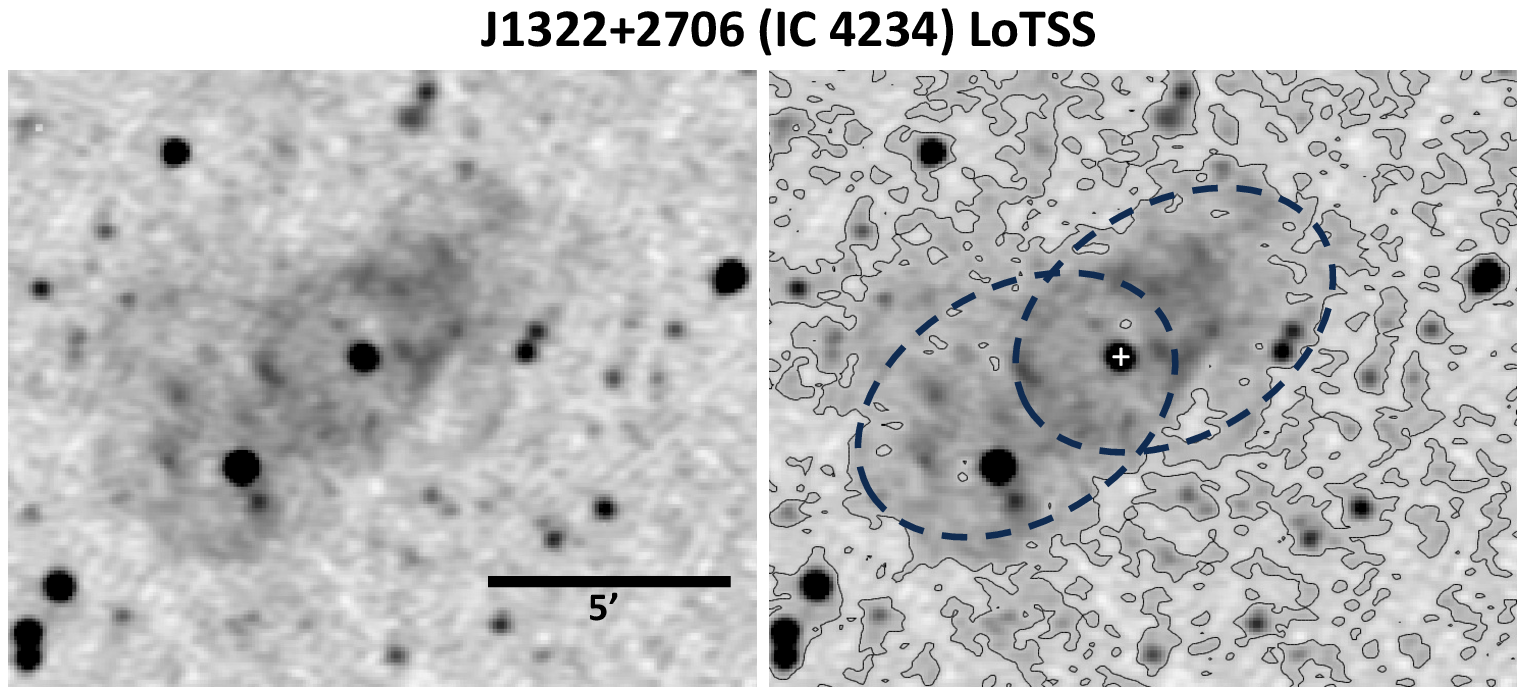}
\caption{Low-resolution LoTSS image of J1322+2706 shown in grayscale to illustrate the radio structure. The left panel shows a negative linear intensity mapping,
with 5-arcminute scale bar. The right panel shows the same mapping with a contour at the local $1-\sigma$ level of 0.125 mJy per beam, and a schematic interpretation
of two elliptical, filamentary lobes seen in projection. The white cross shows the optical and radio nucleus; the size of the optical galaxy detected with HST (see Fig. \ref{fig-HSTMontage}) 
falls within the PSF of the radio nucleus here.} 
\label{fig-ic4234}
\end{figure*}


\subsection{Optical morphologies}
The optical structures are better seen in Fig. \ref{fig-HSTMontage}, without the radio contours and shown with intensity scales
tailored for each galaxy's size and surface brightness. We present morphological classifications for the SDRAGN candidates
in Table \ref{table-Hubbletypes}, on the extended Hubble system, along with orientation and inclination angles derived 
for disk systems from
the HST images, and the morphological characteristics used in the classifications. The derived inclinations are with respect to edge-on, 
so $90^\circ$ indicates a face-on disk. As in the case of 
J0315-1906 \citep{HST0313}, our classifications incorporate  the classic observation (from continuity across galaxy samples with different inclination angles) 
that edge-on spirals have
dust lanes coextensive with the observed stellar disks, while dust lanes in S0 systems are more restricted in radial
extent and form an annulus (\citealt{SandageTammann}, \citealt{SandageBedke}, \citealt{Buta2007}). 
All the hosts we identify as edge-on spirals satisfy this criterion. Spiral features seen mainly via attenuation of background starlight are noted as ``dust arms".

Bars are common among spirals overall, with roughly one-third showing large-scale bars, one-third having bars which are weak or small, and one-third lacking detectable bars altogether.
Bars are clearly not over-represented in SDRAGN hosts.
Only two of these SDRAGNs, both among the previously known subset, host bars large enough to appear in standard Hubble classifications - J0354-1340, type (R)SBa, and J2318+4314 (MCG +07-47-10), type SBc. Among the previous SDRAGNs, J1633+3520 (NGC 6185) shows a small nuclear bar as well as a faint, possibly tidal external arm, while the boxy isophotes in J0315-1906 may
be associated with a central bar seen not only edge-on but end-on. Among the RGZ SDRAGNs, the number of hosts seen close to edge-on or at large redshift limits the recognition of 
potential bar signatures, and we
attempt to classify bars or their absence only for the 8 objects viewed more than $12^\circ$ from edge-on. Only one of the RGZ SDRAGNs shows a small-scale bar, J0806+0624.

\begin{deluxetable*}{lcccl}
\rotate
\tablecaption{Host galaxy Hubble types and orientations\label{table-Hubbletypes}}
 \tablehead{
 \colhead{Galaxy}        &   \colhead{Type} & \colhead{Inclination {\it i}$^\circ$} & \colhead{PA$^\circ$} & \colhead{Features}}
 \startdata
J0209+0750  & Scd & $77\pm 13$  & $134 \pm 15$ & Discontinuous arms, outer one ringlike with knots (like NGC 2805). \\ 
J0219+0155 (UGC 1797)                 & Sab & $13\pm 3$ &$127\pm 2$  & Dust arms, dusty inner disk, outer 20$^\circ$ dust warp, companion.\\
J0806+0624  & Sc & $28\pm 2$ & $153\pm 2$ &  Inconspicuous bulge. Multiple arms, small bar.\\
J0847+1241  & Sbc & $2 \pm 1$ & $113\pm $ & Prominent bulge. Bifurcated dust lane, edge-on.\\
J0855+4204                   & Sab & $6\pm 1$ & $178\pm 2$ & Prominent bulge, uneven dust lane. Edge-on, like M104. \\ 
J0926+4652  & Sb & $2 \pm 2$ & $150\pm 2$ & Prominent bulge. Thick dust lane tilted with respect to starlight.\\
J0941+3126                 & Sab & $1\pm 1$ & $147\pm 2$ & Prominent bulge. Thick, extensive and warped dust lane and plume. Like NGC 4402.\\ 
J0956+1628  & Sb & $24\pm 2$ & $1\pm 1$ & Prominent bulge. Uneven dust lane. Like NGC 891. \\
J1322+2706 (IC 4234)                                 & Sab & $68\pm 22$ & $20\pm 10$ & Prominent bulge. Dusty arms.\\
J1516+0517 & Sb & $9 \pm 3$ & $67.3 \pm 1$ & Dust disk, spiral features, some may be extraplanar.  \\ 
J1633+0847 & Sa & $1\pm 1$  & $48\pm 2$ & Prominent bulge, thick arcuate dust lane with likely star-forming regions.\\
J1636+2432 & Scd &$15\pm 1$ & $125\pm 3$ & Diffuse arms.\\
J1646+3831                    & Sa pec & $40 \pm 3$ & $30 \pm 5$ & Central ring, tail or companion, heavy dust arm plus 3 ordinary ones. \\ 
J1656+6407  & Sab  & $20\pm 1$ & $60\pm 2$ & Prominent bulge, warped dust lanes, disk warp.\\
J1721+2624  & Sc(r) & $28\pm 2$ & $19 \pm 3$ & Prominent bulge, long arms connecting to inner ring. Low surface brightness.\\
 \enddata
\end{deluxetable*}

 \subsection{Galaxy orientations}
 The disk inclinations in Table \ref{table-Hubbletypes} indicate a sample rich in systems seen almost edge-on,
 in contrast to what is found in, for example, optically-selected Seyfert galaxies (\citealt{Keel1980}, \citealt{TovmassianYam},
\citealt{Lagos}). Several selection biases might contribute
 to this. To the extent that the radio-source axes are perpendicular to the disks, projection effects would make the sources'
 angular extent smaller and possibly blend them into single sources at moderate resolution, so they would no longer
 be selected as double sources. For the optical images, selection on the SDSS image parameters for largest exponential-disk fractions is more sensitive for faint galaxies when seen closer to edge-on \citep{Masters2010}. Both attenuation by centrally concentrated dust within the disk,
 and resolution effects for distant galaxies, will enhance this favoritism, making the disk appear more prominent in the outer isophotes (with a larger scale length, and adding less to
 the central surface brightness). This effect acts opposite to the more commonly-encountered dust effect of excluding inclined spirals
 from nominally flux-limited samples due to extinction within the disk. As noted in Section \ref{selectioneffects}, the distribution of projected axial ratios 
 in this sample suggests that this effect dominates the selection for fainter galaxies.

\begin{figure*}
\includegraphics[width=175.mm,angle=0]{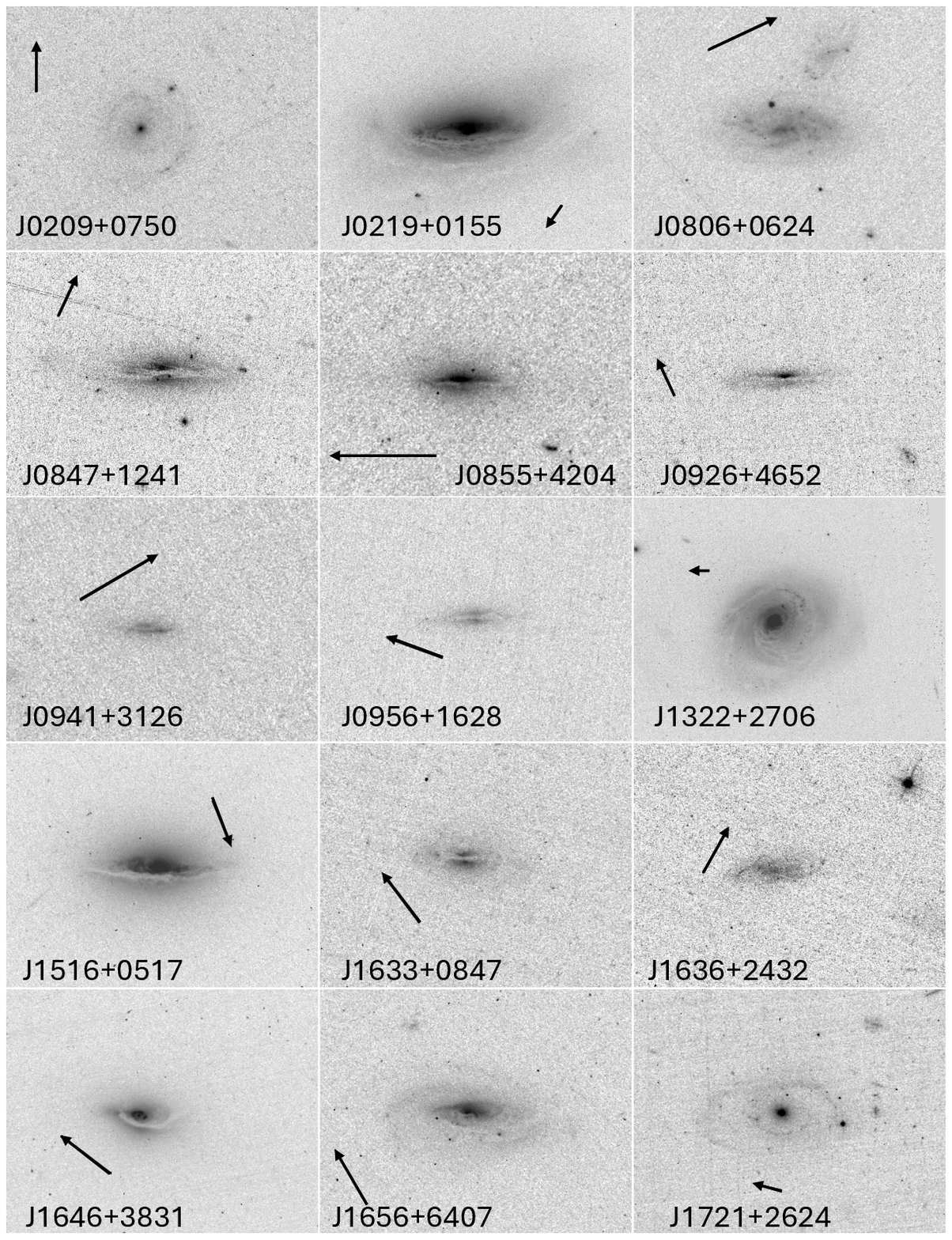}
\caption{ HST ACS F475W images of all 15 RGZ SDRAGNs from the HST sample, in RA order. Various linear scales and intensity mappings, in each case offset log with a common base level,
are used to show the structure of each galaxy. Each scale arrow
shows 5\arcsec and points north; images are rotated to put the projected major axes horizontal for morphological comparison. The image of J1721+2624 was smoothed with a 3-pixel (0.15\arcsec) median to show the galaxy structures more clearly. Some images show low-level residual background structure as streaks along the original row orientations. 
}
\label{fig-HSTMontage}
\end{figure*}

\subsection{Optical spectra and AGN signatures}\label{spectra}


The collected optical spectra of the nuclei of prior SDRAGNs are shown in Fig. \ref{fig-priorspectra}. Similar spectra of the
new SDRAGNs are in Figs. \ref{fig-newspectrahi} and \ref{fig-newspectralo}. In some cases these spectra represent substantial parts of the galaxy;
projected effective aperture diameters (for noncircular extraction regions, the diameter of a circle with the
same area) range from 3--13 kpc for the Zoo Gems SDRAGNs, and from 0.5-11 kpc for the previously-identified objects. This introduces significant dilution of nuclear emission-line signatures both by the surrounding starlight, reducing the equivalent 
widths, and by mixing with emission lines from the star-forming regions common in spiral galaxies as well as gas ionized
by hot evolved stars (Low-Ionization Emission Region or LIERs as defined by \citealt{LIERs} , sometimes known as ``retired galaxies" and distinguished from
low-ionization nuclear emission regions or LINERs which include weak AGN). Classifications as AGN from these
optical spectra are robust, having persisted despite these contrary effects, while more modest optical AGN 
may exist in objects with other spectroscopic classifications. Since LIER emission powered by hot, evolved stars \citep{LIERs} scales with the included
starlight for old populations, we follow \cite{CidFernandes} in adopting an H$\alpha$ equivalent width criterion $> 3$ \AA\  
for ionization contributions 
which clearly come from other sources (star formation and AGN).

\begin{figure*}
\includegraphics[width=138.mm,angle=90]{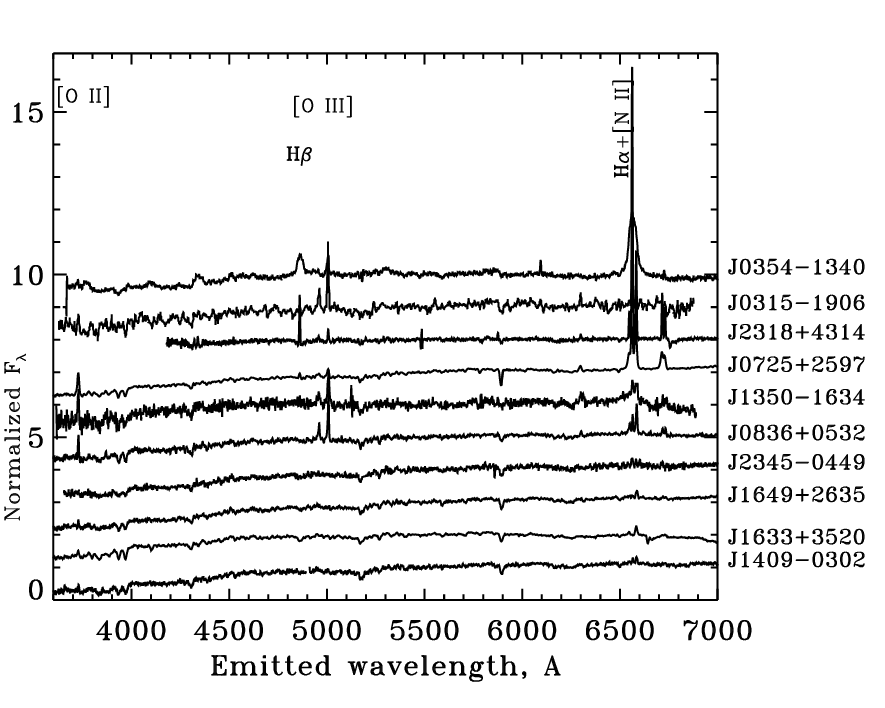}
\caption{Optical spectra of the nuclei of the previously-known SDRAGNs from sources described in Section \ref{prior}.  Each is normalized to the
mean flux from 5400-5600 \AA\  in the emitted frame, and each is offset in zero level by 1.0 unit from the one below for clarity.
Labels at right show the adopted coordinate names as in Table \ref{table-previous} ; objects are ordered vertically to reduce overlap of emission
lines. Locations of the most prominent emission lines are marked at the top. An empirical, approximate sensitivity correction was applied to the 6dFGS spectra.
 }
\label{fig-priorspectra}
\end{figure*}

\begin{figure*}
\includegraphics[width=140.mm,angle=90]{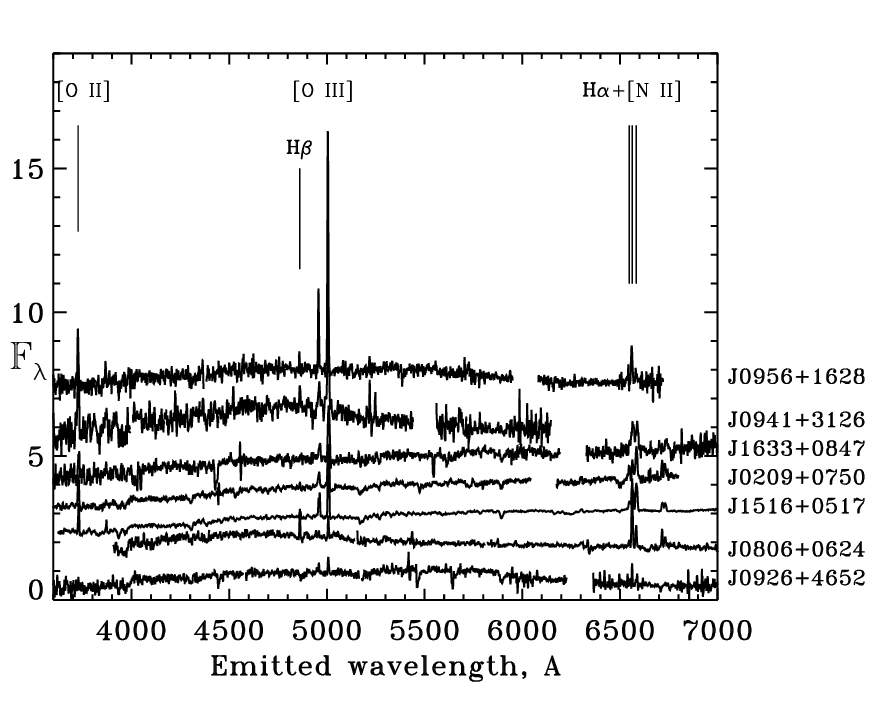}
\caption{Optical spectra of the nuclei of the newly-recognized SDRAGNs from SDSS data and the BTA observations described in Section \ref{newspectroscopy}, showing the seven objects with the strongest line emission. Each is normalized to the
mean flux from 5400-5600 \AA\  in the emitted frame, and each is offset in zero level by 1.0 unit from the one below for clarity
(2 units for J0956+1628 at the top).
Labels at right show the truncated identifiers as in Table \ref{table-crossids}, ordered vertically by [O III] equivalent width
to reduce overlap of emission lines. Gaps in the plotted spectra show the wavelengths affected by uncorrected 
telluric A-band absorption or residuals from sky subtraction
near strong [O I] airglow lines. Wavelengths of the most prominent emission lines are marked at the top.}
\label{fig-newspectrahi}
\end{figure*}

\begin{figure*}
\includegraphics[width=140.mm,angle=90]{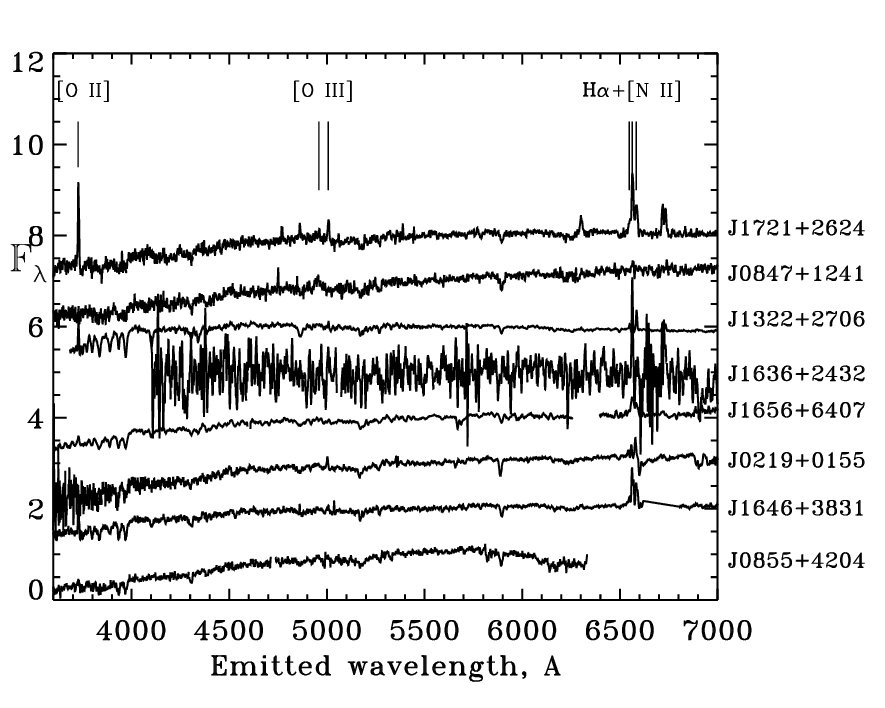}
\caption{Optical spectra of the nuclei of the newly-recognized SDRAGNs from SDSS data and the BTA observations described in Section \ref{newspectroscopy}, showing the seven objects with the weakest line emission. Each is normalized to the
mean flux from 5400-5600 \AA\  in the emitted frame, each offset by one unit in flux from the one below for clarity.
Labels at right show the truncated identifiers as in Table \ref{table-crossids}, ordered vertically by [O III] equivalent width
to reduce overlap of emission lines. Gaps in the plotted spectra show the wavelengths affected by uncorrected 
telluric A-band absorption, residuals from sky subtraction
near strong [O I] airglow lines, or SDSS data gaps. Wavelengths of the most prominent emission lines are marked at the top.}
\label{fig-newspectralo}
\end{figure*}

Table \ref{table-lineratios} summarizes important emission ratios from both prior SDRAGNs and the new RGZ sample, omitting J0354-1349 where the
broad-line region contribution dominates the Balmer lines so narrow-line ratios are essentially indeterminate. For J1350-1634, we measured the archival 6dFGS spectrum; a 
multiple-Gaussian fit to the H$\alpha$+[N II] region suggests that the broad-line region contributes only about 20\% to the blend, which we attempt to
correct in the listed line properties.
Objects observed in SDSS use line measurements from those spectra, either the Portsmouth or {\it GalSpecLine} processing (since these are available for different
subsets of SDSS galaxies).
Values for J0315$-$1906 are taken from \cite{Ledlow1998}. For the objects with 6dFGS spectra, we measured equivalent widths and ratios 
of closely-spaced lines from the data-release spectra, correcting for typical stellar absorption in H$\alpha$ of equivalent width 1.8 \AA . For BTA spectra, 
we used Gaussian fits to 7-pixel slices along the slit, with members of
the H$\alpha$+[N II] and H$\beta$+[O III] sets constrained to have the same FWHM. The redshifts of J0941+3126 and J0855+4204 put the red lines
outside our observed spectral range. The wide redshift range spanned by this sample introduces significant aperture effects in
detecting AGN emission; in particular, surrounding star formation can mask LINER emission at levels which are common
among early-type radio galaxies with low-excitation (LERG) emission \citep{BestHeckman}. The final column tabulates the effective diameter included in each measurement, the diameter of a circular region with the same projected area as the region observed for each object (which is a rectangle for long-slit spectra). We use the classification criteria of \cite{BPT} as updated by \cite{Kauffmann2003} and \cite{Kewley2001}.
 Figure \ref{fig-bpt} shows the location of these nuclei in two of the BPT diagrams, for which the lines involved were strong
enough to be well measured in most of these nuclei. We use a straightforward division at [O III] $\lambda 5007/$H$\beta$=3 to
divide LINERs from traditional AGN. The sample is dominated by classified AGN (plus one not shown here because its broad Balmer lines are so dominant, the
narrow-line Sy 1 J0354$-$1349). For many of these galaxies, H$\beta$ is so weak as to be measured poorly or not at all; in these cases, we estimate its intensity using a typical Balmer
decrement H$\alpha$/H$\beta$=4, following the mean value for narrow-line AGN in the SDSS from \cite{BalmerDecrement}. This estimate gives results which are conservative as to the contribution of AGN, since additional reddening also affects the [O III] lines, making them stronger relative to H$\beta$ and moving their data point upward in Fig. \ref{fig-bpt}. Somewhat surprisingly from initial visual inspection of the spectra, only three likely LINERs are included, one near the boundary with star-forming regions, while the nuclear spectrum of J2318+4314 (MCG +07-47-10) is dominated by star formation. Only one of the new Zoo Gems SDRAGN candidates, J0855+4204, may have emission weak enough to fall in the ``retired galaxy" or LIER category, where the gas can be ionized by
hot, evolved stars (using a criterion of H$\alpha$ equivalent width (EW) $< 3$ \AA\ from \citealt{CidFernandes}). In this object, we do not have data on the red lines of H$\alpha$ and [N II], while the [O II] $\lambda 3727$ EW=9.5 \AA\  suggests EW(H$\alpha$) $\approx 3$ \AA . J0855+4204 does not appear in Fig. \ref{fig-bpt} since only [O II] emission was detected. Among the previous discoveries, only J1649+2635 and J1633+3520 (NGC 6185) fall in the LIER class. 


\begin{deluxetable*}{lcccccccc} 
\tablecaption{Optical emission-line ratios\label{table-lineratios}}
 \tablehead{
\colhead{Galaxy}		& \colhead{${{\rm{[N \ II]} \lambda 6583} \over{\rm{H} \alpha}}$}  & \colhead{${{\rm{[S\ II]} \lambda \lambda 6717, 6731}\over{\rm{H} \alpha}}$}  &  
\colhead{${{\rm{[O\ III]} \lambda 5007} \over{\rm{H} \alpha}}$}  &  \colhead{${{\rm{[O\ III]} \lambda 5007}\over{\rm{H}\beta}}$}   &  
\colhead{${{\rm[O\ II]} \lambda 3727} \over {\rm{[O\ III] \lambda 5007}}$} & \colhead{H$\alpha$ EW (\AA )} & \colhead{Aperture (kpc)}} 
\startdata
Previous SDRAGNs: \\
J0315$-$1906     &     0.20    &     0.20   &   2.33      &       8.18     &    0.055 & 9.6 &  2.1 \\
J0725+2957 & 1.2 & 0.8 & ... & 1.5 & 2.2 & 16.2 & 0.3 \\
J0836+0532  &     1.45      &   0.63   &    1.99   &            9.9     &      0.44  & 5.6 &  5.5 \\
J1350$-$1634 &  0.79    &  0.58  & 1.86  & $>3.0$ &  0.61  & 9.0 & 11.0 \\
J1409$-$0302 (Speca)  &    0.78     &    0.29    &  0.35    & ... &      3.88   & 3.0 & 7.3 \\
J1633+3520 (NGC 6185) &   1.81        &  1.13    &    $<0.29$ &           ... &    $ >9.9 $ &      1.8 & 5.1  \\
J1649+2635   &    1.16    &     0.40  &       0.54     &    ... &     7.90  & 2.1 &  3.2 \\
J2318+4314 (MCG +07-47-10) & 0.32     &  0.28    &   0.04     &   ... &    4.05   & 54.9 &  0.5 \\
J2345$-$0449       &   0.46    & ... &       $<0.05$ & ... & ...      & 4.9 & 9.7 \\
 & & & & & \\
RGZ+HST SDRAGNs: & & & & & \\
J0209+0750     &   1.53    &   0.97    &    1.45       &       7.70    &     0.71 &  8.9 & 13.2\\ 
J0219+0155 &   0.98 & 0.72 & 0.21 & ... & ... & 3.7 & 0.65 \\
J0806+0624 & 0.29 & 0.35 & 0.35 & 0.79 & ... & 25.0 & 5.1  \\
J0847+1241  & 0.22    &    0.36     &      1.39   &          ...       &          0.37 &  3.61 & 8.9 \\ 
J0855+4204 & ... & ... & ... & ... & $>4.0$ & ... & 2.7 \\ 
J0926+4652     &   0.28     &  ...      &        1.16      &       ...     &     0.33 &  8.1 & 6.7 \\ 
J0941+3126      &  ...      &    ...       &          ...      &      5.37     &     0.79 &  ... &10.2  \\ 
J0956+1628      & 0.16    &    ...    &          4.69    &         16.45   &      0.22 & 24.5  & 8.0 \\
J1322+2706 & 1.27 & 0.86 & 0.50 & 1.44 & 3.42 & 5.0 & 2.1 \\ 
J1516+0517 & 0.96 & 0.51 & 2.10 & 4.87 & 0.61 & 11.2 & 3.0 \\ 
J1633+0847      & 0.82     &   0.51   &     1.79      &       15.26     &    0.09  &  15.9 &  6.9 \\ 
J1636+2432 &    0.04        &    0.28         &    0.25    &       ...        &      ...    &   9.0   & 1.8 \\ 
J1646+3831  & 0.59    & ...  & 0.12 & 0.33 &   3.00 & 5.42 & 6.0 \\ 
J1656+6407  &      0.56  &   0.14      &   0.10      &           ...    &      1.85           & 11.7 & 6.6 \\ 
J1721+2624      & 0.56    &    0.73    &      0.35    &         0.89    &      3.14  & 16.2 &  8.7 \\
\enddata
\end{deluxetable*}

\begin{figure*}
\includegraphics[width=180.mm,angle=90]{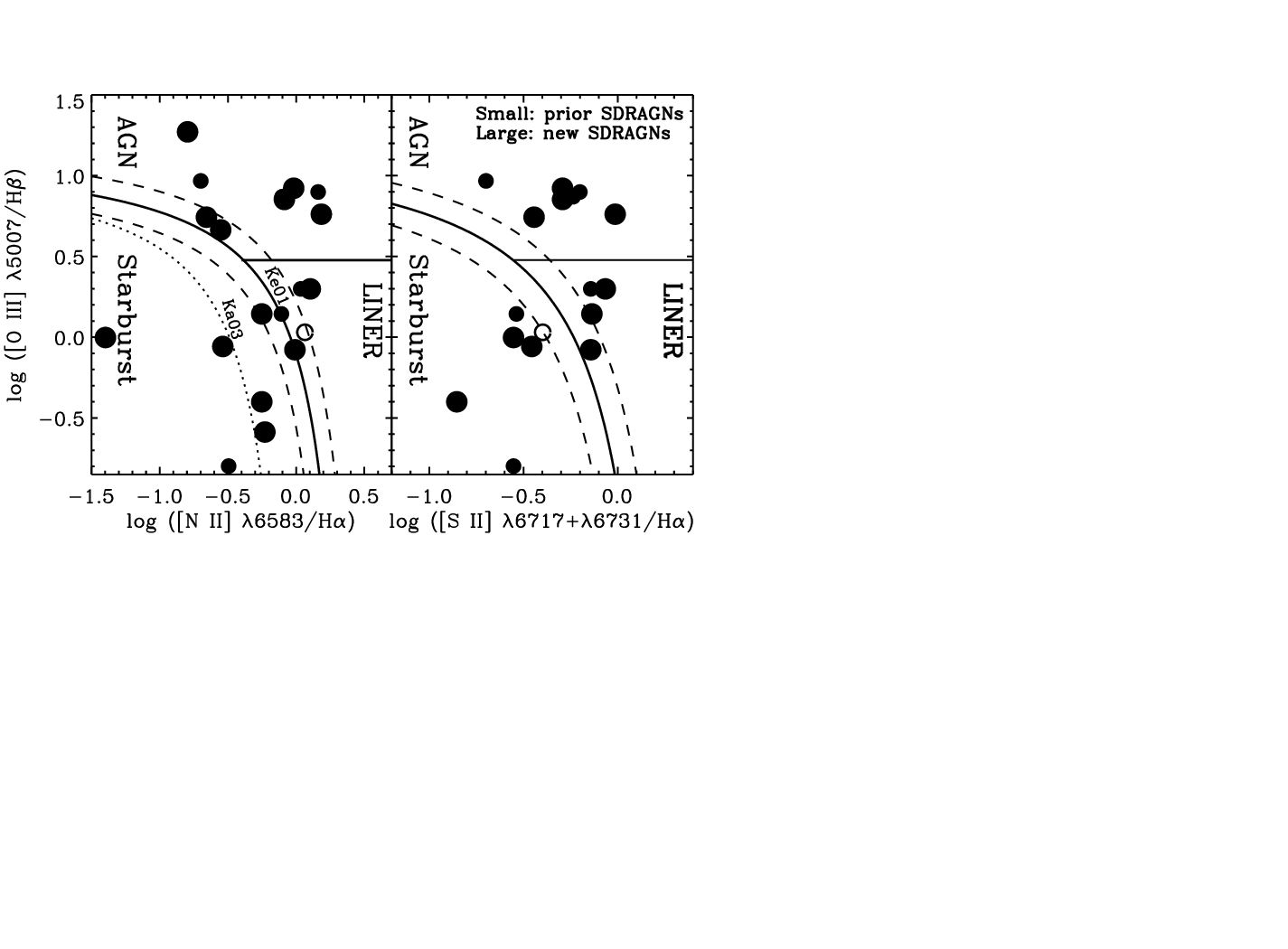}
\caption{Two BPT diagrams illustrating classification of SDRAGN nuclear spectra from optical data.
Ke01 and Ka03 indicate the theoretical starburst lines from Kewley et al. (2001) and Kauffmann et al. (2003). For the Ke01 lines, the stated 10\% uncertain is shown by dashed lines about the reference curve. Data for the one of the two prior SDRAGNs
with H$\alpha$ EW $< 3$ \AA, therefore potential candidates for ``retired galaxy" emission, are shown with open circles; this system lies well
below the maximum-starburst curves, making it unlikely to be confused with an AGN. The second prior SDRAGN with weak emission,
J1633+3520 (NGC 6185), has only an upper limit for [O III] and is not plotted.}
\label{fig-bpt}
\end{figure*}

Roughly half of both prior and ``new" SDRAGNs show emission-line ratios dominated by star formation. Some
AGN are almost certainly hidden in these data by obscuration and the dilution by starlight included
in large projected apertures. For AGN with the low-ionization, 
weak (small equivalent width) emission lines characteristic of many radio galaxies in early-type galaxes (ETGs), the large projected
apertures for systems at larger redshift will mingle the nuclear emission with surrounding star formation in
most spirals, moving the combined emission-line rations into the composite or star-forming regions in the BPT
diagrams. In addition, for a sample with so many spirals viewed edge-on, the effects of dust lanes hiding the 
nuclear regions become important; the intrinsic locations of such nuclei in the BPT diagrams are dominated by surrounding emission, so the fractional
contribution of AGN overall is at least what these diagrams imply and quite possibly greater (since so many
have core radio sources which are often associated with optical AGN signatures).

In both previous and new SDRAGN samples, few broad-line (Seyfert 1) spectra appear. From previous
detections, J0354$-$1340 is a narrow-line Seyfert 1(NLSy 1; \citealt{Vietri} and Fig. \ref{fig-priorspectra}).
From the new SDRAGNs, we identify a weak broad component at H$\alpha$ in J1633+0847, which
would make this a Sy 1.9 following the widely-used criteria from \cite{DEO1981}, noting that recognition
of weak broad components is strongly dependent on the signal-to-noise ratio of the data and the
contrast of the AGN lines against the starlight continuum. Since there are 3 Seyfert spectra among the prior SDRAGNs and six among the new set, this is a fraction 2/9=22\% of Sy 1 (including Sy 1.9).
Among DRAGNs in the usual early-type galaxies (that is, the great majority of radio galaxies in the local Universe), broad-line radio galaxies (BLRGs) are associated
solely with FR II radio structures (\citealt{Tadhunter} and references therein). Since none of the new SDRAGN sources
has FR I morphology (section \ref{radioproperties}), the incidence of BLRG (Sy 1) 
signatures is lower than the 21-36\% quoted by Tadhunter et al. for FR II radio galaxies in the 3CR and 2-Jy samples. 

We can also compare
the fraction of  broad-line AGN to samples selected from hard X-rays (the BASS sample
selected at 14-95 keV, \citealt{BASSAGN}) and 
from spectroscopy of optically complete (or representative) galaxy samples (the CfA and SDSS samples). Both the CfA Seyfert
sample \citep{cfasy} and SDSS sample at $z<0.33$ \citep{SDSSSy} fortuitously match very closely in the
fraction of type 1 AGN among Seyferts, at 30\%. The BASS spectra are of high enough quality to be subdivided more 
finely: summing subtypes, 27\% are in the BLR-dominated Sy 1 or 1.2 categories, and 34\% have
weaker BLR emission in Sy 1.5-1.9. 
In the same way as for recognition of Sy 1.9 nuclei, these fractions are sensitive to signal-to-noise ratio of the optical spectra, and contrast of the
AGN emission lines against the starlight continuum, which in turn declines with increasing measurement aperture.
There might be preferential inclusion of narrow-line objects seen through the surrounding
obscuring torus to the extent that the jets are perpendicular to the current
accretion disk, an effect further amplified by having some host disks edge-on, so disk dust hides all but the outer 
parts of the narrow-line region (NLR). Such effects of distant obscuration within the host galaxy
are seen in narrow-line images of J0315$-$1906 \citep{HST0313}, while
the dust structures seen in HST images of large AGN samples (e.g. \citealt{Malkan}) shows additional examples of what might be termed ``accidental Seyfert 2s" where the BLR is hidden from optical view by distant dust in the host disk rather than circumnuclear structures. The unusual number of edge-on disks seen in the new SDRAGN sample surely plays
a role in hiding their optical AGN signatures from our vantage point: HST isophotes indicate that the true nucleus
is at least partially obscured by disk dust in 8/15 of the new SDRAGNs, including the Sy 1.9 J1633+0847, three Sy 2 nuclei, one
LINER, and one star-forming nucleus. This allows ample scope for a larger number of broad-line nuclei to
be hidden by dust in the disks of these host galaxies.

A few systems' optical spectra show properties not captured in the tabulated data.
The BTA spectrum of J0855+4204
shows a quiescent stellar population with only [O II] securely detected in emission (H$\alpha$ and [N II] fall beyond the observed
spectral range at $z=0.18$).
J1322+2706 and J1656+6407 show high-order Balmer absorption lines, suggesting post-starburst stellar
populations. J0219+0155 and J1721+2624 are LINERS with large equivalent widths and strong [O I] $\lambda 6300$ emission,
falling in the AGN subclass among LINERS.

\subsection{Bulge properties}

While one obvious speculation about spirals hosting powerful double radio sources would entail unusually massive central black holes (\citealt{Ledlow}, \citealt{HST0313}),
that would typically mean luminous bulges, which are not seen among SDRAGNs.
In fact, the GALFIT decompositions by \cite{Wu2022} show very few classical bulges, with the RGZ+HST sample dominated by 
pseudobulges. 

An obvious comparison set of AGN in spiral hosts is comprised of Seyfert nuclei, which are mostly in spirals. They have long been found to be more common
in the ``earlier" parts of the spiral sequence - spirals with Seyfert nuclei have median type index T=0, Sa, for Sy 1 galaxies in the IRAS 12$\mu$ sample \citep{HuntMalkan} and T=2, Sab, for Sy 2, 
for example. Both SDRAGN samples, while spanning the whole range from Sa to Scd, have medians within the Sb (T=3) bin. This fits with image decomposition in finding that
their bulges (classical or pseudobulges) are relatively fainter (in bulge/total luminosity ratio B/T) than typical for Seyfert hosts.  



\subsection{Non-spiral host galaxies}

Some of the host galaxies in the HST images, with secure identifications
from core radio emission, are not spiral galaxies. Since the input sample was selected for
possible spirals and excluding normal elliptical systems, these outliers are morphologically interesting in themselves.
They include elliptical galaxies with dust rings around the projected minor axis, suggesting prolate forms; ellipticals
with other kinds of dust distributions; and tailed merger remnants. J0802+1157 is a two-tailed merger remnant, with multiple bright knots and dust lanes in the main body of the galaxy. VLASS 
data leave the geometry of the radio source unclear.

The combination of improved astrometry using Gaia data, the HST images, and multiband Legacy Survey and 
WISE imaging shows that some of the actual hosts are distant background early-type galaxies; in some of these
cases the extended radio source is seen through the disk of a foreground spiral, potentially allowing Faraday-rotation 
studies of the magnetic fields in the spiral disks. The best cases for such Faraday investigations are
J0813+5520 (foreground spiral at $z=0.2645$),
J0901+1648,
J0903+4328 (foreground $z=0.3736$),
J1136+1252 ( foreground $z=0.0345$), 
J1328+5710 (foreground $z=0.0211$), and
J1509+5152 (likely background $z=0.5789$).

\section{Galaxy Environments}\label{environments}

\subsection{Galaxy distributions}

The structure of large-scale radio sources is shaped by the interaction between the emerging jets and the surrounding diffuse environment, from
circumgalactic to truly intergalactic scales. We examine the environments of SDRAGNs using the galaxy distribution as a proxy for the 
diffuse material which shapes the radio emission. We consider tracers based on spectroscopic redshifts, catalogued galaxy groups and clusters, and photometric redshifts, all within
projected radii 1 Mpc from the SDRAGNs.

The wide redshift range among known SDRAGNS, $z=0.017 - 0.394$, limits our ability to compile uniform information on their Mpc-scale
galaxy surroundings. Using spectroscopic redshifts and cluster identifications tabulated in NED\footnote{https://ned.ipac.caltech.edu/}, or from the Legacy Survey, adopting a criterion $\Delta z < 0.005$ or
$\Delta v < 1500$ km s$^{-1}$ for physical association, within a projected distance 1 Mpc, gives the numbers of associated galaxies listed
in Table \ref{tbl-environment}. 
A 1-Mpc radius
is appropriate for group environments, and sensitive to cluster-scale structure while not diluting the effects of groups. In this table, the number of galaxies within 1 Mpc in projection at matching redshift is N$_{gal}$, and the most distant of these is projected at distance $R_{max}$
from the SDRAGN. The listed cluster or group associations are centered at projected radius $R$ and radial velocity $\Delta v$ from the SDRAGNs. This is our most complete set of data for $z<0.1$, where photometric redshifts in the $ugriz$ system are 
insensitive and deep surveys may give spurious results for bright galaxies either through local background subtraction altering the fluxes,
segmentation breaking single galaxies into multiple detections, or even saturation removing galaxies from a catalog. For $z<0.2$ where group and cluster identifications 
are more complete, 4/10 RGZ SDRAGNs and 4/11 previous SDRAGNs fall in groups or clusters in both location and redshift.

For $z>0.1$,
as spectroscopic redshifts become sparse, we incorporate photometric redshifts, which now have estimated accuracy $\sigma_z = 0.05$
at $r=22$ and $\sigma_z = 0.1$ for $r=23.0$. Using the facilities of NOIRLab's DataLab  (\citealt{Fitzpatrick}, \citealt{Nikutta}), we retrieved
corresponding sets of galaxy coordinates, $r$ magnitudes, photometric redshifts $z_{phot}$ and their $\pm 1 \sigma$ ranges, and spectroscopic redshifts
$z_{spec}$ where known, from the Legacy Survey DR10 products, within cones of angular radius projecting to 1 Mpc at the SDRAGN's distance. After rejecting bright objects with $r<16$ due to contamination by improperly measured very bright systems, we constructed
a redshift distribution in each field, using $z_{spec}$ where available. Respecting the accuracy of $z_{phot}$, we examined these distributions when binned
in widths of $\Delta z = 0.025$. While peaks corresponding to overdensities are apparent in these distributions, we went on to estimate large-scale overdensities (of the set
of galaxies included at each redshift) by treating the number of galaxies at redshifts outside density peaks as a purely heuristic power law in redshift. This 
includes contributions from differing survey depths between pieces of the Legacy Survey, increasing volume elements at higher redshift, and changes in the parts of the galaxy
luminosity function sampled at various redshifts. We divide the distribution of galaxy number with redshift by this power law, ranging from $z^{0.9}$ to $z^{1.5}$, to estimate the
amplitude of galaxy over- or under-densities at the redshifts of SDRAGNs. Legacy Survey data are not available at the locations of four of the Zoo Gems SDRAGNs as well as two of the 
``previous" SDRAGNs. These overdensity amplitudes ($\rho = {{n_{gal}-\bar{n} \over \bar{n}}}$ in Table \ref{tbl-environment}, lacking values for $z<0.1$ or objects outside the Legacy Survey coverage) 
are certainly underestimates given the redshift resolution of photometric redshifts; 
the typical width of density peaks in redshift suggests that typical errors $\pm 0.05$ are consistent with the data.

Even with these limitations, the photometric-redshift distributions show that SDRAGNs are systemically located in regions of galaxy overdensity as measured in redshift, while
the number of these spirals associated with clusters or rich groups also stands out. Mean values of our overdensity parameter $\rho$ are, for the RGZ SDRAGNs, 1.55; for previous SDRAGNs, 1.74; and for all SDRAGNs with $z>0.1$ and  covered in the Legacy Survey, 1.47. Even with the large line-of-sight smoothing imposed by the accuracy of photometric redshifts ($\sigma_z = 0.05$ translates to 215 Mpc in depth), the whole sample of SDRAGNs occurs, in the mean, in the densest third of galaxy environments. The connection between radio luminosity and density of the surrounding medium for
the jets to interact with \citep{Krause} broadly predicts that SDRAGNs preferentially occur in the densest regions occupied by spirals. Similarly, SDRAGNs are likely to be
in gaseous environments less dense than the whole DRAGN population based on their host properties, which would be manifested in lower radio power for SDRAGNs than DRAGNS
as a whole \citep{Turner}, at least consistent with the comparisons we make in section 6.

\begin{deluxetable*}{lccccclcc} 
\tablecaption{Galaxy environments of RGZ and previous SDRAGNs\label{tbl-environment}}
 \tablehead{
 \colhead{Object}       &  \colhead{$z$} &    \colhead{Scale\tablenotemark{a}}  & \colhead{N$_{\rm gal}$} &    \colhead{$R_{max}$,Mpc} & $\rho$\tablenotemark{c} & 
 \colhead{Cluster/Group} &  \colhead{$R$, Mpc} &  \colhead{$\Delta v$, km s$^{-1}$}  }
 \startdata
\multispan2 RGZ/Zoo Gems: \hfil \\
J0209+0750 & 0.2552 & 250 & 0 & -- & -0.8 \\
J0219+0155 & 0.0410 & 1227 & 3 & 0.87 & -- & Unnamed &  0.32 & 400 \\ 
J0806+0624 & 0.0834 &  634 & 1 & 0.88 & 0.2 \\ 
J0847+1241  & 0.1745  &  335 & 1 & 0.91 & 2.9 \\  
J0855+4204 & 0.1804  & 326 & 1\tablenotemark{a} & 0.06 & --  \\ 
J0926+4652 &   0.2181 &  281 & 0 & -- & -- \\ 
J0941+3126  &  0.3940 &  186& 0 & -- & -- \\ 
J0956+1628 &   0.2745 &  237 & 0 & -- & 0.4 \\ 
J1322+2706 &   0.0344 & 1451& 7 & 0.41 & -- & MSPM 1412 & 0.06 & 402 \\ 
J1516+0517 &  0.0512 &  994 & 21 & 0.95  & 1.4 & WHL J151704.0+051522 & 0.16 & 86 \\ 
J1633+0847 &   0.2247 &  275 & 0 & -- & -0.2 \\ 
J1636+2432 &  0.1016 &  531 & 6 & 0.95 & 4.0 \\ 
J1646+3831 &  0.1075 &  505 & 2 & 0.40 & -- \\ 
J1656+6407 &   0.2121 &  287 & 0 & --  & -- \\
J1721+2624 &  0.1696  & 343 & 6 & 0.72 &4.5 & FSVS\_ CL J172110+262919 & 0.84 & 213 \\ 
Previous: \\
J0315$-$1906 &  0.0677  & 766 & 16 & 0.80  & 2.5 & Abell 428 & 0.13 & 253 \\ 
J0354$-$1340 &    0.0772  & 680 & 0 & -- & 1.4 \\ 
J0408$-$6247 &   0.0178 & 2747 & 15 & 0.94 & -- & LGG 110 & 0.24 & 53 \\ 
J0725+2957 & 0.0197 & 2488 & 4 & 0.07 & --\\ 
J0836+0532 &    0.0993 &  542 & 3 & 0.50  & 4.2 \\ 
J1350$-$1634 & 0.0877 & 606 & 0 & -- & -0.5 \\ 
J1409$-$0302  &    0.1376  & 408 & 6 & 0.76  & 1.8 & MPSM 9045 & 0.42 & 185 \\ 
J1633+3520 &   0.0343 & 1455 & 7 & 0.72 & -- \\ 
J1649+2635 &      0.0545 &  937 & 0 & -- & 0.7 & -- \\ 
J2318+4314 & 0.0169 & 2890 & 8 & 0.92 & -- & CIZA J2318.6+4257 & 0.35 & 147 \\ 
J2345$-$0449 &      0.0755 &  693 & 9 & 0.99 & 2.1 \\ 
\enddata
\tablenotetext{a}{Galaxy serendipitously appearing on our spectrograph slit.}
\tablecomments{``Scale" gives the projected image scale in arcsec Mpc$^{-1}$. N$_{\rm gal}$ gives the number of galaxies within $\Delta z < 0.005$ of the SDRAGN from spectroscopic data, of
which the most distant within 1 Mc in  projection is seen at projected separation $R_{max}$ in Mpc.
The density parameter $\rho$ is the overdensity within a photometric-redshift bin of width $\Delta z = 0.025$, from local galaxy density ${{n_{gal}-\bar{n} \over \bar{n}}}$ where $\bar{n}$ is normalized for large-scale variations in depth with $z$. When an associated cluster or group is present, $R$ gives the projected separation of the SDRAGN from the cluster center, and $\Delta v$ gives its redshift offset from the cluster mean.}

\end{deluxetable*}
 
\subsection{Interaction histories}

A galaxy's interaction history is implicitly related to the local galaxy density, and more specifically can affect the presence of bulges and disks.
Among both RGZ and previous sets of SDRAGNs, prominent tidal tails or broader fans are rare. As seen Fig. \ref{fig-HSTMontage}, only J1646+3831 shows bright 
features which are likely tidal or indeed represent a
disrupted companion. We do not expect to see remnants of major mergers in this sample because we select specifically for spiral patterns which
are wiped out by most such mergers, but,
given the common discussions about connections between interactions and occurrence of AGN, it is noteworthy that the RGZ SDRAGNs show
signs of only weak recent encounters. It is equally noteworthy that many of the SDRAGNs seen close to edge-on show warped dust lanes,
which can result from a weaker tidal disturbance and persist for several Gyr. This includes $1^\circ -15^\circ$ offsets between observed orientations of stellar and dust disks,
arcuate dust lanes which depart from the stellar plane in the same sense on both sides, and dust lanes which are split perpendicular to the disk,
a structure which often implies a thin dust distribution with a warp along the line of sight (\citealt{SteimanCameron}, \citealt{Nicholson}, \citealt{Quillen}). Of systems viewed within $24^\circ$ of edge-on so that
dust lanes are well silhouetted, 9/10 in the RGZ SDRAGNs (Table \ref{table-Hubbletypes}) and 2/5 of the previous SDRAGNs show such disturbed dust and hence hint at weak interactions in the
last 1-2 Gyr. (Two of the previous SDRAGNs with undisturbed dust, J0825+2957 and J2345$-$0449, are among the ones with HST imaging available, so the kinds of warps seen in other systems would
be detected in them.) We concentrate on warps visible in the dust lanes, which are detectable only over the front half of the disk where backlighting is most effective. The statistics on
warped {\it stellar} disks are extensive, but our data do not always have the signal-to-noise ratio in the outer parts of disks to detect them even where we can trace dust lanes. Warped dust lanes will be observed less often even when in the same galaxies as stellar warps, because of the front/back dichotomy. This dichotomy makes dust warps prominent only from certain directions, where backlighting
highlights the portions making the largest projected angles with the stellar disks (which, unlike the dust, are seen more nearly integrated along the line of sight).
Among edge-on disks, the incidence of stellar warps is substantial, although occurrence fractions reported in various studies range from 12\% \citep{S4Gwarps} to 65\% (\citealt{ReshetnikovWarps}, \citealt{deGrijs}) with evidence that it grows in denser environments \citep{ReshetnikovWarps}. Very roughly, associated warped dust disks should be four times less frequent than warped stellar disks in
random samples of edgewise spirals, due to these front/back and angle effects.

The prevalence of pseudobulges among SDRAGNs, which suggests that the galaxies have not undergone a major merger, contrasts with
the incidence of minor interactions 1-2 Gyr ago which would leave the kinds of disturbed dust distributions we see.

\section{Radio-source properties}\label{radioproperties}

\subsection{Radio power and size comparisons}

We measured spatially-integrated fluxes for the 15 RGZ SDRAGN candidates (section 3.1), and the spectral index between 150 and 1400 MHz
using lower-frequency  LoTSS, TGSS, and RACS data. Table \ref{tbl-radiopower} gives the 1400-MHz flux density, spectral index $\alpha_{150-1400}$ (in the
sense ${\rm S}_\nu \propto \nu^\alpha$), and corresponding power at 1400 MHz in the emitted frame. For J1350-1634 and J1516+0517, the spectral-index fits
also incorporate 151-158 MHz data from GLEAM (\citealt{GLEAM1}, \citealt{GLEAM2}). For comparison, we list the LLS and 1400-MHz power values we adopt for the previous SDRAGNs,
where appropriate values are available, in Table \ref{tbl-radiopowerold}. As a representative comparison set of ``ordinary" DRAGNs, overwhelmingly with
host galaxies of early Hubble type, we use the FIRST sample from \cite{MiraghaeiBest}, comprising 1329 double sources with known host redshifts and 1400-MHz flux
density greater than 40 mJy.

\begin{deluxetable*}{lcccc} 
\tablecaption{Integrated radio properties of RGZ SDRAGNs\label{tbl-radiopower}}
 \tablehead{
 \colhead{Object}       &  \colhead{$z$} &    \colhead{S$_{1400 ~ \rm MHz}$, NVSS (mJy)}  & \colhead{$\alpha_{150-1400}$} &    \colhead{log P$_{1400 ~\rm MHz}$ (W Hz$^{-1}$)}}
 \startdata
J0209+0750  & 0.2552   & $ 395 \pm 10  $ &  -0.71 & 25.87 \\
J0219+0155  & 0.0410   & $ 555 \pm  20  $ & -0.59 & 24.33 \\
J0806+0624  & 0.0834   &  $ 5.3  \pm 0.5  $ & -0.30 & 22.94 \\
J0847+1241  & 0.1745   & $ 11.3  \pm 0.6  $ & -1.0  & 23.98 \\
J0855+4204  & 0.1804   & $ 177  \pm  6.0   $ &   -0.66 & 25.18 \\
J0926+4652  & 0.2181   & $ 10.7  \pm 0.9  $ & -0.75 & 24.15 \\
J0941+3126  & 0.3940   & $ 225 \pm  6   $ &   -0.76 & 26.05 \\
J0956+1628  & 0.2745   & $ 88  \pm 3  $ &  -0.64 & 25.28 \\
J1322+2706  & 0.0344   & $ 50 \pm 15 $  & $<-1.0$ &   23.14 \\
J1516+0517  & 0.0512   & $ 456 \pm  15   $ & -0.70 & 24.45 \\
J1633+0847  & 0.2247   & $ 40 \pm   5   $ & -0.48  & 24.73 \\
J1636+2432  & 0.1016   & $  60 \pm  2   $ & -0.71  & 24.18 \\
J1646+3831  & 0.1075   & $ 390 \pm 13  $ &  -0.62  & 25.05 \\
J1656+6407  & 0.2121   & $  74 \pm   3   $ & -0.63  & 24.96 \\
J1721+2624  & 0.1696   & $ 193 \pm   8  $ &  -0.67 & 25.16 \\
\enddata
\end{deluxetable*}

\begin{deluxetable*}{lcccl} 
\tablecaption{Integrated radio properties of previous SDRAGNs\label{tbl-radiopowerold}}
 \tablehead{
 \colhead{Object}       &  \colhead{$z$} &    \colhead{log P$_{1400 ~\rm MHz}$ (W Hz$^{-1}$)} &  \colhead{LLS, kpc} & \colhead{Data source}}
 \startdata
J0315$-$1906 & 0.0677 & 24.00  & 350  &  \cite{J0836paper} \\ 
J0354$-$1350 & 0.0722  & 23.18  & 177  &  log L$_{5\ \rm GHz}$  (W Hz$^{-1}$) $= 22.87$, assume $\alpha=-0.8$  \citep{Vietri} \\ 
J0408$-$6247 & 0.0178  & 22.65  & 755   &  log L$_{325\ \rm MHz}$ (W Hz$^{-1}$) $= 23.99$, $\alpha=-2.1$ \citep{HurleyWalker} \\ 
J0725+2957 & 0.0197   & 23.69  &  12.5 & log L$_{408\ \rm MHz}$ (W Hz$^{-1}$) $=24.11$, assume $\alpha = -0.8$ \citep{Emonts} \\ 
J0836+0532 & 0.0993 & 24.18  &  280   &   \cite{J0836paper} \\ 
J1350$-$1634 & 0.0877  & 24.72    &  832 &\cite{Sethi}  \\ 
J1409$-$0302 & 0.1376  & 24.85 & 1300  &  log L$_{325\ \rm MHz}$ (W Hz$^{-1}$) $= 25.36$, $\alpha=-0.8$  \citep{Hota}\\ 
J1633+3520 & 0.0343 &  ...  &   2450 & \cite{Oei} \\ 
J1649+2635 & 0.0545 & 24.03 &   85   &   \cite{J0836paper} \\ 
J2318+4314 & 0.0169  &   22.05 &     207 & \cite{Mulcahy} \\
J2345$-$0449 & 0.0755  & 24.40 & 1600  &   \cite{J0836paper} \\ 
\enddata
\end{deluxetable*}

Broadly, the three-way comparison among typical DRAGNs, previous SDRAGNs, and RGZ-HST SDRAGNs shows that the distributions of SDRAGNs in power and LLS are similar to the overall
DRAGN population, the RGZ SDRAGNs extend to larger $z$ and P$_{1400 ~\rm MHz}$ than the previous SDRAGNs,
and the boundary between FR I and FR II sources appears at lower  P$_{1400 ~\rm MHz}$ for SDRAGNs than for general samples of double sources.

As also found by \cite{Wu2022}, the new Zoo Gems SDRAGN sample
is dominated by radio-source morphologies of \cite{FanaroffRiley} type II (FR II). We find no FR I structures, and two
radio morphologies which lack both hot spots and center-brightened jets and might be regarded as
transitional between FR I and FR II. These are J1322+2706 (IC 4234), where LOFAR data show what may be two
filamentary lobes partially overlapping in projection, much like the structure of Fornax A; and  J1646+3831,
with similar structure detected by VLASS data. Neither of these shows the kind of center-brightened twin jets which define FR I sources. 
J0219+0155 (UGC 1797) shows symmetric hot spots with
additional, collimated but more diffuse emission at larger radii. All previous SDRAGNs, except the FR I case J0725+2957 (B2 0722+30) have
FR II morphologies, based on published maps and
our re-examination of the 1400-MHz VLA data on J0315-1906 from \cite{Ledlow1998}; its jet is one-sided and the peak
lobe emission on each side is well-separated from the core. 
The FR I/II difference has long
been viewed as mostly driven by jet power, reflected in radio output, with FR II sources typically at log P$_{1400 ~\rm MHz}$ (W Hz$^{-1}$) $>  23.93$ ; examining this division, \cite{LedlowOwen} found that the dividing value depends on
properties of the host galaxies as well, occurring at higher radio power for more optically luminous host galaxies. However, the extensive LoTSS sample analyzed by \cite{Clews} brings 
such a straightforward distinction into question, while still suggesting a role for local environment in whether jets disrupt within the host galaxy or not.  
Excluding from either FR category the two RGZ SDRAGNs which do not
fit the typical FR I/II description, FR II morphologies are seen to unusually low power P$_{1400 ~\rm MHz}$; the only distinct FR I source is the previous SDRAGN J0725+2957 at
log P$_{1400 ~\rm MHz}$  (W Hz$^{-1}$) $= 23.69$. There are an additional 3 FR II sources (out of a total of 23 of both FR types) among SDRAGNs below this nominal division, one as low as 
log P$_{1400 ~\rm MHz} $ (W Hz$^{-1}$) $=  22.65$.
While the FR I/II luminosity division is not as well-defined as once believed (\citealt{Mingo}, \citealt{Clews}), having such low-power FR II sources in this small sample suggests that in this way SDRAGNs 
differ from
the whole population of DRAGNS. It may be relevant, as \cite{Mingo} speculate under a paradigm where jet disruption in the surrounding medium shapes the FR I/II division, that FR II
sources could be occur at lower power in unusually low-density environments. We may address this possibility using galaxy environments in section \ref{environments};
since SDRAGNs occur preferentially in denser environments than average, this mechanism by itself will not account for their preponderance of FR II structures even at lower power.
 The numerous Seyfert nuclei in spirals form a useful comparison sample of AGN in similar host galaxies. Their extended radio emission has been
 identified either as broad outflows, sometimes with a component driven by a circumnuclear starburst, or jets reaching only to scales of a few kiloparsecs and oriented randomly
 with respect to the galaxy disk \citep{Kinney}.  
 We might speculate (joining previous workers such as \citealt{Gallimore}) that disruption of the jets in gas-rich disks, particularly for jets
propagating near the disk plane, can be so complete that few FR I structures survive, while those jets which do persist outside the host galaxy ISM remain highly collimated and produce FR II structures.


The ``previous" SDRAGNs are found at lower redshifts, and have correspondingly lower radio power, than our RGZ+HST sample which were found in a specific search. Both span the same
wide range in LLS as FIRST sources in general (where redshifts are known). We attribute this to the RGZ project carrying out a systematic visual as well as algorithmic search for spiral hosts, 
and the use of HST images to verify spiral structure (or eliminate candidates). The RGZ SDRAGNS have mean $z=0.163$ (median 0.169) compared to the previous SDRAGNs with
mean $z=0.063$ (median 0.068). Similarly, the RGZ SDRAGNs have mean  P$_{1400 ~\rm MHz}$  (W Hz$^{-1}$) $ =24.63$, while previous SDRAGNs have mean  log P$_{1400 ~\rm MHz}$  (W Hz$^{-1}$) $= 23.87$, 
a luminosity ratio 5.7 which is close to 
the ratio 6.7 expected for the same flux limits simply based on the mean redshifts.

In contrast, the RGZ SDRAGNs have smaller source sizes (LLS) in
several straightforward metrics, although the small sample sizes mean that the formal probability of coming from
the same distribution, using the very conservative two-sample Kolmogorov-Smirnov test, is significant ($\approx 30$\%
when the
direction of the difference between samples is not specified beforehand). The median projected separations
are 280 kpc for the 11 previously identified SDRAGNs versus 146 kpc for the new HST sample of 15, while if a mean is taken
in the log, the values are 319 kpc for old versus 144 kpc for new. The extreme values differ in the same direction - the
three of the  four smallest-separation sources occur in the new HST sample (the outlier is J0725+2957, an FR I source where the lobe
separation is small and ill-defined), while the two largest were previously found. Similar behavior occurs for the largest projected size (median LLS 212 kpc for new ones versus 350 for old), with the three largest values among old SDRAGNs and two of the three smallest in the new sample. To some extent
these differences result from selection at different frequency bands - several of the previous objects showed radio lobes 
only when observed at lower frequencies than the 3-GHz band used for the VLASS. Indeed, of the two previous
SDRAGNS with lobe separations exceeding 1 Mpc, the lobes of J1633+3520 (NGC 6185) were detected in LOFAR data at
144 MHz \citep{Oei}, while those of J2345-0449 were found in NVSS data at 45\arcsec resolution \citep{Andernach2012}, and further examined using GMRT data at 325 MHz,
\citep{Bagchi}. 
Furthermore, the outermost ``relic" lobes of J1409$-$0302 (Speca), not included in the sizes from 
Table  \ref{table-previous}) for better consistency, are detected only at 325 MHz \citep{Hota}, and the outer pair of lobes in J1350$-$1634
was detected at 200 and 887 MHz  \citep{Sethi}.

For samples as small as even the whole set of known SDRAGNs, we do not attempt to calculate a radio luminosity function, especially in view of the optical selection effects
shown in section 2.5. We do note that the ranges of P$_{1400~\rm MHz}$ and LLS are similar between SDRAGNs and ordinary DRAGNs. We illustrate this in Fig. \ref{fig-sizepower},
a comparison of these samples in the power-size plane. The FIRST sample's 40-mJy flux limit gives a small sampling volume at the lowest P$_{1400 ~\rm MHz}$ values represented
by SDRAGNs, as seen by the points at the bottom of the diagram. Otherwise, the joint ranges of P$_{1400 ~\rm MHz}$ and LLS are quite similar, with a hint that
large LLS values are more common among SDRAGNs. 

\begin{figure*}
\includegraphics[width=120.mm,angle=90]{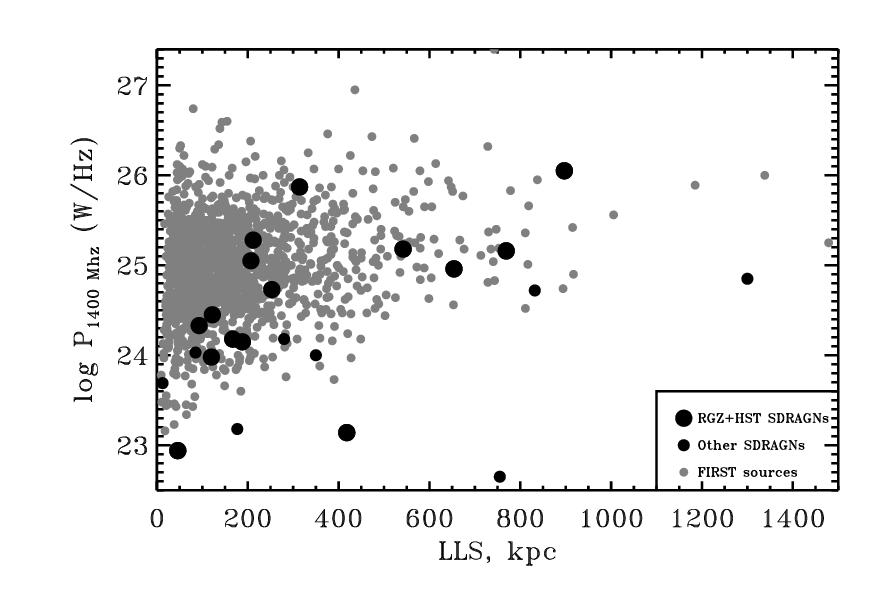}
\caption{Distribution of the SDRAGN samples in 1400-MHz power and 
LLS. The new RGZ+HST SDRAGNs are shown as large black circles, smaller black circles show those
previous SDRAGNs for which 
flux information is available, and a comparison set of 1329 FIRST sources from \cite{MiraghaeiBest} is shown as small grey circles.
Other SDRAGNs are systematically at lower power than our new sample, fitting with their lower redshifts and brighter optical magnitudes. 
Two previous SDRAGNs are not shown:
J1633+2635 would lie beyond the right edge at 2450 kpc, but its 1400-Mhz power remains very poorly constrained, while J2318+4314
lies below the bottom at 207 kpc, log $P_{1400 ~ \rm Mhz} = 22.05$.}
\label{fig-sizepower}
\end{figure*}

\subsection{Orientations of jets and host-galaxy disks}

The dense interstellar gas in spirals compared to ellipticals might suggest that jets escape from spirals preferentially where the column
densities are lowest, along the poles of the disk, 
a conjecture which we can test with the combined SDRAGN samples. For the RGZ SDRAGNs,
Table \ref{table-radiosources} shows geometrical properties of the radio-lobe sources (separation of peaks from the core and position angle
of this separation) along with the angles $\theta_1, \theta_2$ between poles of the galaxy disks and each radio source (in the sense that
zero indicates radio sources exactly aligned with the galaxy poles in projection). As indicated, some core positions use the
optical location of the galaxy nucleus, when the radio core is undetected or potentially confused with interferometric artifacts. 
The orientations are summarized visually, along with the previous SDRAGNs, in Fig. \ref{fig-diskjet}. These data
allow us to examine the alignments of radio sources and host galaxies, since projection effects operate in different ways for various
combinations of disk orientation and (unknown) angle between the radio axis and line of sight. When we see the disks edge-on,
the second angle can be treated as random but independent of the disk inclination.

\begin{figure*}
\includegraphics[width=175.mm,angle=90]{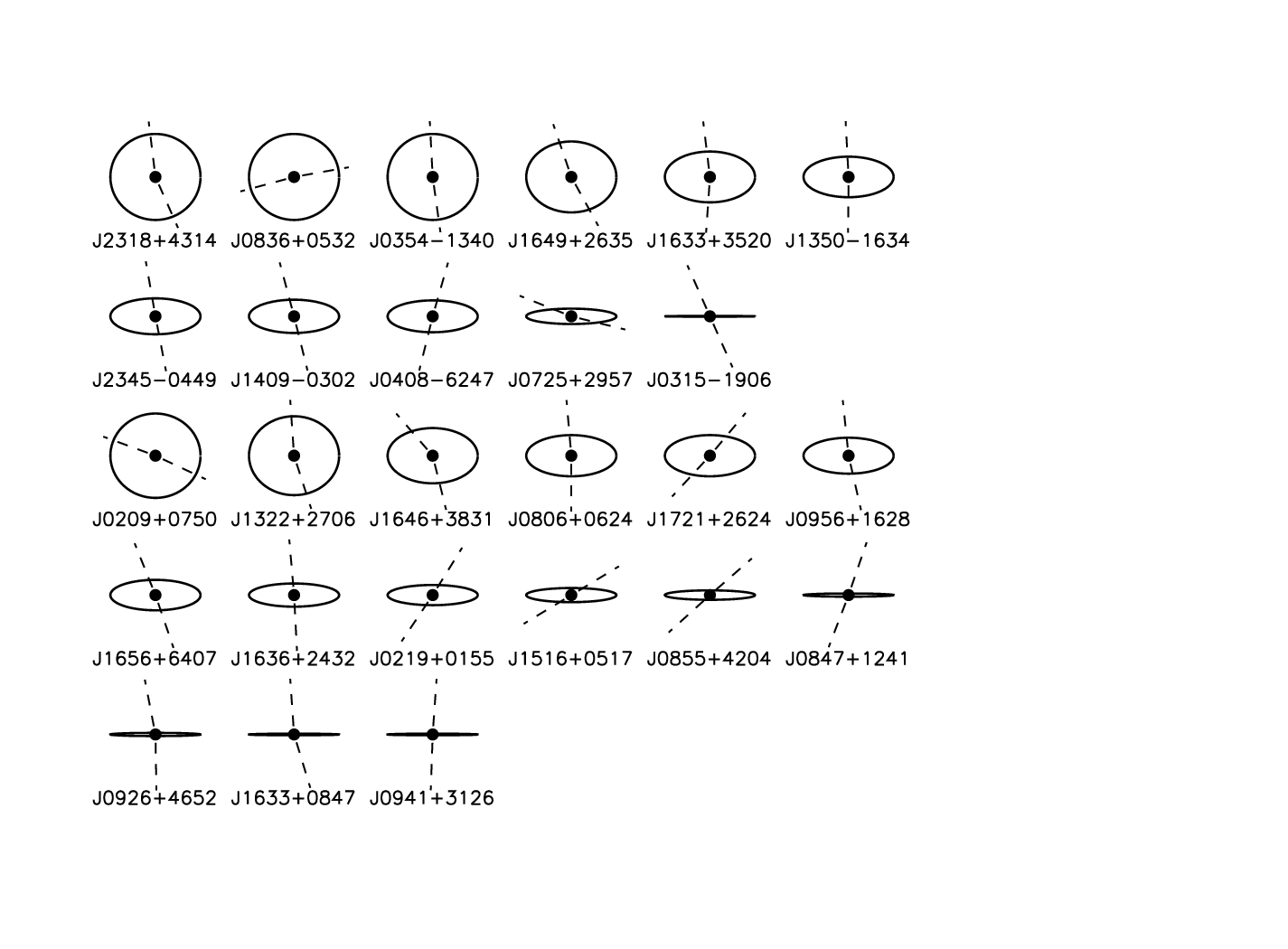}
\caption{Graphical summary of the galaxy-disk/radio-jet orientations in the previous SDRAGN sample (upper two rows) and the new RGZ SDRAGNs (lower three rows). Ellipses indicate the viewing geometry of each galaxy disk (showing the adopted disk 
viewing angle, not the isophotal shapes) and dashed lines show directions of the radio jets, inferred from the radio structures, pointing toward hot spots where present even if not centered in the lobes as an indicator of the most recent outflow direction. For clarity, each
sample is sorted by observed disk inclination, from face-on to edge-on. }
\label{fig-diskjet}
\end{figure*}

\begin{deluxetable*}{lccccccccc} 
\tablecaption{Core-lobe separations and orientations\label{table-radiosources}}
 \tablehead{
 \colhead{Object}       &  \colhead{Core ID} &    \colhead{PA$^\circ$}  & \colhead{Sep$_1$ \arcsec} &    \colhead{PA$^\circ$} &  \colhead{Sep$_2$ \arcsec} & 
 \colhead{${\theta_1}^\circ$} & \colhead{${\theta_2}^\circ$} & \colhead{Sep (kpc)} & \colhead{LLS, kpc}}
 \startdata
J0209+0750  & Radio   & $-60.7\pm1$ & 36.1  & $124.3\pm1$ & 32.0 & --- & ---& 269 & 314 \\
J0219+0155                        & Radio    & $4.7\pm1$ &  22.0  & $182.6\pm1$ & 19.7 &  32  & 35 & 32.7 & 93   \\
J0806+0624  & Optical  & $68.1\pm5$ &  5.2  & $243.0\pm5$ &  9.1 & 5  & 0  & 28.2 &46 \\
J0847+1241  &Radio     & $4\pm2$ &   16.1  & $182 \pm 2$ & 15.2 &  19 & 21&  94.5 & 120\\
J0855+4204                       & Optical   & $39\pm2$ &   4.6   & $220.3\pm3$ &   3.8 & 49 & 48 & 25 & 542 \\ 
J0926+4652  & Radio    & $-29 \pm2$ & 18.4 & $162.6\pm2$ &  18.4 & 11 & 1 &132 & 188\\
J0941+3126                    & Radio    & $53.5\pm1$ & 71.9 & $234.8\pm1$ &   83.4 & 3  & 3 & 829 & 897 \\ 
J0956+1628  & Radio    & $-49.8\pm1$ & 20.6  & $117.7\pm1$ & 17.3 & 8 & 19 & 159 & 212 \\
J1322+2706   & Radio &  $ 114 \pm 5$ & 137 &   $ -52 \pm 5$  & 198 & 7  & 7 & 190 & 418 \\ 
J1516+0517     & Radio & $ -82 \pm 2$ & 58.7  & $98 \pm 1$ &  38.6 & 59 & 59 & 99 & 122 \\ 
J1633+0847  & Radio    & $-24.7\pm1$ & 26.6 & $142.4\pm1$ &  28.9 & 17 & 4 & 197 & 253 \\
J1636+2432 & Optical    & $39.7\pm2$ & 43.4 & $218.4\pm1$ & 35.3  & 4 & 3 & 146 & 167  \\
J1646+3831 & Optical & $161 \pm 4$ & 30 & $ -46 \pm 3$ & 54 & 41  & 16  & 116 & 207 \\ 
J1656+6407  & Radio     & $-10.6\pm1$ & 70.1 & $170.7\pm1$ & 88.0 & 19 & 10 & 572 & 654 \\
J1721+2624  & Radio     & $68.8\pm1$ &  103.2 & $246.4\pm1$ & 121.5 & 41 & 43 & 654 & 769 \\
\enddata
\tablecomments{For each radio lobe, we give the position angle PA in degrees, north through east, and associated projected separation of the lobe peak from the galaxy nucleus. The projected
angle between the implied poles of each galaxy disk and the jets (1 and 2) are shown as $\theta_1$ and $\theta_2$. Projected lobe separations ``Sep" and LLS values are shown for reference.}
\end{deluxetable*}


Especially for the edge-on host galaxies, inspection shows that the orientation of the double source (and by
implication jets) favors directions roughly perpendicular to the galaxy disks. We can examine this quantitatively using 
geometry similar to the
approach of \cite{Clarke} as extended by \cite{Kinney}. We define the inclination $i$ in Table \ref{table-Hubbletypes} to
be zero for edge-on galaxies, making the smaller inclination uncertainties in
these systems clear, and $90^\circ$ for face-on galaxies. Since the distribution of host-galaxy inclinations shows clear selection effects,
we adopt a numerical approach to generate the distribution of probability of the projected angle $\theta$
between radio jets and poles of the host galaxy disks as simple parameters of the proposed distribution of
the true angle $\phi$, using the probabilities at the observed values of $i, \theta$ to estimate the range
of acceptable distributions of $\phi$. In doing this, we can combine old and new SDRAGN samples as well
as compare their separate outcomes for consistency. (The ranges of lobe separation and LAS are so wide for both old and new
sets of SDRAGNs that neither quantity shows any correlation with disk inclination, nor would we plausibly see such
correlations with this sample size and radio-size range.)

Those galaxies seen nearly edge-on convey the most information on the distribution of $\phi$. If the distribution is populated only up to a limiting angle $\phi_{max}$, there will be no systems in a region 
\footnote{Derived in response to another query, by user {\it intelligenti pauca} at https://math.stackexchange.com/questions/3728462/cone-projected-tangent-angle.}
with $\theta > \theta_c$ where
$$ \theta_c = \tan^{-1} {{\tan( \phi_{max})} \over {\sqrt{\cos^2 (i) - \tan^2( \phi_{max}) \sin^2 (i)}}} $$
which ranges from $\theta_c = \phi_{max}$ at $i=0$ to
$\theta_c = 90^\circ$ at $i = 90^\circ - \phi_{max}$. Comparison with the observed distribution of $\theta, i$ for both old and new SDRAGNs (Fig. \ref{fig-thetaphi})
gives a robust conclusion that $\phi_{max} \approx 60^\circ$ for both old and new sets of SDRAGNs. While not a strong constraint in 
angle itself, this implies that half of the possible solid angles for jet escape from the AGN are not seen in these samples. At face value,
the previous sample satisfies $\phi_{max} < 35^\circ$. The bunching of points at small $\theta$ suggests that most
systems are better aligned than these limits.

In greater detail, we can constrain the distribution of $\phi$ by comparing the distributions in Fig. \ref{fig-thetaphi}
to numerical realizations of simple models for the probability distribution P($\phi$).
To do this, we use the HEALPIX \citep{Gorski} divisions of the sphere into
equal-area regions, providing convenient sets of unit-vector directions evenly sampling the sphere in solid angle
\footnote{We used the FITS files listing the coordinates from https://lambda.gsfc.nasa.gov/toolbox/pixelcoords.html.}.
We use these unit vectors to represent possible jet directions from the AGN, associating each with a weight given by a 
desired probability distribution in $\phi$. We rotate
these unit vectors to correspond to observed galaxy disk inclination $i$, project them onto the sky plane, and derive the predicted 
distribution of projected pole-jet angle $\theta$ incorporating the weightings corresponding to trial distributions $P(\phi)$
and, for some purposes, differing solid angles $\propto \cos (i)$ where galaxy disks are seen with inclination $i$. It is 
most convenient, remaining close to observed quantities, to consider the distributions in the ($i,\theta$) plane. Populating $1^\circ$ bins in 
both $i$ and $\theta$, we found it sufficient to use the HEALPIX order 7 sampling (196608 points with spacing $0.45^\circ$).
Slices of this plane in restricted ranges of $i$ can be examined to isolate regimes most sensitive to $P(\phi)$ or to allow for selection effects depending on galaxy inclination. 


\begin{figure*}
\includegraphics[width=140.mm,angle=90]{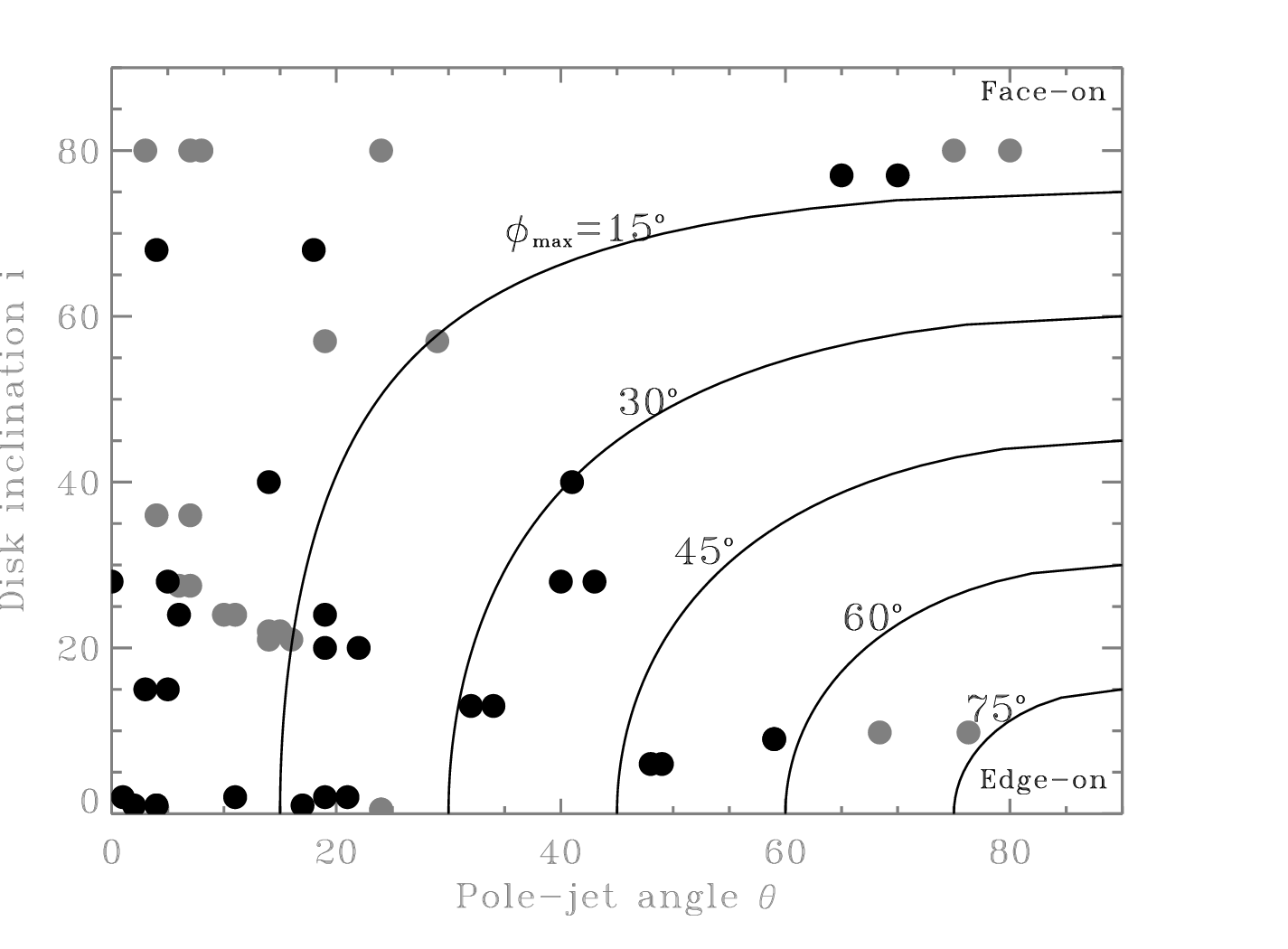}
\caption{Distribution of the SDRAGN samples in disk inclination $i$ and projected angle $\theta$ between jet direction and poles of the
galaxy disk. New RGZ SDRAGNs are shown as black-filled circles, and jets in previously-known SDRAGNS are
marked in grey. Individual jets are shown, usually giving two symbols per galaxy except where the two
jet directions are very close to $180^\circ$ apart. Overlaid curves show the edges of distributions with a cutoff
$\phi_{max}$ in the physical distribution of pole-jet angle $\phi$; regions below and to the right of each curve will be empty
for the indicated $\phi_{max}$. We analyze the objects with $i < 30^\circ$ where projection effects are weakest. The ``outlier" jets
beyond $\phi_{max}=60^\circ$ belong to J0725+2957, the only clear FR I source in either SDRAGN sample; implications of including it or excluding this system in the analysis are described in the text.}
\label{fig-thetaphi}
\end{figure*}

Without assuming a form for $P(\phi )$, a simple analysis shows a grouping of jet alignments toward small $\phi$.
Annular ranges of $\phi$ project to areas in the $i, \theta$ plane which are weighted toward the largest occupied values of $\theta$, so
inward reconstruction from the largest occupied $\theta$ is feasible for systems near edge-on (we adopt $i < 30^\circ$). One
takes the outermost occupied annulus, matches the density of galaxies to its peak, and subtracts the expected contribution at smaller 
$\theta$ from the relevant values of $\phi$, and steps inward in annuli of $\phi$. Combining previous and new sets of SDRAGNs,
and considering each jet rather than a mean for each galaxy, we have 30 (32 including the lone FR I source) measures with $i < 30^\circ$, and divide these
into three bins of nearly equal population: $\theta < 15^\circ$, $ 15^\circ < \theta < 30^\circ$, and $30^\circ < \theta < 60^\circ$. We initially consider the
sample without the single, outlying FR I source J0725+2957 (which is also the physically smallest source in projection, and for
these reasons might possibly belong to a different population).
Matching the number of jets in the outermost bin of $\theta$ and carrying out this procedure for the three bins yields 
fractions of jets in the three bins of $\phi$ as follows: 0.13 : 0.42 : 0.44 (and zero for $\phi > 60^\circ$). These ratios imply that the fraction of jets per unit solid angle
is peaked toward small values, since the solid angle included in an annulus on a sphere with inner (outer) radius $\phi_1$ ($\phi_2$)
is $2 \pi (\cos \phi_1 - \cos \phi_2)$. Specifically, for the same three annuli, the relative numbers of jets per unit solid angle
are 1: 0.95 : 0.7. Given the sparse statistics, this result is consistent with a simple model in which the probability $P(\phi)=$ constant for $\phi < 30^\circ$ and dropping linearly to zero over the range 30--60$^\circ$. 
Fitting only those systems where the host galaxy is seen nearly edge-on avoids issues with the selection effects described in Section \ref{selectioneffects}; it is this part of the
galaxy distribution which is more sensitive to the distribution of jet/pole angles. Including the lone FR I system J0725+2957 in this fit has somewhat paradoxical results. In order to have significant
probability of seeing a galaxy at $\theta=75^\circ$ in a sample this small, without overpredicting the fraction of sources between $\theta = 30^\circ - 60^\circ$, not only
does $\phi_{max}$ have to be at least $75^\circ$, but the dropoff for $\phi > 30^\circ$ has to be much steeper than our assumed linear ramp, or start for $\theta < 30^\circ$.
The current sample size does not allow a more detailed fit to the distribution $P(\phi)$, but does clearly indicate a preference for jets to emerge within $30^\circ$ of
the poles of the galaxy disks.

%


This alignment of the large-scale emission with the poles of the host galaxy disk matches the alignment of
(a small sample of) spirals with VLBI-scale jets analyzed by \cite{KSDM}, but contrasts with the essentially
random alignments of kpc-scale emission (well within the disk radius) found for Seyfert galaxies in spiral
hosts by \cite{Kinney}. The preferential occurrence of near edge-on galaxies in this SDRAGN sample contrasts with the
lack of such systems (especially for broad-line AGN) in optically-selected samples (\citealt{Keel1980}, \citealt{TovmassianYam},
\citealt{Lagos})

The analysis above, using a disk-inclination limit which respects the selection factors going into the sample, suggests a
stronger trend toward radio jets perpendicular to the disks than found by \cite{Wu2022}. They note that this trend
is strongest in galaxies with pseudobulges rather than classical bulges, since two of the three least-aligned systems in their subsample were the only ones where 
they measured a core Sersic index $n>2$ indicating classical bulges and histories including major mergers). The combination of pseudobulges and
alignment of the radio jets with poles of the host disk may indicate that important factors in the occurrence of SDRAGNs relate 
more to directly the SMBH spin than to (the more intuitively influential) SMBH mass. Simulations show that the spins of central black holes which have grown without major mergers 
are more aligned with the host axes than the general population, though the alignment
for the whole sample of merger-free hosts is weaker than might have been expected
(\citealt{Smethurst2024}, \citealt{Beckmann}); the angular momentum changes of gas infalling from directions
out of the host plane, though substantial, may be outweighed by effects during the merger of two black holes originally
at the cores of the merging galaxies.

 These results also fit with simulations showing that jets of moderate power, in the range affected by buoyancy in a stratified interstellar medium, can be reoriented
 toward the poles of a disk (\citealt{YatesJones}, \citealt{HodgesKluck}). In this case, the host galaxies contribute to the orientation of double radio sources in a more active way than
 disruption of jets within the disk, and reorient some of them in directions favoring escape from the host ISM. Some of the SDRAGNs
 may show evidence of this process, where the radio-source structures on the smallest resolved scales near the nuclei  in the VLA maps (Fig. \ref{fig-sdragnmontagebw},
 especially in J0219+0155, J0855+4204, and J0941+3126) are less aligned with the host-disk poles than
 on larger scales.

\section{Conclusions}

We have analyzed HST images, radio structures, and optical spectra for spiral galaxies hosting large-scale double radio sources (Spiral Double Radio-source AGN or SDRAGNs), including eleven
previously known and 15 newly identified from RGZ and HST short-exposure data. Thirteen of the new
identifications are of high probability, incorporating astrometric association of a radio core with the optical galaxy nucleus, while the remaining
two use location between radio lobes in cases where a radio core has not been detected.

We identify selection effects with galaxy inclination operating within the SDRAGN samples, in opposite directions for brighter galaxies (seen
preferentially face-on) and fainter ones (preferentially edge-on). We attribute these differences to the ability to resolve spiral structure in face-on galaxies compared to the sensitivity of bulge/disk decomposition improving for edge-on galaxies. Subdividing the plane of galaxy inclination and angular offset between
host galaxy orientation and radio jets to isolate combinations most sensitive to the intrinsic distribution of angles $\phi$ between radio
jets and the axes of the galaxy disks, we find that radio jets in SDRAGNs preferentially escape the galaxies near their disk poles. The 
admittedly sparse statistics are consistent with a probability distribution P($\phi$) = constant from $0-30^\circ$, ramping down to zero for $\phi > 60^\circ$.

Both old and new subsets of SDRAGNs show distinct properties. SDRAGNs are seen preferentially close to edge-on,
partly explained by selection effects.  They occur throughout the sequence of Hubble types from Sa-Scd, and most of them have pseudobulges rather than classical bulges \citep{Wu2022}. 
FR II radio structures predominate, as do narrow-line (type 2) Seyfert spectra among their optical AGN. Strong interactions or mergers are rare,
but disturbed dust lanes are not. SDRAGNs are found in denser galaxy environments than average, even with the limited accuracy of
photometric redshifts. 

The prevalence of pseudobulges suggests that the central supermassive black holes (SMBHs) have grown via internal secular processes rather than through
major mergers, leading to closer alignment between their spin axes and those of the host disks (both leading to accretion disks and jets
aligned with the host properties as well) than found for random samples of galaxies. As a result, jets emerging near the poles of the host disk and its interstellar
material will encounter the least material to disrupt them (contrasting with the kpc-scale jets in many spirals with Seyfert nuclei). Warped dust disks may suggest that weak interactions a few Gyr ago are
common, perhaps increasing the flow of material to the SMBH accretion disk. The wide range of Hubble stages, and low bulge luminosities in
some systems, suggest that SMBH mass in itself is not the most important factor in the occurrence of SDRAGNs.

Put together, these factors
suggest that the rarity of SDRAGNs compared to DRAGNS may result from a combination of individually rare properties, beginning with
alignment between the jets and poles of the host galaxy disk. Further large and well-quantified samples, such as
that from cross-correlating Euclid and LOFAR sources \citep{EuclidLOFAR}, should be able to test this idea, and further refine our understanding of how galaxies produce these large-scale radio sources.

\section*{acknowledgments}
This research is based on observations made with the NASA/ESA Hubble Space Telescope obtained from the Space Telescope Science Institute, which is operated by the Association of Universities for Research in Astronomy, Inc., under NASA contract NAS 5-26555. These observations are associated with program 15445.

Some SDRAGN candidates were identified by additional Radio Galaxy Zoo participants
1001GLEN, 958bacsal, A1001, aeolus, andymarrison, antikodon, aufelipe, axrldn, AZooKeeper0001, bartinhogoool, 
basst82, BoazWildcat , Brucea, cafeeciencia, Cardiffian, civilsparky, Claude Cornen, 
Corcaroli, cserfalvi, csunjoto, dude1818, ElisabethB, Emmabray , equidad1, Explorer15, firejuggler, gavinrider.
graham d, Gweilouk, HannaViolet, Hruk, infoservador, jesse.rehm, jessicamh, jiipee, jo-luc, joconnell,
leonie\_van\_vliet, lethalparadox0, Martin\_Sexton, mbstone12, mdwilber, Milkybear, nico775, Olly314, ongole.geeth,
OrlandoGR, Peter van Zuylen, planetari7, planetaryscience, Ptd, rezoloot, Rickss, SG1966, sharqua, southernclaw,
tcwang, teamaynard, theyak330, Tony Wei, Ushiromiya Xyrius , WeBs in space, WizardHowl, wpatelunas, 
xDocR, and zutopian. 
This publication has been made possible by the participation of more than 8700 volunteers in the 
Radio Galaxy Zoo project. Their contributions are individually acknowledged at http://rgzauthors.galaxyzoo.org. We thank Rolf
Jansen for a FITS version of the NGC 6185 optical spectrum and supporting information, B. Emonts for FITS versions of WHT spectra of B2 0722+30, and Frazer Owen for finding copies of the
optical spectra of J0315-1906. This research has made use of the NASA/IPAC Extragalactic Database (NED), which is funded by the National Aeronautics and Space Administration and operated by the California Institute of Technology.

The National Radio Astronomy Observatory is a facility of the National Science Foundation operated under cooperative agreement 
by Associated Universities, Inc. This research has made use of the CIRADA cutout service at URL cutouts.cirada.ca, operated by the Canadian Initiative for Radio Astronomy Data Analysis (CIRADA). CIRADA is funded by a grant from the Canada Foundation for Innovation 2017 Innovation Fund (Project 35999), as well as by the Provinces of Ontario, British Columbia, Alberta, Manitoba and Quebec, in collaboration with the National Research Council of Canada, the US National Radio Astronomy Observatory and Australia’s Commonwealth Scientific and Industrial Research Organisation. LOFAR data products were provided by the LOFAR Surveys Key Science project (LSKSP; https://lofar-surveys.org/) and were derived from observations with the International LOFAR Telescope (ILT). LOFAR (van Haarlem et al. 2013) is the Low Frequency Array designed and constructed by ASTRON. It has observing, data processing, and data storage facilities in several countries, which are owned by various parties (each with their own funding sources), and which are collectively operated by the ILT foundation under a joint scientific policy. The efforts of the LSKSP have benefited from funding from the European Research Council, NOVA, NWO, CNRS-INSU, the SURF Co-operative, the UK Science and Technology Funding Council and the J\"uich Supercomputing Centre.

This work is partly based on observations obtained with the 6-m telescope of the Special Astrophysical Observatory of the Russian Academy of Sciences carried out with the financial support of the Ministry of Science and Higher Education of the Russian Federation. The renovation of the  telescope equipment is currently provided within the national project "Science and Universities". The work on AGN spectroscopy was performed as part of the SAO RAS government contract approved by the Ministry of Science and Higher Education of the Russian Federation.

The Legacy Surveys project is honored to be permitted to conduct astronomical research on Iolkam Du'ag (Kitt Peak), a mountain with 
particular significance to the Tohono O'odham Nation.
The Legacy Surveys consist of three individual and complementary projects: the Dark Energy Camera Legacy Survey (DECaLS; Proposal ID \#2014B-0404; PIs: David Schlegel and Arjun Dey), the Beijing-Arizona Sky Survey (BASS; NOAO Prop. ID \#2015A-0801; PIs: Zhou Xu and Xiaohui Fan), and the Mayall $z$-band Legacy Survey (MzLS; Prop. ID \#2016A-0453; PI: Arjun Dey). DECaLS, BASS and MzLS together include data obtained, respectively, at the Blanco telescope, Cerro Tololo Inter-American Observatory, NSF's NOIRLab; the Bok telescope, Steward Observatory, University of Arizona; and the Mayall telescope, Kitt Peak National Observatory, NOIRLab. 
NOIRLab is operated by the Association of Universities for Research in Astronomy (AURA) under a cooperative agreement with the National Science Foundation.

This project used data obtained with the Dark Energy Camera (DECam), which was constructed by the Dark Energy Survey (DES) collaboration. Funding for the DES Projects has been provided by the U.S. Department of Energy, the U.S. National Science Foundation, the Ministry of Science and Education of Spain, the Science and Technology Facilities Council of the United Kingdom, the Higher Education Funding Council for England, the National Center for Supercomputing Applications at the University of Illinois at Urbana-Champaign, the Kavli Institute of Cosmological Physics at the University of Chicago, Center for Cosmology and Astro-Particle Physics at the Ohio State University, the Mitchell Institute for Fundamental Physics and Astronomy at Texas A\& M University, Financiadora de Estudos e Projetos, Fundacao Carlos Chagas Filho de Amparo, Financiadora de Estudos e Projetos, Fundacao Carlos Chagas Filho de Amparo a Pesquisa do Estado do Rio de Janeiro, Conselho Nacional de Desenvolvimento Cientifico e Tecnologico and the Ministerio da Ciencia, Tecnologia e Inovacao, the Deutsche Forschungsgemeinschaft and the Collaborating Institutions in the Dark Energy Survey. The Collaborating Institutions are Argonne National Laboratory, the University of California at Santa Cruz, the University of Cambridge, Centro de Investigaciones Energeticas, Medioambientales y Tecnologicas-Madrid, the University of Chicago, University College London, the DES-Brazil Consortium, the University of Edinburgh, the Eidgenossische Technische Hochschule (ETH) Zurich, Fermi National Accelerator Laboratory, the University of Illinois at Urbana-Champaign, the Institut de Ciencies de l'Espai (IEEC/CSIC), the Institut de Fisica de Altes Energies, Lawrence Berkeley National Laboratory, the Ludwig Maximilians Universitat Munchen and the associated Excellence Cluster Universe, the University of Michigan, NSF's NOIRLab, the University of Nottingham, the Ohio State University, the University of Pennsylvania, the University of Portsmouth, SLAC National Accelerator Laboratory, Stanford University, the University of Sussex, and Texas A\& M University.

BASS is a key project of the Telescope Access Program (TAP), which has been funded by the National Astronomical Observatories of China, the Chinese Academy of Sciences (the Strategic Priority Research Program ``The Emergence of Cosmological Structures" Grant \#XDB09000000), and the Special Fund for Astronomy from the Ministry of Finance. The BASS is also supported by the External Cooperation Program of Chinese Academy of Sciences (Grant \# 114A11KYSB20160057), and Chinese National Natural Science Foundation (Grant \# 11433005).

The Legacy Survey team makes use of data products from the Near-Earth Object Wide-field Infrared Survey Explorer (NEOWISE), which is a project of the Jet Propulsion Laboratory/California Institute of Technology. NEOWISE is funded by the National Aeronautics and Space Administration. The Legacy Surveys imaging of the DESI footprint is supported by the Director, Office of Science, Office of High Energy Physics of the U.S. Department of Energy under Contract No. DE-AC02-05CH1123, by the National Energy Research Scientific Computing Center, a DOE Office of Science User Facility under the same contract; and by the U.S. National Science Foundation, Division of Astronomical Sciences under Contract No. AST-0950945 to NOAO. This research uses services or data provided by the Astro Data Lab, which is part of the Community Science and Data Center (CSDC) Program of NSF NOIRLab. NOIRLab is operated by the Association of Universities for Research in Astronomy (AURA), Inc. under a cooperative agreement with the U.S. National Science Foundation.

%

\vspace{5mm}
\facility{HST(ACS)}
\facility{BTA} 
\facility{SARA}
\facility{Astro Data Lab}


\software{IRAF (\citealt{Tody1986}, \citealt{Tody1993}), IDL
          }



\appendix

\section{Radio Galaxy Zoo cross-identifications}
 
To facilitate searches in the RGZ forum discussions, Table \ref{table-crossids} lists SDSS identifiers (used in the form SDSS J001627.47+022602.1) for all RGZ candidate SDRAGN host galaxies along with the
Radio Galaxy Zoo identifiers, used in the format RGZ J020904.7+075004, and the internal RGZ alphanumeric identifiers, where these exist. RGZ coordinate names are
based on KD-tree collection of locations input by volunteers; in some cases the algorithm did not converge so there is no
such identifier. Due to galaxy structure, the final digits of the SDSS and RGZ coordinates do not always coincide, so we list both sets here. The table also lists the short identifier used in the text for convenience.

Precedence of catalogs used for other object names was guided by STScI coordinate resolution during APT target input for Zoo Gems, deleting the hyphens required by that tool to avoid imbedded spaces. This follows the common convention of
preference for NGC and IC catalogs over SDSS names, and B2 and B3 radio catalogs over longer coordinate-based names.

\begin{deluxetable*}{lcccl} 
\tablecaption{Host galaxy cross-identications and names\label{table-crossids}}
 \tablehead{
 \colhead{Short name} & \colhead{SDSS} & \colhead{Other name} & \colhead{RGZ coordinates} & \colhead{Identifier} }
 \startdata
J0016+0226 & J001627.47+022602.1	& & -- & b001 \\ 
J0209+0750 & J020904.75+075004.5 &  & J020904.7+075004 & b010 \\ 
J0219+0155 & J021958.73+015548.7 &  UGC 1797 & J021958.8+015555& c042 \\ 
J0802+1157 & J080259.73+115709.7 &   & J080259.7+115710 & a301 \\ 
J0806+0624 & J080658.46+062453.4 &  & J080658.4+062453 & c125 \\ 
J0813+5520 & J081303.10+552050.7 &   &   J081302.9+552051 & b016 \\ 
J0823+0330 & J082312.91+033301.3 &    &  J082312.9+033301 & d170 \\ 
J0832+1848 & J083224.82+184855.4 &   &  J083224.8+184855 & c102 \\ 
J0833+0457 & J083351.28+045745.4 &    &  J083352.3+045814 & b027 \\ 
J0847+1241 & J084759.90+124159.3 &   &  J084759.8+124159 & c043 \\ 
J0855+4204 & J085549.15+420420.1 &  B3 0852+422  & J085549.1+420420 & b039  \\ 
J0901+1648 & J090147.17+164851.3 &   & J090147.1+164851 & c084 \\ 
J0903+4328 & J090305.84+432820.4 &  &  J090306.2+432817 & c034 \\ 
J0914+4137 & J091445.53+413714.3 &  B3 0911+418  &   J091445.5+413714& a072  \\
J0919+1359 & J091949.07+135910.7 &    & J091948.1+135951 & a205 \\ 
J0926+4652 & J092605.17+465233.9 &   &  J092605.2+465232 & c118 \\ 
J0941+3126 & J094103.62+312618.7 & B2 0938+31A   & J094103.6+312618 & a248 \\ 
J0956+1628 & J095605.87+162829.9 &   & J095605.8+162830 & a221 \\ 
J0958+5619 & J095833.44+561937.8 &   &  J095833.4+561937 & d093 \\ 
J1128+2417 & J112811.63+241746.9 &    & J112811.6+241746 & a300 \\ 
J1136+1252 & J113648.57+125239.7 &     & --  & c114  \\ 
J1303+5119 & J130300.80+511954.7 &     &  J130300.7+511954 & d110 \\ 
J1322+2706 & J132259.87+270659.1 & IC 4234   &  -- & s008 \\ 
J1328+5710 & J132809.31+571023.3 &   &  J132809.1+571025 & a346 \\ 
J1349+4542 & J134900.13+454256.5 &    &  J134901.5+454259 & c103 \\ 
J1354+4657 & J135436.02+465701.4 & B3 1352+471    & J135435.8+465658 & b177 \\ 
J1457+2832 & J145753.81+283218.7 & 4C +28.38 & -- & b203 \\ 
J1509+5152 & J150903.21+515247.9 &  & -- & a154 \\ 
J1516+0517 & J151659.24+051751.5	 & & -- & c051  \\ 
J1613+3018 & J161358.61+301809.4  & B2 1611+30 & -- & a325 \\ 
J1633+0847 & J163300.85+084736.4 &    & J163300.8+084736 & a287 \\ 
J1636+2432 & J163624.97+243230.8 &   & J163625.5+243226 & b238 \\ 
J1646+3831 & J164628.41+383116.0 & B2 1644+38    & J164628.4+383115 & s014 \\ 
J1656+6407 & J165620.58+640752.9 &   & J165622.0+640633 &  s015 \\ 
J1721+2624 & J172107.89+262432.1 &     & -- & b252 \\ 
J2141+0821 & J214110.61+082132.6 &   & J214110.6+082132 & d106 \\ 
\enddata
\end{deluxetable*}

\section{Notes on individual sources and identifications}


SDSS J021958.73+015548.7 (UGC 1797) - 
Z-symmetry in radio suggests precessing jets.

SDSS J080259.73+115709.7 - VLASS data show complex radio structure, not a symmetric core+double. The galaxy is a 2-tailed merger or merger remnant.

SDSS J080658.46+062453.4 - No radio core detected, late-type spiral well centered between lobes, double radio 
source perpendicular to disk. Probable SDRAGN. However, a blue starlike object SDSS J080658.59+062453.2, r=20.05
might be a background QSO hosting that radio double.

SDSS J081303.10+552050.7 -  Radio core outside the spiral galaxy. The actual host galaxy may not appear in HST image, while a very red object
is detected just west of the spiral in Legacy Survey data.

SDSS J082312.91+033301.3 - Elliptical galaxy with multiple dust rings. Late merger remnant? Radio core, no clear double source.

SDSS J083224.82+184855.4 - Double source is very asymmetric. HST image shows galaxy merger with tails and dust lanes.

SDSS J083351.28+045745.4 - Complex or multiple radio sources, targeted galaxy is not obviously the AGN host.

SDSS J084759.90+124159.3 - core source only in VLASS, well aligned with edge-on spiral core. 
RACS shows diffuse
sources $\approx ~18$\arcsec N of the spiral and $\approx 1$\arcmin E of it,
too faint to say whether they may be separate sources. Diffuse emission 
leaves the possobility that two double are superimposed.

SDSS J085549.15+420420.1  (B3 0852+422) - 
Z-like radio symmetry suggests precessing jets.

SDSS J090147.17+164851.3 - No radio core associated with spiral. FIRST+VLASS source 0.4\arcsec from compact galaxy NW (similarly close to the interlobe line), making it the likely host identification. It is catalogued as CWISE J090146.92+164853.1 by \cite{CatWISE2020}, with WISE W12 color index +0.84.

{\emph SDSS J091445.53+413714.3 (B3 0911+418)  - the host galaxy shares features with Centaurus A, so we infer that it represents an advanced merger with
one spiral precursor rather than an SDRAGN. The stellar component appears nearly circular, with asymmetric outer isophotes, and is crossed by a thick dust lane which is twisted and accompanied by
additional dust patches on the  southeastern side. The VLASS radio structure shows an FR II source, unlike Centaurus A, with an axis projected about 47$^\circ$ to the pole of the dust lane. At $z=0.140$,
the projected size of the radio source corresponds to 204 kpc.}

SDSS J091949.07+135910.7 - Weak radio core associated with dim elliptical galaxy.

SDSS J092605.17+465233.9  - Symmetric double source, radio core coincident with nucleus of highly inclined spiral.

SDSS J094103.62+312618.7  (B2 0938+31A) - Radio core coincident with nucleus of edge-on spiral, twisted dust lane. Double source is very prominent in LoTSS.


SDSS J095833.44+561937.8 - Complex radio source, galaxy appears prolate with dust lane.

SDSS J112811.63+241746.9 - No radio core, face-on spiral galaxy not quite on source axis.

SDSS J113648.57+125239.7 - Nearest galaxy has very late Hubble type. Starlike object projected within 1\arcsec of radio core,
possibly contributing to SDSS fiber spectrum, is a more likely host identification.

SDSS J130300.80+511954.7 - Two superimposed double sources?



SDSS J134900.13+454256.5 - Asymmetric double, host galaxy unclear.

SDSS J135436.02+465701.4 (B3 1352+471) - Late-type spiral or disk, bulge very faint. 1.8\arcsec from centerline of coreless double. Faint red object ($r$=24.10, $z_{ph}=1.068$)
DESI J208.6491+46.9510 is about 3\arcsec WNW of the candidate SDRAGN and
is a much more likely host for the E-W FR II source. The DESI
object is also detected as CWISE J135435.81+465703.8  with WISE W12 color index
16.42-15.96=0.46.


SDSS J163300.85+084736.4 - Radio core matches nucleus of M104-like edge-on spiral. Double source very asymmetric or 
superimposed with another source. There is a faint radio bridge in RACS, and the SE lobe is extended close
to the source major axis. VLASS shows the lobe
to consist of several peaks, gradually bending from the source
major axis due E. Further SE there is a separate
FR II source oriented perpendicular to the SDRAGN.

SDSS J163624.97+243230.8 - Inclined late-type spiral right on axis at center of double source, no detected radio core.
There is another possible host closer to the NE lobe, WISEA
which is faint, but uncatalogued in Pan-STARRS, but listed as DESI J249.1075+24.5446
on legacysurvey.org in DESI DR9 with $z_{ph}=0.873$, aka WISEA J163625.79+243240.4 with
W12=16.67-16.54=0.13,  which would make it an FR II of 600 kpc, not unusual.



SDSS J172107.89+262432.1 - 
RACS shows a diffuse radio bridge connecting the outer hotspots.

SDSS J214110.61+082132.6 - Small double source, not centered on distorted galaxy.

\section{Extended emission-line regions in SDSS J095833.44+561937.8}\label{AppendixEELR}

Our BTA long-slit spectrum of the radio galaxy SDSS J095833.44+561937.8 (J0958+5619) shows high-ionization emission
lines over the large span of 22\arcsec (85 kpc at $z=0.24$). The emission regions, roughly symmetric east and west of 
the nucleus, are detectable in the $g$ image from the Legacy Survey, and overlap in angle from the nucleus with the inner
double radio source but not the other double and hot spots. The host is a normal-looking early-type galaxy with very close 
companion in the HST image; we speculate that the emission clouds contributed to making it appear disklike in the
original selection from SDSS images. At this redshift our spectrum does not extend tp the H$\alpha$ region so we cannot
examine the clouds' location in the common BPT diagrams, but the high [O III]/H$\beta$ ratios and He II/H$\beta$ ratio in
the brightest parts suggest either AGN photoionization or interaction with radio jets. Two velocity components overlap
in projection near the nucleus, separated by 180 km s$^{-1}$. The emission properties are shown in Fig. \ref{fig-sdss0958}.

We generated an [O III] image from Legacy Survey data, using the $g$ and $z$ images to estimate the continuum
to be subtracted from the $r$ image containing [O III] and varying the scaling to minimize residual starlight, following the approach from
\cite{MNRAS2012}. The detailed spatial correspondence of faint filaments in the
ACS image with [O III] emission suggests that some of the extended component in the
HST F475W image comes from [O II] emission in this extended emission-line region (EELR), redshifted to 4627 \AA . 

\begin{figure*}
\includegraphics[width=145.mm,angle=90]{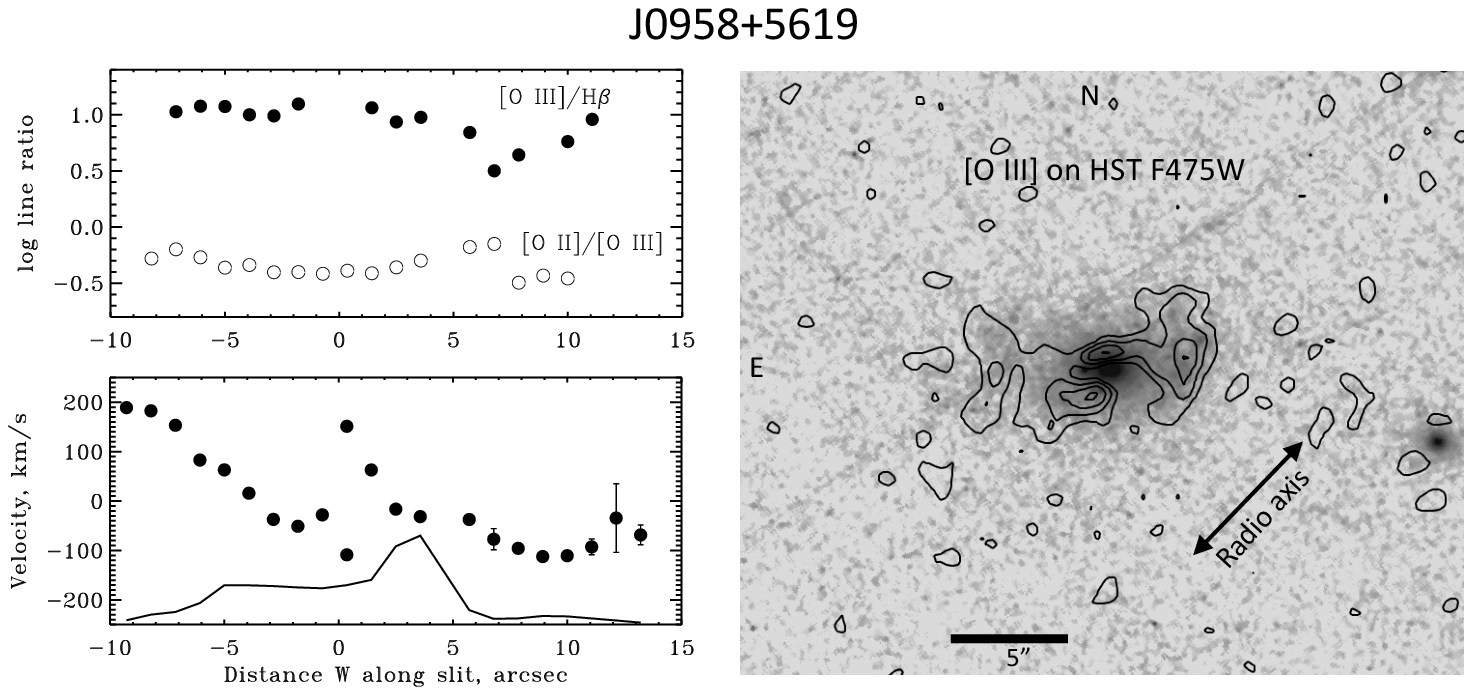}
\caption{Extended ionized-gas structures in J0958+5619. Emission-line ratios (upper left) and radial-velocity changes relative to the mean (left) are shown along the east-west spectroscopic slit. 
For orientation, the velocity panel includes a trace of [O III] intensity along the slit, with zero at the bottom. Radial velocities combine both
[O III] lines; the line profile is double
at the center and was deblended as a pair of Gaussian profiles. Velocity error bars in most locations are smaller than the plotted points.
The spectrum was summed over 3-pixel (1.07\arcsec) spans along the slit for these measurements. The image to the right overlays contours of [O III] emission derived from 
Legacy Survey broadband images on the HST ACS F475W image, showing how the brightest [O III] features off the nucleus align with structures in the HST image. The
double-headed arrow indicates the orientation of the radio-source axis, offset from the nucleus for clarity.}
\label{fig-sdss0958}
\end{figure*}

The emission lines are double-peaked near the nucleus, where we sum two [O III] components  for line ratios 
since we do not have enough information to deblend components for [O II]. The velocity profile could be interpreted as showing a small
central counter-rotating zone.

\section{Starting sample of Radio Galaxy Zoo SDRAGN candidates}\label{AppendixC}

This table presents the list of SDRAGN candidates from RGZ which were input to the voting
used in selecting the final Zoo Gems target objects. Since the RGZ Talk discussion remains 
available online, we include the internal identifiers assigned by Jean Tate wherever possible. Broadly, 
these identifiers beginning with $a,b,c$ are in that order of priority, with $s$ used for some special cases.
An object's status is listed as {\it Targ} if it was selected for the Zoo Gems target list and $Obs$ if it
was observed by the completion of this project. For galaxies covered by the SDSS imaging survey,
this table also lists the quantities used to evaluate observation selection effects. The {\it r} magnitude is taken from the
SDSS {\it modelMag\_r} parameter, axial ratio $b/a$ used the SDSS {\it expAB\_r} value from an exponential-disk fit to each image, and the column labelled Bulge gives the
bulge intensity fraction using {\it fracDev\_r}. 

\startlongtable
\begin{deluxetable*}{lrrrrrlll}
\tablecaption{Candidate SDRAGN systems and data status\label{tbl-candidates}}
 \tablehead{
 \colhead{Identifier}	& \colhead{RA (degrees)}	& \colhead{Dec (degrees)}	& \colhead{$r$}	& \colhead{$b/a$} & \colhead{Bulge}	& \colhead{Status}	
 & \colhead{Internal}	& \colhead{Cross-ID}	
} 
\startdata
SDSS  J001627.47+022602.1	&	4.11447	&	2.43394	&	15.65	&	0.71	&	1.00	&	Obs	&		&		\\
2MASX J01352563$-$0044467	&	23.85693	&	$-$0.74649	&	16.43	&	0.40	&	0.72	&	Targ	&	s023	&		\\
SDSS J014200.38+054204.7	&	25.50160	&	5.70131	&	20.46	&	0.69	&	1.00	&		&	c132	&		\\
SDSS J015744.20+024044.9	&	29.43420	&	2.67916	&	19.48	&	0.33	&	0.35	&		&	a316	&		\\
SDSS J020904.75+075004.5	&	32.26980	&	7.83459	&	17.60	&	0.92	&	1.00	&	Obs	&	b010	&		\\
						&	34.99472	&	1.93022	&	13.70	&	0.48	&	0.73	&	Obs	&	c046	&	UGC 1797		\\
SDSS J023832.67+023349.1	&	39.63614	&	2.56364	&	17.13	&	0.53	&	1.00	&		&	a098	&		\\
						&	52.15210	&	$-$28.69870	&		&		&		&	Targ	&		&	PKS 0236$-$288	\\
2MASX J03283652$-$2841554 	&	53.62400	&	$-$27.29090	&		&		&		&		&	s017	&		\\
2MASX J07523474+3550235	&	118.14466		&	35.83983	&	16.31	&	0.50	&	0.54	&		&	s002	&		\\
SDSS J075529.95+520450.6	&	118.87480		&	52.08072	&	17.86	&	1.00	&	0.11	&		&	b012	&		\\
SDSS J080217.94+112535.0	&	120.57476	&	11.42641	&	17.02	&	0.90	&	0.23	&		&	a036	&		\\
SDSS J080259.73+115709.7	&	120.74889	&	11.95270	&	17.98	&	0.33	&	0.76	&	Obs	&	a301	&		\\
SDSS J080658.46+062453.4	&	121.74362	&	6.41484	&	18.26	&	0.47	&	0.00	&	Obs	&	c125	&		\\
SDSS J081300.95+555235.1	&	123.25022	&	55.87772	&	16.32	&	0.52	&	0.21	&		&	a112	&		\\
SDSS J081303.10+552050.7	&	123.26293	&	55.34743	&	19.62	&	0.59	&	0.10	&	Obs	&	b016	&		\\
2MASX J08193601+5717523	&	124.90023	&	57.29787	&	16.33	&	0.36	&	0.53	&		&	s004	&		\\
SDSS J082227.34+314059.2	&	125.61395	&	31.68313	&	19.73	&	0.73	&	0.00	&		&	a263	&		\\
						&	125.80381	&	3.55038	&	15.66	&	0.86	&	1.00	&	Obs	&		&		\\
SDSS J082333.58+112946.1	&	125.88992	&	11.49616	&	19.43	&	0.60	&	0.46	&		&	b020	&		\\
SDSS J083132.73+384940.3	&	127.88640	&	38.82788	&	19.85	&	0.49	&	0.34	&		&	b025	&		\\
SDSS J083224.82+184855.4	&	128.10342	&	18.81540	&	16.13	&	0.63	&	1.00	&	Obs	&	c102	&		\\
SDSS J083351.28+045745.4	&	128.46368	&	4.96262	&	19.07	&	0.80	&	0.00	&	Obs	&	b027	&		\\
SDSS J083405.37+453439.2	&	128.52240	&	45.57757	&	18.96	&	0.50	&	0.58	&		&	a142	&		\\
SDSS J083454.61+452949.2	&	128.72757	&	45.49701	&	19.53	&	0.67	&	0.15	&		&	a231	&		\\
SDSS J084438.68+364307.8	&	131.16117		&	36.71884	&	17.85	&	0.47	&	0.06	&		&	b032	&		\\
SDSS J084759.90+124159.3	&	131.99959	&	12.69983	&	17.21	&	0.47	&	0.17	&	Obs	&	c043	&		\\
SDSS J084948.35+425536.9	&	132.45148	&	42.92692	&	20.16	&	0.33	&	0.00	&		&	b035	&		\\
SDSS J085138.23+564224.9	&	132.90932	&	56.70692	&	21.73	&	0.05	&	0.00	&		&	b036	&		\\
SDSS J085512.85+442532.1	&	133.80358	&	44.42560	&	19.73	&	0.40	&	0.22	&		&	a299	&		\\
SDSS J085549.15+420420.1	&	133.95479	&	42.07225	&	18.21	&	0.46	&	0.60	&	Obs	&	b039	&	B3 0852+422	\\
SDSS J085935.14+371023.1	&	134.89645	&	37.17310	&	19.45	&	0.46	&	0.34	&		&	c131	&		\\
SDSS J090147.17+164851.3	&	135.44655	&	16.81425	&	18.33	&	0.57	&	0.04	&	Obs	&	c084	&		\\
SDSS J090305.84+432820.4	&	135.77435	&	43.47233	&	19.21	&	0.28	&	0.28	&	Obs	&	c034	&		\\
SDSS J091129.36+284014.3	&	137.87234	&	28.67064	&	19.78	&	0.64	&	0.10	&		&	b040	&		\\
SDSS J091249.24+544509.4	&	138.20518	&	54.75264	&	19.67	&	0.45	&	0.00	&		&	a256	&		\\
SDSS J091251.98+210218.9	&	138.21659	&	21.03859	&	20.03	&	0.21	&	0.00	&		&	c122	&		\\
						&	138.68975	&	41.62065	&	15.26	&	0.73	&	0.93	&	Obs	&	a072	&	B3 0911+418	\\
SDSS J091523.47+084737.7	&	138.84782	&	8.79381	&	20.77	&	0.11	&	0.00	&		&	a254	&		\\
SDSS J091949.07+135910.7	&	139.95449	&	13.98632	&	20.59	&	0.66	&	0.00	&	Obs	&	a205	&		\\
SDSS J092605.17+465233.9	&	141.52156	&	46.87609	&	18.25	&	0.30	&	0.55	&	Obs	&	c118	&		\\
SDSS J092659.94+405804.1	&	141.74979	&	40.96782	&	20.96	&	0.15	&	1.00	&		&	b046	&		\\
SDSS J092754.96+351040.8	&	141.97904	&	35.17803	&	20.14	&	0.15	&	0.08	&		&	a310	&		\\
SDSS J092929.50+195349.7	&	142.37292	&	19.89715	&	18.82	&	0.65	&	0.13	&		&	b047	&		\\
SDSS J093852.70+524943.9	&	144.71960	&	52.82888	&	17.68	&	0.73	&	0.06	&		&	a025	&		\\
SDSS J093856.16+092801.6	&	144.73401	&	9.46713	&	18.28	&	0.78	&	0.00	&		&	b050	&		\\
SDSS J093856.30+151923.7	&	144.73460	&	15.32326	&	20.01	&	0.45	&	0.00	&		&	c098	&		\\
SDSS J094103.62+312618.7	&	145.26510	&	31.43854	&	19.61	&	0.33	&	0.21	&	Obs	&	a248	&	B2 0938+31A	\\
SDSS J094124.02+394441.8	&	145.35009	&	39.74496	&	16.12	&	0.51	&	1.00	&	Targ	&	s005	&		\\
SDSS J094348.17+363540.8	&	145.95075	&	36.59468	&	18.44	&	0.40	&	0.24	&		&	a309	&		\\
SDSS J095605.87+162829.9	&	149.02446	&	16.47500	&	19.26	&	0.39	&	0.24	&	Obs	&	a221	&		\\
SDSS J095833.44+561937.8	&	149.63935	&	56.32717	&	17.81	&	0.64	&	1.00	&	Obs	&	gp102	&		\\
SDSS J095846.43+260951.9	&	149.69349	&	26.16442	&	19.31	&	0.34	&	0.01	&		&	a271	&		\\
SDSS J101316.03+171715.0	&	153.31681	&	17.28752	&	20.97	&	0.45	&	0.00	&		&	b063	&		\\
SDSS J101529.72+550401.2	&	153.87385	&	55.06700	&	19.99	&	0.70	&	0.44	&		&	c117	&		\\
						&	153.99977	&	4.95476	&	14.49	&	0.67	&	0.51	&	Targ	&		&	Mkn 719	\\
SDSS J101704.05+081744.9	&	154.26690	&	8.29582	&	16.92	&	0.45	&	0.38	&	Targ	&	a056	&		\\
SDSS J102148.49+171926.3	&	155.45207	&	17.32398	&	18.74	&	0.28	&	0.00	&	Targ	&	c019	&		\\
SDSS J102444.01+460011.4	&	156.18341	&	46.00318	&	20.03	&	0.29	&	0.23	&		&	b066	&		\\
SDSS J102609.45+144932.7	&	156.53941	&	14.82576	&	17.41	&	0.56	&	0.43	&	Targ	&	b067	&		\\
SDSS J102747.03+100600.7	&	156.94599	&	10.10020	&	21.07	&	0.26	&	0.00	&		&	a174	&		\\
SDSS J103258.48+475437.3	&	158.24369	&	47.91037	&	19.93	&	0.21	&	0.00	&		&	b071	&		\\
SDSS J103409.34+073610.9	&	158.53894	&	7.60304	&	21.11	&	0.51	&	0.00	&		&	a345	&		\\
SDSS J103514.94+023203.6	&	158.81229	&	2.53434	&	17.90	&	0.52	&	0.55	&		&	a275	&		\\
						&	159.13283	&	2.36225	&	15.51	&	0.74	&	0.09	&	Targ	&		&	VII Zw 090	\\
SDSS J103635.49+483425.5	&	159.14791	&	48.57376	&	19.13	&	0.46	&	0.17	&		&	a183	&		\\
SDSS J103932.12+461205.3	&	159.88384	&	46.20149	&	16.54	&	0.63	&	0.38	&	Targ	&	a090	&	B3 1036+464	\\
SDSS J104018.62+281518.5	&	160.07761	&	28.25515	&	19.57	&	0.57	&	0.28	&		&	b076	&		\\
SDSS J104151.75+483149.1	&	160.46564	&	48.53033	&	20.26	&	0.72	&	0.00	&		&	b077	&		\\
SDSS J104201.36+090606.1	&	160.50569	&	9.10170	&	19.73	&	0.67	&	0.00	&		&	b079	&		\\
SDSS J104459.45+280459.1	&	161.24773	&	28.08308	&	19.17	&	0.22	&	0.74	&		&	a222	&		\\
SDSS J104616.66+181842.8	&	161.56945	&	18.31189	&	19.62	&	0.24	&	0.00	&		&	c101	&		\\
SDSS J105023.13+035250.1	&	162.59639	&	3.88061	&	19.78	&	0.67	&	1.00	&		&	b085	&		\\
SDSS J105452.02+552112.2	&	163.71677	&	55.35340	&	17.65	&	0.83	&	0.40	&		&	b091	&		\\
						&	165.56534	&	29.12368	&	15.41	&	0.71	&	1.00	&	Targ	&	b095	& B2 1059+29		\\
SDSS J111801.98+193209.1	&	169.50826	&	19.53586	&	18.37	&	0.83	&	0.00	&		&	b099	&		\\
SDSS J111956.19+072710.2	&	169.98416	&	7.45286	&	20.24	&	0.46	&	1.00	&		&	b100	&		\\
SDSS J112526.53+014301.5	&	171.36056	&	1.71711	&	19.31	&	0.72	&	1.00	&	Targ	&		&		\\
SDSS J112626.44+083821.4	&	171.61017	&	8.63930	&	17.56	&	0.53	&	0.00	&		&	b102	&		\\
SDSS J112811.63+241746.9	&	172.04847	&	24.29636	&	17.38	&	0.87	&	0.41	&	Obs	&	a300	&		\\
SDSS J112932.83+595518.3	&	172.38682	&	59.92175	&	18.97	&	0.29	&	0.02	&		&	a312	&		\\
SDSS J113032.02+104725.7	&	172.63345	&	10.79049	&	19.56	&	0.65	&	0.13	&		&	b104	&		\\
SDSS J113231.69+250352.7	&	173.13208	&	25.06466	&	19.97	&	0.31	&	0.31	&		&	a217	&		\\
SDSS J113648.57+125239.7	&	174.20239	&	12.87772	&	17.09	&	0.73	&	0.14	&	Obs	&	c114	&		\\
SDSS J113711.82+263335.5	&	174.29927	&	26.55988	&	18.58	&	0.93	&	0.00	&		&	c073	&		\\
SDSS J113814.89+323238.6	&	174.56206	&	32.54408	&	15.58	&	0.55	&	1.00	&		&	b107	&		\\
SDSS J114252.21+580814.6	&	175.71755	&	58.13741	&	18.64	&	0.27	&	0.00	&		&	b109	&		\\
						&	177.49463	&	41.20262	&	17.78	&	0.44	&	1.00	&	Targ	&		&	B3 1147+414	\\
SDSS J115029.33$-$021918.3	&	177.62222	&	$-$2.32177	&	18.11	&	0.46	&	0.84	&		&	s007	&		\\
SDSS J115259.71+332531.1	&	178.24880	&	33.42531	&	18.07	&	0.67	&	0.33	&		&	a329	&		\\
SDSS J115412.09+302241.7	&	178.55039	&	30.37826	&	19.49	&	0.66	&	0.24	&		&	b116	&		\\
SDSS J115437.43+114858.9	&	178.65599	&	11.81638	&	18.92	&	0.74	&	1.00	&		&	b117	&		\\
SDSS J115449.12+450006.4	&	178.70471	&	45.00179	&	19.94	&	0.72	&	0.51	&		&	a344	&		\\
SDSS J115634.73+081652.5	&	179.14473	&	8.28125	&	19.53	&	0.77	&	0.00	&		&	b118	&		\\
SDSS J115935.45+323749.7	&	179.89772	&	32.63050	&	19.70	&	0.38	&	0.28	&		&	b121	&		\\
SDSS J120058.72+313321.7   &	180.24467	&	31.55604	&	18.52	&	0.38	&	0.90	&	Targ	&		&	3C 268.2	\\
SDSS J120204.64+081240.8	&	180.51935	&	8.21134	&	19.32	&	0.39	&	0.00	&		&	b122	&		\\
SDSS J120232.58+361402.8	&	180.63579	&	36.23413	&	20.44	&	0.68	&	0.22	&		&	a261	&		\\
SDSS J120301.43+235319.9	&	180.75599	&	23.88887	&	17.94	&	0.63	&	1.00	&	Targ	&	a087	&		\\
SDSS J120339.20+275537.2	&	180.91335	&	27.92700	&	19.28	&	0.15	&	0.09	&	Targ	&	a297	&		\\
SDSS J120603.08+635731.4	&	181.51284	&	63.95872	&	19.14	&	0.71	&	0.00	&		&	b123	&		\\
SDSS J120625.91+210039.0	&	181.60796	&	21.01085	&	18.76	&	0.46	&	0.55	&		&	c039	&		\\
SDSS J120733.36+633540.4	&	181.88903	&	63.59457	&	20.20	&	0.52	&	0.12	&		&	b126	&		\\
SDSS J121124.78+062531.9	&	182.85326	&	6.42553	&	19.87	&	0.39	&	0.15	&		&	c025	&		\\
SDSS J121135.87+354417.4	&	182.89949	&	35.73818	&	15.14	&	0.62	&	1.00	&	Targ	&		&	KUG 1209+36	\\
						&	183.18710	&	37.52882	&	17.98	&	0.63	&	1.00	&	Targ	&		&	B3 1210+378	\\
SDSS J121257.95+250925.4	&	183.24147	&	25.15706	&	18.14	&	0.76	&	0.64	&	Targ	&	a188	&		\\
SDSS J122324.74+070301.9	&	185.85311		&	7.05054	&	17.93	&	0.44	&	0.34	&	Targ	&		&		\\
SDSS J122423.61+204937.9	&	186.09841	&	20.82721	&	19.55	&	0.83	&	1.00	&		&	b130	&		\\
SDSS J122640.22+253855.5	&	186.66760	&	25.64876	&	17.47	&	0.96	&	0.00	&	Targ	&	a069	&		\\
SDSS J122640.85+325937.2	&	186.67024	&	32.99369	&	20.24	&	0.28	&	0.13	&		&	b131	&		\\
SDSS J122705.12+194917.9	&	186.77135	&	19.82165	&	19.59	&	0.38	&	0.26	&		&	a303	&		\\
SDSS J123119.74+112242.9	&	187.83227	&	11.37859	&	18.84	&	0.58	&	0.70	&		&	a149	&		\\
SDSS J123500.21+214436.4	&	188.75089	&	21.74347	&	18.31	&	0.41	&	0.23	&		&	c126	&		\\
SDSS J123705.70+331559.6	&	189.27379	&	33.26658	&	20.10	&	0.20	&	0.00	&		&	a225	&		\\
SDSS J123825.49+075209.1	&	189.60622	&	7.86920	&	19.84	&	0.58	&	0.25	&		&	b138	&		\\
SDSS J123842.36+031825.2	&	189.67651	&	3.30701	&	19.67	&	0.21	&	0.00	&		&	b139	&		\\
SDSS J124127.34+435147.1	&	190.36395	&	43.86309	&	17.34	&	0.47	&	0.31	&		&	a078	&		\\
SDSS J125312.97+484235.0	&	193.30406	&	48.70975	&	18.68	&	0.95	&	0.00	&		&	b143	&		\\
SDSS J125503.07+433442.3	&	193.76281	&	43.57842	&	18.14	&	0.76	&	0.01	&		&	a293	&		\\
SDSS J125606.71+110027.6	&	194.02797	&	11.00769	&	19.82	&	0.73	&	0.00	&		&	a223	&		\\
SDSS J125743.34+385722.3	&	194.43059	&	38.95621	&	18.58	&	0.79	&	0.17	&		&	b144	&		\\
SDSS J125755.81+045852.3	&	194.48255	&	4.98120	&	18.59	&	0.47	&	0.50	&	Targ	&		&		\\
SDSS J125825.39+514226.0	&	194.60580	&	51.70724	&	19.61	&	0.60	&	1.00	&		&	b147	&		\\
SDSS J130122.06+052353.0	&	195.34195	&	5.39806	&	20.18	&	0.93	&	0.00	&		&	b150	&		\\
SDSS J130300.80+511954.7	&	195.75336	&	51.33188	&	18.37	&	0.60	&	0.09	&	Obs	&		&		\\
SDSS J130427.11+060319.7	&	196.11297		&	6.05549	&	20.24	&	0.43	&	0.14	&		&	b153	&		\\
SDSS J130849.99+481852.8	&	197.20831	&	48.31468	&	17.54	&	0.70	&	0.60	&		&	a094	&		\\
SDSS J131103.53+595504.0	&	197.76473	&	59.91778	&	18.88	&	0.65	&	0.53	&		&	a093	&		\\
SDSS J132259.87+270659.1	&	200.74948	&	27.11643	&	13.69	&	0.82	&	0.89	&	Obs	&	s008	&	IC 4234	\\
SDSS J132809.31+571023.3	&	202.03882	&	57.17315	&	16.81	&	0.38	&	0.14	&	Obs	&		&		\\
SDSS J132816.05+351051.9	&	202.06688	&	35.18109	&	19.78	&	0.72	&	0.61	&		&	b161	&		\\
SDSS J132821.46+453045.1	&	202.08943	&	45.51255	&	18.81	&	0.31	&	0.37	&		&	a253	&		\\
SDSS J133150.72+082720.3	&	202.96137	&	8.45566	&	19.89	&	0.69	&	0.00	&		&	b163	&		\\
SDSS J133529.16+430802.8	&	203.87153	&	43.13412	&	19.99	&	0.70	&	0.00	&		&	b164	&		\\
SDSS J133538.98+073000.3	&	203.91243	&	7.50009	&	20.85	&	0.20	&	0.00	&	Targ	&	a185	&		\\
SDSS J134147.14+043319.5	&	205.44646	&	4.55543	&	19.84	&	0.48	&	0.49	&		&	b170	&		\\
SDSS J134849.88+353213.1	&	207.20786	&	35.53699	&	20.48	&	0.17	&	0.97	&		&	b172	&		\\
SDSS J134900.13+452456.5	&	207.25058	&	45.71571	&	19.49	&	0.29	&	0.50	&	Obs	&	c103	&		\\
SDSS J135027.27+500925.0	&	207.61363	&	50.15695	&	19.38	&	0.26	&	0.12	&		&	a232	&		\\
SDSS J135055.29+333207.4	&	207.73040	&	33.53542	&	20.77	&	0.44	&	0.00	&	Targ	&		&		\\
SDSS J135414.65+153118.5	&	208.56107	&	15.52182	&	18.94	&	0.23	&	0.77	&		&	b175	&		\\
SDSS J135436.02+465701.4	&	208.65009	&	46.95042	&	20.01	&	0.31	&	0.17	&	Obs	&	b177	&	B3 1352+471	\\
SDSS J135706.06+491844.7	&	209.27528	&	49.31244	&	18.06	&	0.85	&	0.00	&		&	b179	&		\\
2MASX J13582118+0329136.0	&	209.58814	&	3.48704	&	15.48	&	0.82	&	0.84	&		&	s009	&		\\
SDSS J135856.20+432708.6	&	209.73420	&	43.45239	&	19.06	&	0.37	&	0.00	&		&	c035	&		\\
SDSS J140917.76+325925.4	&	212.32403	&	32.99040	&	18.31	&	0.71	&	0.25	&		&	b183	&		\\
SDSS J140932.69+415107.0	&	212.38622	&	41.85195	&	20.13	&	0.21	&	0.00	&		&	b184	&		\\
SDSS J141558.81+132023.7	&	213.99505	&	13.33992	&	19.05	&	0.41	&	0.43	&		&	a109	&		\\
SDSS J142305.86+215735.5	&	215.77444	&	21.95988	&	18.18	&	0.63	&	0.14	&		&	a314	&		\\
SDSS J142558.12+300931.2	&	216.49217	&	30.15869	&	18.22	&	0.56	&	0.24	&		&	a229	&		\\
SDSS J142606.19+402432.0	&	216.52580	&	40.40889	&	19.85	&	0.34	&	1.00	&		&	b191	&		\\
SDSS J142647.80+581756.3	&	216.69918	&	58.29900	&	20.08	&	0.46	&	0.00	&		&	c071	&		\\
SDSS J143340.75+635830.4	&	218.41979	&	63.97513	&	16.20	&	0.56	&	0.98	&		&	s010	&		\\
SDSS J143447.95+165631.8	&	218.69981	&	16.94218	&	21.21	&	0.80	&	0.68	&		&	b195	&		\\
SDSS J143507.46+225414.0	&	218.78112	    	&	22.90389	&	20.62	&	0.66	&	0.00	&		&	b196	&		\\
SDSS J143554.21+360853.0	&	218.97591	&	36.14807	&	18.66	&	0.93	&	0.06	&		&	a337	&		\\
SDSS J143643.83+080728.3	&	219.18263	&	8.12454	&	14.49	&	0.68	&	1.00	&	Targ	&	s011	&	CGCG 047$-$115	\\
SDSS J143854.46$-$015646.9	&	219.72696	&	$-$1.94637	&	18.52	&	0.30	&	0.01	&		&	s012	&		\\
SDSS J143934.48+531437.0	&	219.89368	&	53.24363	&	19.34	&	0.35	&	0.00	&		&	b197	&		\\
2MASX J14425854+5209018	&	220.74427	&	52.15058	&	17.92	&	0.67	&	1.00	&		&	s013	&		\\
SDSS J144714.49+424723.4	&	221.81040	&	42.78985	&	19.18	&	0.39	&	0.01	&		&	b198	&		\\
SDSS J144915.64+335835.7	&	222.31517	&	33.97660	&	19.87	&	0.46	&	0.16	&		&	b201	&		\\
SDSS J145218.00+052105.5	&	223.07501	&	5.35155	&	20.06	&	0.38	&	0.00	&		&	a212	&		\\
SDSS J145753.81+283218.7	&	224.47424	&	28.53854	&	16.22	&	0.68	&	0.48	&	 Obs	&	b203	&	4C +28.38	\\
SDSS J145847.52+561035.6	&	224.69800	&	56.17658	&	19.32	&	0.83	&	1.00	&		&	b204	&		\\
SDSS J145914.09+315740.9	&	224.80874	&	31.96139	&	18.09	&	0.23	&	0.27	&		&	b205	&		\\
SDSS J150326.21+304817.1	&	225.85924	&	30.80477	&	21.21	&	0.11	&	0.47	&		&	b206	&		\\
SDSS J150440.06+071109.6	&	226.16693	&	7.18602	&	18.95	&	1.00	&	1.00	&		&	b207	&		\\
SDSS J150543.38+195304.0	&	226.43077	&	19.88447	&	20.01	&	0.53	&	0.00	&		&	a317	&		\\
SDSS J150903.21+515247.9	&	227.26340	&	51.87999	&	19.59	&	0.67	&	1.00	&	Obs	&	a154	&		\\
SDSS J151356.33+572523	&	228.48475	&	57.42310	&	18.04	&	0.42	&	0.82	&		&	b210	&		\\
SDSS J151659.24+051751.5	&	229.24684	&	5.29764	&	15.29	&	0.57	&	0.93	&	Obs	&	c051	&		\\
SDSS J151918.35+103505.1	&	229.82646	&	10.58477	&	20.02	&	0.55	&	0.07	&		&	c127	&		\\
SDSS J155046.42+173630.2	&	229.87631	&	6.56453	&	19.57	&	0.51	&	0.00	&	Targ	&		&		\\
SDSS J152207.70+325113.3	&	230.53209	&	32.85372	&	22.54	&	0.60	&	0.00	&		&	s021	&		\\
SDSS J152838.43+242025.6	&	232.15475	&	24.23822	&	18.95	&	0.66	&	0.00	&	Targ	&		&		\\
SDSS J152955.47+344955.2	&	232.48115		&	34.83201	&	19.92	&	0.11	&	0.00	&		&	a250	&		\\
SDSS J152955.55+344943.0	&	232.48146	&	34.82862	&	20.63	&	1.00	&	1.00	&		&	a283	&		\\
SDSS J153200.95+272528.9	&	233.00397	&	27.42470	&	18.66	&	0.91	&	0.16	&	Targ	&	b217	&		\\
SDSS J153233.17+241525.7	&	233.13823	&	24.25716	&	19.80	&	0.71	&	0.61	&		&	b218	&		\\
SDSS J153250.68+573424.9	&	233.21118		&	57.57359	&	17.39	&	0.83	&	0.15	&		&	a080J	&		\\
SDSS J153637.22+120746.1	&	234.15512	&	12.12948	&	19.43	&	0.55	&	0.05	&		&	a286	&		\\
SDSS J153917.23+075953.1	&	234.82183	&	7.99809	&	19.69	&	0.22	&	0.00	&		&	a276	&		\\
SDSS J154053.15+533002.0	&	235.22150	&	53.50057	&	18.15	&	0.88	&	0.00	&		&	a339	&		\\
SDSS J154502.82+513500.8	&	236.26176	&	51.58356	&	17.55	&	0.68	&	1.00	&		&	b224	&		\\
SDSS J154814.62+483502.0	&	237.06093	&	48.58390	&	19.86	&	0.46	&	1.00	&		&	a192	&		\\
SDSS J154826.32+581313.9	&	237.10968	&	58.22055	&	19.06	&	0.72	&	1.00	&		&	b226	&		\\
SDSS J155046.42+173630.2	&	237.69342	&	17.60840	&	17.28	&	0.80	&	0.68	&	Targ	&		&		\\
SDSS J155126.62+534814.7	&	237.86092	&	53.80410	&	20.55	&	0.06	&	0.00	&		&	b228	&		\\
SDSS J155203.47+073141.9	&	238.01446	&	7.52833	&	18.73	&	0.53	&	0.20	&		&	a331	&		\\
SDSS J155549.80+241409.9	&	238.95751	&	24.23610	&	21.52	&	0.44	&	0.00	&		&	b229	&		\\
SDSS J155619.77+511848.8	&	239.08240	&	51.31356	&	18.32	&	0.83	&	0.60	&		&	a139	&		\\
SDSS J155641.93+245941.1	&	239.17475	&	24.99477	&	20.58	&	0.48	&	0.96	&		&	a177	&		\\
SDSS J160057.30+051438.4	&	240.23876	&	5.24402	&	21.24	&	0.11	&	0.00	&		&	c092	&		\\
SDSS J161358.61+301809.4	&	243.49423	&	30.30263	&	19.00	&	0.40	&	0.00	&	 Obs	&	a325	&	B2 1611+30	\\
SDSS J161449.71+321054.5	&	243.70714	&	32.18181	&	20.45	&	0.75	&	0.61	&		&	b233	&		\\
SDSS J161948.86+300607.8	&	244.95360	&	30.10218	&	17.94	&	0.93	&	0.58	&	Targ	&		&		\\
SDSS J163218.08+475616.3	&	248.07534	&	47.93787	&	22.63	&	0.09	&	1.00	&		&	b237	&		\\
SDSS J163300.85+084736.4	&	248.25358	&	8.79346	&	17.84	&	0.58	&	0.49	&	Obs	&	a287	&		\\
SDSS J163624.97+243230.8	&	249.10407	&	24.54191	&	19.69	&	0.38	&	0.00	&	Obs	&	b238	&		\\
SDSS J163901.65+243139.8	&	249.75690	&	24.52775	&	21.86	&	0.74	&	0.00	&		&	b239	&		\\
SDSS J164351.20+424953.6	&	250.96333	&	42.83157	&	19.34	&	0.66	&	0.00	&		&	b240	&		\\
SDSS J164604.10+513024.0	&	251.51710	&	51.50668	&	21.65	&	0.05	&	0.96	&		&	b241	&		\\
2MASX J16462838+3831156	&	251.61841	&	38.52112	&	16.66	&	0.64	&	0.95	&	Obs	&	s014	&	B2 1644+38	\\
SDSS J165304.66+240409.2	&	253.26944	&	24.06925	&	17.93	&	0.41	&	0.25	&	Targ	&	b245	&		\\
2MASX J16562058+6407529	&	254.08586	&	64.13138	&	16.96	&	0.44	&	0.53	&	Obs	&	s015	&		\\
SDSS J170008.72+291903.7	&	255.03634	&	29.31771	&	16.52	&	0.73	&	0.00	&		&	s022	&		\\
SDSS J170953.87+410503.8	&	257.47449	&	41.08441	&	18.34	&	0.49	&	0.38	&		&	b248	&		\\
SDSS J171845.25+364911.0	&	259.68856	&	36.81973	&	21.84	&	0.29	&	0.95	&		&	b250	&		\\
SDSS J172107.89+262432.1	&	260.28291	&	26.40894	&	17.77	&	0.94	&	1.00	&	Obs	&		&		\\
SDSS J172853.38+342348.1	&	262.22242	&	34.39671	&	19.71	&	0.69	&	0.29	&		&	a307	&		\\
SDSS J214110.61+082132.6	&	325.29424	&	8.35906	&	19.25	&	0.45	&	0.70	&	Obs	&		&		\\
SDSS J214530.21+104949.2	&	326.37591	&	10.83036	&	19.03	&	0.30	&	1.00	&		&	b255	&		\\
SDSS J231620.15$-$010207.3	&	349.08399	&	$-$1.03537	&	18.79	&	0.62	&	0.63	&		&	s016	&		\\
\enddata
\tablecomments{Objects observed in the HST filler program are denoted by ``Obs"; ``Targ" indicates objects that
were selected for the HST program but not observed due to the essentially random selection of targets for
scheduling. Both these subsets reflect the observational selections going into the HST sample.}
\end{deluxetable*}



\end{document}